%% file: 2024_congedo_lensmc.tex
\documentclass[longauth]{aa}

\usepackage{euclid}
\usepackage{macros}

\begin{document}

\title{\Euclid preparation}
\subtitle{LIII. LensMC, weak lensing cosmic shear measurement with forward modelling and Markov Chain Monte Carlo sampling}

\input{tex/authors}


\abstract{\lensmc is a weak lensing shear measurement method developed for \euclid and Stage-IV surveys.
It is based on forward modelling in order to deal with
convolution by a point spread function (PSF) with comparable size to many
galaxies,
sampling the posterior distribution of galaxy parameters via Markov chain Monte Carlo, and marginalisation over nuisance parameters for each of the 1.5 billion galaxies observed by \euclid.
We quantified the scientific performance through high-fidelity images
based on the \euclid Flagship simulations and emulation of the \euclid VIS images,
realistic clustering with a mean surface number density of $\qty{250}{arcmin^{-2}}$ ($\IE<29.5$) for galaxies,
and $\qty{6}{arcmin^{-2}}$ ($\IE<26$) for stars,
and a diffraction-limited chromatic PSF with a full width at half maximum of $\ang{;;0.2}$
and spatial variation across the field of view.
\lensmc measured objects with a density of $\qty{90}{arcmin^{-2}}$ ($\IE<26.5$) in $4500\,\unit{deg^2}$.
The total shear bias was broken down into measurement (our main focus here) and selection effects (which will be addressed in future work).
We found
measurement multiplicative and additive biases of $m_1=(-3.6\pm0.2)\times10^{-3}$, $m_2=(-4.3\pm0.2)\times10^{-3}$, 
$c_1=(-1.78\pm0.03)\times10^{-4}$, and $c_2=(0.09\pm0.03)\times10^{-4}$;
a large detection bias with a multiplicative component of $1.2\times10^{-2}$ and an additive component of $-3\times10^{-4}$;
and a measurement PSF leakage
of $\alpha_1=(-9\pm3)\times10^{-4}$ and $\alpha_2=(2\pm3)\times10^{-4}$.
When model bias is suppressed, the obtained measurement biases are close to \euclid requirement and largely dominated by undetected faint galaxies ($-5\times10^{-3}$).
Although significant, model bias will be straightforward to calibrate given its weak sensitivity on galaxy morphology parameters.
\lensmc is publicly available at \href{https://gitlab.com/gcongedo/LensMC}{gitlab.com/gcongedo/LensMC}.}


\titlerunning{\Euclid\ preparation LIII. \lensmc}
\authorrunning{Euclid Collaboration: G. Congedo et al.}

\maketitle


\input{tex/sec1_introduction}
\input{tex/sec2_method}
\input{tex/sec3_simulations}
\input{tex/sec4_results}
\input{tex/sec5_conclusions}

\begin{acknowledgements}
\input{tex/acknowledgements.tex}
\end{acknowledgements}

\bibliographystyle{aa_url}
\bibliography{references}

\begin{appendix}
\input{tex/app1_templates}
\input{tex/app2_likelihood}
\input{tex/app3_mcmc}
\input{tex/app4_validation}
\input{tex/app5_shear_bias}
\end{appendix}


\end{document}

%% file: tex/authors.tex
\author{Euclid Collaboration: G.~Congedo\orcid{0000-0003-2508-0046}\thanks{\email{giuseppe.congedo@ed.ac.uk}}\inst{\ref{aff1}}
\and L.~Miller\orcid{0000-0002-3376-6200}\inst{\ref{aff2}}
\and A.~N.~Taylor\inst{\ref{aff1}}
\and N.~Cross\inst{\ref{aff1}}
\and C.~A.~J.~Duncan\inst{\ref{aff3},\ref{aff2}}
\and T.~Kitching\orcid{0000-0002-4061-4598}\inst{\ref{aff4}}
\and N.~Martinet\orcid{0000-0003-2786-7790}\inst{\ref{aff5}}
\and S.~Matthew\inst{\ref{aff1}}
\and T.~Schrabback\orcid{0000-0002-6987-7834}\inst{\ref{aff6}}
\and M.~Tewes\orcid{0000-0002-1155-8689}\inst{\ref{aff7}}
\and N.~Welikala\inst{\ref{aff1}}
\and N.~Aghanim\inst{\ref{aff8}}
\and A.~Amara\inst{\ref{aff9}}
\and S.~Andreon\orcid{0000-0002-2041-8784}\inst{\ref{aff10}}
\and N.~Auricchio\inst{\ref{aff11}}
\and M.~Baldi\orcid{0000-0003-4145-1943}\inst{\ref{aff12},\ref{aff11},\ref{aff13}}
\and S.~Bardelli\orcid{0000-0002-8900-0298}\inst{\ref{aff11}}
\and R.~Bender\orcid{0000-0001-7179-0626}\inst{\ref{aff14},\ref{aff15}}
\and C.~Bodendorf\inst{\ref{aff14}}
\and D.~Bonino\inst{\ref{aff16}}
\and E.~Branchini\orcid{0000-0002-0808-6908}\inst{\ref{aff17},\ref{aff18},\ref{aff10}}
\and M.~Brescia\orcid{0000-0001-9506-5680}\inst{\ref{aff19},\ref{aff20},\ref{aff21}}
\and J.~Brinchmann\orcid{0000-0003-4359-8797}\inst{\ref{aff22}}
\and S.~Camera\orcid{0000-0003-3399-3574}\inst{\ref{aff23},\ref{aff24},\ref{aff16}}
\and V.~Capobianco\orcid{0000-0002-3309-7692}\inst{\ref{aff16}}
\and C.~Carbone\orcid{0000-0003-0125-3563}\inst{\ref{aff25}}
\and V.~F.~Cardone\inst{\ref{aff26},\ref{aff27}}
\and J.~Carretero\orcid{0000-0002-3130-0204}\inst{\ref{aff28},\ref{aff29}}
\and S.~Casas\orcid{0000-0002-4751-5138}\inst{\ref{aff30}}
\and F.~J.~Castander\orcid{0000-0001-7316-4573}\inst{\ref{aff31},\ref{aff32}}
\and M.~Castellano\orcid{0000-0001-9875-8263}\inst{\ref{aff26}}
\and S.~Cavuoti\orcid{0000-0002-3787-4196}\inst{\ref{aff20},\ref{aff21}}
\and A.~Cimatti\inst{\ref{aff33}}
\and C.~J.~Conselice\inst{\ref{aff3}}
\and L.~Conversi\orcid{0000-0002-6710-8476}\inst{\ref{aff34},\ref{aff35}}
\and Y.~Copin\orcid{0000-0002-5317-7518}\inst{\ref{aff36}}
\and F.~Courbin\orcid{0000-0003-0758-6510}\inst{\ref{aff37}}
\and H.~M.~Courtois\orcid{0000-0003-0509-1776}\inst{\ref{aff38}}
\and M.~Cropper\inst{\ref{aff4}}
\and A.~Da~Silva\orcid{0000-0002-6385-1609}\inst{\ref{aff39},\ref{aff40}}
\and H.~Degaudenzi\orcid{0000-0002-5887-6799}\inst{\ref{aff41}}
\and A.~M.~Di~Giorgio\orcid{0000-0002-4767-2360}\inst{\ref{aff42}}
\and J.~Dinis\inst{\ref{aff40},\ref{aff39}}
\and F.~Dubath\orcid{0000-0002-6533-2810}\inst{\ref{aff41}}
\and X.~Dupac\inst{\ref{aff35}}
\and M.~Farina\inst{\ref{aff42}}
\and S.~Farrens\orcid{0000-0002-9594-9387}\inst{\ref{aff43}}
\and S.~Ferriol\inst{\ref{aff36}}
\and P.~Fosalba\orcid{0000-0002-1510-5214}\inst{\ref{aff32},\ref{aff44}}
\and M.~Frailis\orcid{0000-0002-7400-2135}\inst{\ref{aff45}}
\and E.~Franceschi\orcid{0000-0002-0585-6591}\inst{\ref{aff11}}
\and S.~Galeotta\orcid{0000-0002-3748-5115}\inst{\ref{aff45}}
\and B.~Garilli\orcid{0000-0001-7455-8750}\inst{\ref{aff25}}
\and B.~Gillis\orcid{0000-0002-4478-1270}\inst{\ref{aff1}}
\and C.~Giocoli\orcid{0000-0002-9590-7961}\inst{\ref{aff11},\ref{aff46}}
\and A.~Grazian\orcid{0000-0002-5688-0663}\inst{\ref{aff47}}
\and F.~Grupp\inst{\ref{aff14},\ref{aff15}}
\and S.~V.~H.~Haugan\orcid{0000-0001-9648-7260}\inst{\ref{aff48}}
\and M.~S.~Holliman\inst{\ref{aff49}}
\and W.~Holmes\inst{\ref{aff50}}
\and F.~Hormuth\inst{\ref{aff51}}
\and A.~Hornstrup\orcid{0000-0002-3363-0936}\inst{\ref{aff52},\ref{aff53}}
\and P.~Hudelot\inst{\ref{aff54}}
\and K.~Jahnke\orcid{0000-0003-3804-2137}\inst{\ref{aff55}}
\and E.~Keih\"anen\orcid{0000-0003-1804-7715}\inst{\ref{aff56}}
\and S.~Kermiche\orcid{0000-0002-0302-5735}\inst{\ref{aff57}}
\and A.~Kiessling\orcid{0000-0002-2590-1273}\inst{\ref{aff50}}
\and M.~Kilbinger\orcid{0000-0001-9513-7138}\inst{\ref{aff58}}
\and B.~Kubik\inst{\ref{aff36}}
\and K.~Kuijken\orcid{0000-0002-3827-0175}\inst{\ref{aff59}}
\and M.~K\"ummel\orcid{0000-0003-2791-2117}\inst{\ref{aff15}}
\and M.~Kunz\orcid{0000-0002-3052-7394}\inst{\ref{aff60}}
\and H.~Kurki-Suonio\orcid{0000-0002-4618-3063}\inst{\ref{aff61},\ref{aff62}}
\and S.~Ligori\orcid{0000-0003-4172-4606}\inst{\ref{aff16}}
\and P.~B.~Lilje\orcid{0000-0003-4324-7794}\inst{\ref{aff48}}
\and V.~Lindholm\orcid{0000-0003-2317-5471}\inst{\ref{aff61},\ref{aff62}}
\and I.~Lloro\inst{\ref{aff63}}
\and D.~Maino\inst{\ref{aff64},\ref{aff25},\ref{aff65}}
\and E.~Maiorano\orcid{0000-0003-2593-4355}\inst{\ref{aff11}}
\and O.~Mansutti\orcid{0000-0001-5758-4658}\inst{\ref{aff45}}
\and O.~Marggraf\orcid{0000-0001-7242-3852}\inst{\ref{aff7}}
\and K.~Markovic\orcid{0000-0001-6764-073X}\inst{\ref{aff50}}
\and F.~Marulli\orcid{0000-0002-8850-0303}\inst{\ref{aff66},\ref{aff11},\ref{aff13}}
\and R.~Massey\orcid{0000-0002-6085-3780}\inst{\ref{aff67}}
\and S.~Maurogordato\inst{\ref{aff68}}
\and H.~J.~McCracken\orcid{0000-0002-9489-7765}\inst{\ref{aff54}}
\and E.~Medinaceli\orcid{0000-0002-4040-7783}\inst{\ref{aff11}}
\and S.~Mei\orcid{0000-0002-2849-559X}\inst{\ref{aff69}}
\and M.~Melchior\inst{\ref{aff70}}
\and M.~Meneghetti\orcid{0000-0003-1225-7084}\inst{\ref{aff11},\ref{aff13}}
\and E.~Merlin\orcid{0000-0001-6870-8900}\inst{\ref{aff26}}
\and G.~Meylan\inst{\ref{aff37}}
\and M.~Moresco\orcid{0000-0002-7616-7136}\inst{\ref{aff66},\ref{aff11}}
\and B.~Morin\inst{\ref{aff58}}
\and L.~Moscardini\orcid{0000-0002-3473-6716}\inst{\ref{aff66},\ref{aff11},\ref{aff13}}
\and E.~Munari\orcid{0000-0002-1751-5946}\inst{\ref{aff45}}
\and S.-M.~Niemi\inst{\ref{aff71}}
\and J.~W.~Nightingale\orcid{0000-0002-8987-7401}\inst{\ref{aff72},\ref{aff67}}
\and C.~Padilla\orcid{0000-0001-7951-0166}\inst{\ref{aff28}}
\and S.~Paltani\inst{\ref{aff41}}
\and F.~Pasian\inst{\ref{aff45}}
\and K.~Pedersen\inst{\ref{aff73}}
\and W.~J.~Percival\orcid{0000-0002-0644-5727}\inst{\ref{aff74},\ref{aff75},\ref{aff76}}
\and V.~Pettorino\inst{\ref{aff77}}
\and S.~Pires\orcid{0000-0002-0249-2104}\inst{\ref{aff43}}
\and G.~Polenta\orcid{0000-0003-4067-9196}\inst{\ref{aff78}}
\and M.~Poncet\inst{\ref{aff79}}
\and L.~A.~Popa\inst{\ref{aff80}}
\and L.~Pozzetti\orcid{0000-0001-7085-0412}\inst{\ref{aff11}}
\and F.~Raison\orcid{0000-0002-7819-6918}\inst{\ref{aff14}}
\and R.~Rebolo\inst{\ref{aff81},\ref{aff82}}
\and A.~Renzi\orcid{0000-0001-9856-1970}\inst{\ref{aff83},\ref{aff84}}
\and J.~Rhodes\inst{\ref{aff50}}
\and G.~Riccio\inst{\ref{aff20}}
\and E.~Romelli\orcid{0000-0003-3069-9222}\inst{\ref{aff45}}
\and M.~Roncarelli\orcid{0000-0001-9587-7822}\inst{\ref{aff11}}
\and E.~Rossetti\inst{\ref{aff12}}
\and R.~Saglia\orcid{0000-0003-0378-7032}\inst{\ref{aff15},\ref{aff14}}
\and D.~Sapone\orcid{0000-0001-7089-4503}\inst{\ref{aff85}}
\and B.~Sartoris\inst{\ref{aff15},\ref{aff45}}
\and P.~Schneider\orcid{0000-0001-8561-2679}\inst{\ref{aff7}}
\and A.~Secroun\orcid{0000-0003-0505-3710}\inst{\ref{aff57}}
\and G.~Seidel\orcid{0000-0003-2907-353X}\inst{\ref{aff55}}
\and S.~Serrano\orcid{0000-0002-0211-2861}\inst{\ref{aff32},\ref{aff31},\ref{aff86}}
\and C.~Sirignano\orcid{0000-0002-0995-7146}\inst{\ref{aff83},\ref{aff84}}
\and G.~Sirri\orcid{0000-0003-2626-2853}\inst{\ref{aff13}}
\and L.~Stanco\orcid{0000-0002-9706-5104}\inst{\ref{aff84}}
\and P.~Tallada-Cresp\'{i}\orcid{0000-0002-1336-8328}\inst{\ref{aff87},\ref{aff29}}
\and D.~Tavagnacco\orcid{0000-0001-7475-9894}\inst{\ref{aff45}}
\and I.~Tereno\inst{\ref{aff39},\ref{aff88}}
\and R.~Toledo-Moreo\orcid{0000-0002-2997-4859}\inst{\ref{aff89}}
\and F.~Torradeflot\orcid{0000-0003-1160-1517}\inst{\ref{aff29},\ref{aff87}}
\and I.~Tutusaus\orcid{0000-0002-3199-0399}\inst{\ref{aff90}}
\and E.~A.~Valentijn\inst{\ref{aff91}}
\and L.~Valenziano\orcid{0000-0002-1170-0104}\inst{\ref{aff11},\ref{aff92}}
\and T.~Vassallo\orcid{0000-0001-6512-6358}\inst{\ref{aff15},\ref{aff45}}
\and A.~Veropalumbo\orcid{0000-0003-2387-1194}\inst{\ref{aff10},\ref{aff18}}
\and Y.~Wang\orcid{0000-0002-4749-2984}\inst{\ref{aff93}}
\and J.~Weller\orcid{0000-0002-8282-2010}\inst{\ref{aff15},\ref{aff14}}
\and G.~Zamorani\orcid{0000-0002-2318-301X}\inst{\ref{aff11}}
\and J.~Zoubian\inst{\ref{aff57}}
\and E.~Zucca\orcid{0000-0002-5845-8132}\inst{\ref{aff11}}
\and A.~Biviano\orcid{0000-0002-0857-0732}\inst{\ref{aff45},\ref{aff94}}
\and M.~Bolzonella\orcid{0000-0003-3278-4607}\inst{\ref{aff11}}
\and A.~Boucaud\orcid{0000-0001-7387-2633}\inst{\ref{aff69}}
\and E.~Bozzo\orcid{0000-0002-8201-1525}\inst{\ref{aff41}}
\and C.~Burigana\orcid{0000-0002-3005-5796}\inst{\ref{aff95},\ref{aff92}}
\and C.~Colodro-Conde\inst{\ref{aff81}}
\and D.~Di~Ferdinando\inst{\ref{aff13}}
\and J.~Graci\'{a}-Carpio\inst{\ref{aff14}}
\and N.~Mauri\orcid{0000-0001-8196-1548}\inst{\ref{aff33},\ref{aff13}}
\and C.~Neissner\inst{\ref{aff28},\ref{aff29}}
\and A.~A.~Nucita\inst{\ref{aff96},\ref{aff97},\ref{aff98}}
\and Z.~Sakr\orcid{0000-0002-4823-3757}\inst{\ref{aff99},\ref{aff90},\ref{aff100}}
\and V.~Scottez\inst{\ref{aff101},\ref{aff102}}
\and M.~Tenti\orcid{0000-0002-4254-5901}\inst{\ref{aff13}}
\and M.~Viel\orcid{0000-0002-2642-5707}\inst{\ref{aff94},\ref{aff45},\ref{aff103},\ref{aff104},\ref{aff105}}
\and M.~Wiesmann\inst{\ref{aff48}}
\and Y.~Akrami\orcid{0000-0002-2407-7956}\inst{\ref{aff106},\ref{aff107}}
\and V.~Allevato\inst{\ref{aff20}}
\and S.~Anselmi\orcid{0000-0002-3579-9583}\inst{\ref{aff83},\ref{aff84},\ref{aff108}}
\and C.~Baccigalupi\orcid{0000-0002-8211-1630}\inst{\ref{aff103},\ref{aff45},\ref{aff104},\ref{aff94}}
\and M.~Ballardini\inst{\ref{aff109},\ref{aff110},\ref{aff11}}
\and S.~Borgani\orcid{0000-0001-6151-6439}\inst{\ref{aff111},\ref{aff94},\ref{aff45},\ref{aff104}}
\and A.~S.~Borlaff\orcid{0000-0003-3249-4431}\inst{\ref{aff112},\ref{aff113},\ref{aff114}}
\and S.~Bruton\orcid{0000-0002-6503-5218}\inst{\ref{aff115}}
\and R.~Cabanac\orcid{0000-0001-6679-2600}\inst{\ref{aff90}}
\and A.~Cappi\inst{\ref{aff11},\ref{aff68}}
\and C.~S.~Carvalho\inst{\ref{aff88}}
\and G.~Castignani\orcid{0000-0001-6831-0687}\inst{\ref{aff66},\ref{aff11}}
\and T.~Castro\orcid{0000-0002-6292-3228}\inst{\ref{aff45},\ref{aff104},\ref{aff94},\ref{aff105}}
\and G.~Ca\~{n}as-Herrera\orcid{0000-0003-2796-2149}\inst{\ref{aff71},\ref{aff116}}
\and K.~C.~Chambers\orcid{0000-0001-6965-7789}\inst{\ref{aff117}}
\and A.~R.~Cooray\orcid{0000-0002-3892-0190}\inst{\ref{aff118}}
\and J.~Coupon\inst{\ref{aff41}}
\and S.~Davini\inst{\ref{aff18}}
\and G.~De~Lucia\orcid{0000-0002-6220-9104}\inst{\ref{aff45}}
\and G.~Desprez\inst{\ref{aff119}}
\and S.~Di~Domizio\orcid{0000-0003-2863-5895}\inst{\ref{aff17},\ref{aff18}}
\and H.~Dole\orcid{0000-0002-9767-3839}\inst{\ref{aff8}}
\and A.~D\'{i}az-S\'{a}nchez\orcid{0000-0003-0748-4768}\inst{\ref{aff120}}
\and J.~A.~Escartin~Vigo\inst{\ref{aff14}}
\and S.~Escoffier\orcid{0000-0002-2847-7498}\inst{\ref{aff57}}
\and I.~Ferrero\orcid{0000-0002-1295-1132}\inst{\ref{aff48}}
\and F.~Finelli\orcid{0000-0002-6694-3269}\inst{\ref{aff11},\ref{aff92}}
\and L.~Gabarra\inst{\ref{aff83},\ref{aff84}}
\and J.~Garc\'ia-Bellido\orcid{0000-0002-9370-8360}\inst{\ref{aff106}}
\and E.~Gaztanaga\orcid{0000-0001-9632-0815}\inst{\ref{aff31},\ref{aff32},\ref{aff9}}
\and F.~Giacomini\orcid{0000-0002-3129-2814}\inst{\ref{aff13}}
\and G.~Gozaliasl\orcid{0000-0002-0236-919X}\inst{\ref{aff121},\ref{aff61}}
\and D.~Guinet\orcid{0000-0002-8132-6509}\inst{\ref{aff36}}
\and A.~Hall\orcid{0000-0002-3139-8651}\inst{\ref{aff1}}
\and H.~Hildebrandt\orcid{0000-0002-9814-3338}\inst{\ref{aff122}}
\and S.~Ili\'c\orcid{0000-0003-4285-9086}\inst{\ref{aff123},\ref{aff79},\ref{aff90}}
\and A.~Jimenez~Mu\~noz\orcid{0009-0004-5252-185X}\inst{\ref{aff124}}
\and S.~Joudaki\inst{\ref{aff9},\ref{aff74},\ref{aff75}}
\and J.~J.~E.~Kajava\orcid{0000-0002-3010-8333}\inst{\ref{aff125},\ref{aff126}}
\and V.~Kansal\inst{\ref{aff127},\ref{aff128},\ref{aff129}}
\and D.~Karagiannis\inst{\ref{aff130}}
\and C.~C.~Kirkpatrick\inst{\ref{aff56}}
\and L.~Legrand\orcid{0000-0003-0610-5252}\inst{\ref{aff60}}
\and J.~Macias-Perez\inst{\ref{aff124}}
\and G.~Maggio\orcid{0000-0003-4020-4836}\inst{\ref{aff45}}
\and M.~Magliocchetti\orcid{0000-0001-9158-4838}\inst{\ref{aff42}}
\and R.~Maoli\orcid{0000-0002-6065-3025}\inst{\ref{aff131},\ref{aff26}}
\and M.~Martinelli\orcid{0000-0002-6943-7732}\inst{\ref{aff26},\ref{aff27}}
\and C.~J.~A.~P.~Martins\orcid{0000-0002-4886-9261}\inst{\ref{aff132},\ref{aff22}}
\and M.~Maturi\orcid{0000-0002-3517-2422}\inst{\ref{aff99},\ref{aff133}}
\and L.~Maurin\orcid{0000-0002-8406-0857}\inst{\ref{aff8}}
\and R.~B.~Metcalf\orcid{0000-0003-3167-2574}\inst{\ref{aff66},\ref{aff11}}
\and M.~Migliaccio\inst{\ref{aff134},\ref{aff135}}
\and P.~Monaco\inst{\ref{aff111},\ref{aff45},\ref{aff104},\ref{aff94}}
\and G.~Morgante\inst{\ref{aff11}}
\and S.~Nadathur\orcid{0000-0001-9070-3102}\inst{\ref{aff9}}
\and L.~Patrizii\inst{\ref{aff13}}
\and A.~Peel\orcid{0000-0003-0488-8978}\inst{\ref{aff37}}
\and A.~Pezzotta\inst{\ref{aff14}}
\and V.~Popa\inst{\ref{aff80}}
\and C.~Porciani\orcid{0000-0002-7797-2508}\inst{\ref{aff7}}
\and D.~Potter\orcid{0000-0002-0757-5195}\inst{\ref{aff136}}
\and M.~P\"{o}ntinen\orcid{0000-0001-5442-2530}\inst{\ref{aff61}}
\and P.~Reimberg\orcid{0000-0003-3410-0280}\inst{\ref{aff101}}
\and P.-F.~Rocci\inst{\ref{aff8}}
\and A.~G.~S\'anchez\orcid{0000-0003-1198-831X}\inst{\ref{aff14}}
\and J.~A.~Schewtschenko\inst{\ref{aff1}}
\and A.~Schneider\orcid{0000-0001-7055-8104}\inst{\ref{aff136}}
\and E.~Sefusatti\orcid{0000-0003-0473-1567}\inst{\ref{aff45},\ref{aff104},\ref{aff94}}
\and M.~Sereno\orcid{0000-0003-0302-0325}\inst{\ref{aff11},\ref{aff13}}
\and P.~Simon\inst{\ref{aff7}}
\and A.~Spurio~Mancini\orcid{0000-0001-5698-0990}\inst{\ref{aff4}}
\and J.~Stadel\orcid{0000-0001-7565-8622}\inst{\ref{aff136}}
\and J.~Steinwagner\inst{\ref{aff14}}
\and G.~Testera\inst{\ref{aff18}}
\and R.~Teyssier\orcid{0000-0001-7689-0933}\inst{\ref{aff137}}
\and S.~Toft\orcid{0000-0003-3631-7176}\inst{\ref{aff53},\ref{aff138},\ref{aff139}}
\and S.~Tosi\orcid{0000-0002-7275-9193}\inst{\ref{aff17},\ref{aff18},\ref{aff10}}
\and A.~Troja\orcid{0000-0003-0239-4595}\inst{\ref{aff83},\ref{aff84}}
\and M.~Tucci\inst{\ref{aff41}}
\and C.~Valieri\inst{\ref{aff13}}
\and J.~Valiviita\orcid{0000-0001-6225-3693}\inst{\ref{aff61},\ref{aff62}}
\and D.~Vergani\orcid{0000-0003-0898-2216}\inst{\ref{aff11}}}

\institute{Institute for Astronomy, University of Edinburgh, Royal Observatory, Blackford Hill, Edinburgh EH9 3HJ, UK\label{aff1}
\and
Department of Physics, Oxford University, Keble Road, Oxford OX1 3RH, UK\label{aff2}
\and
Jodrell Bank Centre for Astrophysics, Department of Physics and Astronomy, University of Manchester, Oxford Road, Manchester M13 9PL, UK\label{aff3}
\and
Mullard Space Science Laboratory, University College London, Holmbury St Mary, Dorking, Surrey RH5 6NT, UK\label{aff4}
\and
Aix-Marseille Universit\'e, CNRS, CNES, LAM, Marseille, France\label{aff5}
\and
Universit\"at Innsbruck, Institut f\"ur Astro- und Teilchenphysik, Technikerstr. 25/8, 6020 Innsbruck, Austria\label{aff6}
\and
Universit\"at Bonn, Argelander-Institut f\"ur Astronomie, Auf dem H\"ugel 71, 53121 Bonn, Germany\label{aff7}
\and
Universit\'e Paris-Saclay, CNRS, Institut d'astrophysique spatiale, 91405, Orsay, France\label{aff8}
\and
Institute of Cosmology and Gravitation, University of Portsmouth, Portsmouth PO1 3FX, UK\label{aff9}
\and
INAF-Osservatorio Astronomico di Brera, Via Brera 28, 20122 Milano, Italy\label{aff10}
\and
INAF-Osservatorio di Astrofisica e Scienza dello Spazio di Bologna, Via Piero Gobetti 93/3, 40129 Bologna, Italy\label{aff11}
\and
Dipartimento di Fisica e Astronomia, Universit\`a di Bologna, Via Gobetti 93/2, 40129 Bologna, Italy\label{aff12}
\and
INFN-Sezione di Bologna, Viale Berti Pichat 6/2, 40127 Bologna, Italy\label{aff13}
\and
Max Planck Institute for Extraterrestrial Physics, Giessenbachstr. 1, 85748 Garching, Germany\label{aff14}
\and
Universit\"ats-Sternwarte M\"unchen, Fakult\"at f\"ur Physik, Ludwig-Maximilians-Universit\"at M\"unchen, Scheinerstrasse 1, 81679 M\"unchen, Germany\label{aff15}
\and
INAF-Osservatorio Astrofisico di Torino, Via Osservatorio 20, 10025 Pino Torinese (TO), Italy\label{aff16}
\and
Dipartimento di Fisica, Universit\`a di Genova, Via Dodecaneso 33, 16146, Genova, Italy\label{aff17}
\and
INFN-Sezione di Genova, Via Dodecaneso 33, 16146, Genova, Italy\label{aff18}
\and
Department of Physics "E. Pancini", University Federico II, Via Cinthia 6, 80126, Napoli, Italy\label{aff19}
\and
INAF-Osservatorio Astronomico di Capodimonte, Via Moiariello 16, 80131 Napoli, Italy\label{aff20}
\and
INFN section of Naples, Via Cinthia 6, 80126, Napoli, Italy\label{aff21}
\and
Instituto de Astrof\'isica e Ci\^encias do Espa\c{c}o, Universidade do Porto, CAUP, Rua das Estrelas, PT4150-762 Porto, Portugal\label{aff22}
\and
Dipartimento di Fisica, Universit\`a degli Studi di Torino, Via P. Giuria 1, 10125 Torino, Italy\label{aff23}
\and
INFN-Sezione di Torino, Via P. Giuria 1, 10125 Torino, Italy\label{aff24}
\and
INAF-IASF Milano, Via Alfonso Corti 12, 20133 Milano, Italy\label{aff25}
\and
INAF-Osservatorio Astronomico di Roma, Via Frascati 33, 00078 Monteporzio Catone, Italy\label{aff26}
\and
INFN-Sezione di Roma, Piazzale Aldo Moro, 2 - c/o Dipartimento di Fisica, Edificio G. Marconi, 00185 Roma, Italy\label{aff27}
\and
Institut de F\'{i}sica d'Altes Energies (IFAE), The Barcelona Institute of Science and Technology, Campus UAB, 08193 Bellaterra (Barcelona), Spain\label{aff28}
\and
Port d'Informaci\'{o} Cient\'{i}fica, Campus UAB, C. Albareda s/n, 08193 Bellaterra (Barcelona), Spain\label{aff29}
\and
Institute for Theoretical Particle Physics and Cosmology (TTK), RWTH Aachen University, 52056 Aachen, Germany\label{aff30}
\and
Institute of Space Sciences (ICE, CSIC), Campus UAB, Carrer de Can Magrans, s/n, 08193 Barcelona, Spain\label{aff31}
\and
Institut d'Estudis Espacials de Catalunya (IEEC),  Edifici RDIT, Campus UPC, 08860 Castelldefels, Barcelona, Spain\label{aff32}
\and
Dipartimento di Fisica e Astronomia "Augusto Righi" - Alma Mater Studiorum Universit\`a di Bologna, Viale Berti Pichat 6/2, 40127 Bologna, Italy\label{aff33}
\and
European Space Agency/ESRIN, Largo Galileo Galilei 1, 00044 Frascati, Roma, Italy\label{aff34}
\and
ESAC/ESA, Camino Bajo del Castillo, s/n., Urb. Villafranca del Castillo, 28692 Villanueva de la Ca\~nada, Madrid, Spain\label{aff35}
\and
Universit\'e Claude Bernard Lyon 1, CNRS/IN2P3, IP2I Lyon, UMR 5822, Villeurbanne, F-69100, France\label{aff36}
\and
Institute of Physics, Laboratory of Astrophysics, Ecole Polytechnique F\'ed\'erale de Lausanne (EPFL), Observatoire de Sauverny, 1290 Versoix, Switzerland\label{aff37}
\and
UCB Lyon 1, CNRS/IN2P3, IUF, IP2I Lyon, 4 rue Enrico Fermi, 69622 Villeurbanne, France\label{aff38}
\and
Departamento de F\'isica, Faculdade de Ci\^encias, Universidade de Lisboa, Edif\'icio C8, Campo Grande, PT1749-016 Lisboa, Portugal\label{aff39}
\and
Instituto de Astrof\'isica e Ci\^encias do Espa\c{c}o, Faculdade de Ci\^encias, Universidade de Lisboa, Campo Grande, 1749-016 Lisboa, Portugal\label{aff40}
\and
Department of Astronomy, University of Geneva, ch. d'Ecogia 16, 1290 Versoix, Switzerland\label{aff41}
\and
INAF-Istituto di Astrofisica e Planetologia Spaziali, via del Fosso del Cavaliere, 100, 00100 Roma, Italy\label{aff42}
\and
Universit\'e Paris-Saclay, Universit\'e Paris Cit\'e, CEA, CNRS, AIM, 91191, Gif-sur-Yvette, France\label{aff43}
\and
Institut de Ciencies de l'Espai (IEEC-CSIC), Campus UAB, Carrer de Can Magrans, s/n Cerdanyola del Vall\'es, 08193 Barcelona, Spain\label{aff44}
\and
INAF-Osservatorio Astronomico di Trieste, Via G. B. Tiepolo 11, 34143 Trieste, Italy\label{aff45}
\and
Istituto Nazionale di Fisica Nucleare, Sezione di Bologna, Via Irnerio 46, 40126 Bologna, Italy\label{aff46}
\and
INAF-Osservatorio Astronomico di Padova, Via dell'Osservatorio 5, 35122 Padova, Italy\label{aff47}
\and
Institute of Theoretical Astrophysics, University of Oslo, P.O. Box 1029 Blindern, 0315 Oslo, Norway\label{aff48}
\and
Higgs Centre for Theoretical Physics, School of Physics and Astronomy, The University of Edinburgh, Edinburgh EH9 3FD, UK\label{aff49}
\and
Jet Propulsion Laboratory, California Institute of Technology, 4800 Oak Grove Drive, Pasadena, CA, 91109, USA\label{aff50}
\and
von Hoerner \& Sulger GmbH, Schlossplatz 8, 68723 Schwetzingen, Germany\label{aff51}
\and
Technical University of Denmark, Elektrovej 327, 2800 Kgs. Lyngby, Denmark\label{aff52}
\and
Cosmic Dawn Center (DAWN), Denmark\label{aff53}
\and
Institut d'Astrophysique de Paris, UMR 7095, CNRS, and Sorbonne Universit\'e, 98 bis boulevard Arago, 75014 Paris, France\label{aff54}
\and
Max-Planck-Institut f\"ur Astronomie, K\"onigstuhl 17, 69117 Heidelberg, Germany\label{aff55}
\and
Department of Physics and Helsinki Institute of Physics, Gustaf H\"allstr\"omin katu 2, 00014 University of Helsinki, Finland\label{aff56}
\and
Aix-Marseille Universit\'e, CNRS/IN2P3, CPPM, Marseille, France\label{aff57}
\and
AIM, CEA, CNRS, Universit\'{e} Paris-Saclay, Universit\'{e} de Paris, 91191 Gif-sur-Yvette, France\label{aff58}
\and
Leiden Observatory, Leiden University, Einsteinweg 55, 2333 CC Leiden, The Netherlands\label{aff59}
\and
Universit\'e de Gen\`eve, D\'epartement de Physique Th\'eorique and Centre for Astroparticle Physics, 24 quai Ernest-Ansermet, CH-1211 Gen\`eve 4, Switzerland\label{aff60}
\and
Department of Physics, P.O. Box 64, 00014 University of Helsinki, Finland\label{aff61}
\and
Helsinki Institute of Physics, Gustaf H{\"a}llstr{\"o}min katu 2, University of Helsinki, Helsinki, Finland\label{aff62}
\and
NOVA optical infrared instrumentation group at ASTRON, Oude Hoogeveensedijk 4, 7991PD, Dwingeloo, The Netherlands\label{aff63}
\and
Dipartimento di Fisica "Aldo Pontremoli", Universit\`a degli Studi di Milano, Via Celoria 16, 20133 Milano, Italy\label{aff64}
\and
INFN-Sezione di Milano, Via Celoria 16, 20133 Milano, Italy\label{aff65}
\and
Dipartimento di Fisica e Astronomia "Augusto Righi" - Alma Mater Studiorum Universit\`a di Bologna, via Piero Gobetti 93/2, 40129 Bologna, Italy\label{aff66}
\and
Department of Physics, Institute for Computational Cosmology, Durham University, South Road, DH1 3LE, UK\label{aff67}
\and
Universit\'e C\^{o}te d'Azur, Observatoire de la C\^{o}te d'Azur, CNRS, Laboratoire Lagrange, Bd de l'Observatoire, CS 34229, 06304 Nice cedex 4, France\label{aff68}
\and
Universit\'e Paris Cit\'e, CNRS, Astroparticule et Cosmologie, 75013 Paris, France\label{aff69}
\and
University of Applied Sciences and Arts of Northwestern Switzerland, School of Engineering, 5210 Windisch, Switzerland\label{aff70}
\and
European Space Agency/ESTEC, Keplerlaan 1, 2201 AZ Noordwijk, The Netherlands\label{aff71}
\and
School of Mathematics, Statistics and Physics, Newcastle University, Herschel Building, Newcastle-upon-Tyne, NE1 7RU, UK\label{aff72}
\and
Department of Physics and Astronomy, University of Aarhus, Ny Munkegade 120, DK-8000 Aarhus C, Denmark\label{aff73}
\and
Waterloo Centre for Astrophysics, University of Waterloo, Waterloo, Ontario N2L 3G1, Canada\label{aff74}
\and
Department of Physics and Astronomy, University of Waterloo, Waterloo, Ontario N2L 3G1, Canada\label{aff75}
\and
Perimeter Institute for Theoretical Physics, Waterloo, Ontario N2L 2Y5, Canada\label{aff76}
\and
Universit\'e Paris-Saclay, Universit\'e Paris Cit\'e, CEA, CNRS, Astrophysique, Instrumentation et Mod\'elisation Paris-Saclay, 91191 Gif-sur-Yvette, France\label{aff77}
\and
Space Science Data Center, Italian Space Agency, via del Politecnico snc, 00133 Roma, Italy\label{aff78}
\and
Centre National d'Etudes Spatiales -- Centre spatial de Toulouse, 18 avenue Edouard Belin, 31401 Toulouse Cedex 9, France\label{aff79}
\and
Institute of Space Science, Str. Atomistilor, nr. 409 M\u{a}gurele, Ilfov, 077125, Romania\label{aff80}
\and
Instituto de Astrof\'isica de Canarias, Calle V\'ia L\'actea s/n, 38204, San Crist\'obal de La Laguna, Tenerife, Spain\label{aff81}
\and
Departamento de Astrof\'isica, Universidad de La Laguna, 38206, La Laguna, Tenerife, Spain\label{aff82}
\and
Dipartimento di Fisica e Astronomia "G. Galilei", Universit\`a di Padova, Via Marzolo 8, 35131 Padova, Italy\label{aff83}
\and
INFN-Padova, Via Marzolo 8, 35131 Padova, Italy\label{aff84}
\and
Departamento de F\'isica, FCFM, Universidad de Chile, Blanco Encalada 2008, Santiago, Chile\label{aff85}
\and
Satlantis, University Science Park, Sede Bld 48940, Leioa-Bilbao, Spain\label{aff86}
\and
Centro de Investigaciones Energ\'eticas, Medioambientales y Tecnol\'ogicas (CIEMAT), Avenida Complutense 40, 28040 Madrid, Spain\label{aff87}
\and
Instituto de Astrof\'isica e Ci\^encias do Espa\c{c}o, Faculdade de Ci\^encias, Universidade de Lisboa, Tapada da Ajuda, 1349-018 Lisboa, Portugal\label{aff88}
\and
Universidad Polit\'ecnica de Cartagena, Departamento de Electr\'onica y Tecnolog\'ia de Computadoras,  Plaza del Hospital 1, 30202 Cartagena, Spain\label{aff89}
\and
Institut de Recherche en Astrophysique et Plan\'etologie (IRAP), Universit\'e de Toulouse, CNRS, UPS, CNES, 14 Av. Edouard Belin, 31400 Toulouse, France\label{aff90}
\and
Kapteyn Astronomical Institute, University of Groningen, PO Box 800, 9700 AV Groningen, The Netherlands\label{aff91}
\and
INFN-Bologna, Via Irnerio 46, 40126 Bologna, Italy\label{aff92}
\and
Infrared Processing and Analysis Center, California Institute of Technology, Pasadena, CA 91125, USA\label{aff93}
\and
IFPU, Institute for Fundamental Physics of the Universe, via Beirut 2, 34151 Trieste, Italy\label{aff94}
\and
INAF, Istituto di Radioastronomia, Via Piero Gobetti 101, 40129 Bologna, Italy\label{aff95}
\and
Department of Mathematics and Physics E. De Giorgi, University of Salento, Via per Arnesano, CP-I93, 73100, Lecce, Italy\label{aff96}
\and
INAF-Sezione di Lecce, c/o Dipartimento Matematica e Fisica, Via per Arnesano, 73100, Lecce, Italy\label{aff97}
\and
INFN, Sezione di Lecce, Via per Arnesano, CP-193, 73100, Lecce, Italy\label{aff98}
\and
Institut f\"ur Theoretische Physik, University of Heidelberg, Philosophenweg 16, 69120 Heidelberg, Germany\label{aff99}
\and
Universit\'e St Joseph; Faculty of Sciences, Beirut, Lebanon\label{aff100}
\and
Institut d'Astrophysique de Paris, 98bis Boulevard Arago, 75014, Paris, France\label{aff101}
\and
Junia, EPA department, 41 Bd Vauban, 59800 Lille, France\label{aff102}
\and
SISSA, International School for Advanced Studies, Via Bonomea 265, 34136 Trieste TS, Italy\label{aff103}
\and
INFN, Sezione di Trieste, Via Valerio 2, 34127 Trieste TS, Italy\label{aff104}
\and
ICSC - Centro Nazionale di Ricerca in High Performance Computing, Big Data e Quantum Computing, Via Magnanelli 2, Bologna, Italy\label{aff105}
\and
Instituto de F\'isica Te\'orica UAM-CSIC, Campus de Cantoblanco, 28049 Madrid, Spain\label{aff106}
\and
CERCA/ISO, Department of Physics, Case Western Reserve University, 10900 Euclid Avenue, Cleveland, OH 44106, USA\label{aff107}
\and
Laboratoire Univers et Th\'eorie, Observatoire de Paris, Universit\'e PSL, Universit\'e Paris Cit\'e, CNRS, 92190 Meudon, France\label{aff108}
\and
Dipartimento di Fisica e Scienze della Terra, Universit\`a degli Studi di Ferrara, Via Giuseppe Saragat 1, 44122 Ferrara, Italy\label{aff109}
\and
Istituto Nazionale di Fisica Nucleare, Sezione di Ferrara, Via Giuseppe Saragat 1, 44122 Ferrara, Italy\label{aff110}
\and
Dipartimento di Fisica - Sezione di Astronomia, Universit\`a di Trieste, Via Tiepolo 11, 34131 Trieste, Italy\label{aff111}
\and
NASA Ames Research Center, Moffett Field, CA 94035, USA\label{aff112}
\and
Kavli Institute for Particle Astrophysics \& Cosmology (KIPAC), Stanford University, Stanford, CA 94305, USA\label{aff113}
\and
Bay Area Environmental Research Institute, Moffett Field, California 94035, USA\label{aff114}
\and
Minnesota Institute for Astrophysics, University of Minnesota, 116 Church St SE, Minneapolis, MN 55455, USA\label{aff115}
\and
Institute Lorentz, Leiden University, Niels Bohrweg 2, 2333 CA Leiden, The Netherlands\label{aff116}
\and
Institute for Astronomy, University of Hawaii, 2680 Woodlawn Drive, Honolulu, HI 96822, USA\label{aff117}
\and
Department of Physics \& Astronomy, University of California Irvine, Irvine CA 92697, USA\label{aff118}
\and
Department of Astronomy \& Physics and Institute for Computational Astrophysics, Saint Mary's University, 923 Robie Street, Halifax, Nova Scotia, B3H 3C3, Canada\label{aff119}
\and
Departamento F\'isica Aplicada, Universidad Polit\'ecnica de Cartagena, Campus Muralla del Mar, 30202 Cartagena, Murcia, Spain\label{aff120}
\and
Department of Computer Science, Aalto University, PO Box 15400, Espoo, FI-00 076, Finland\label{aff121}
\and
Ruhr University Bochum, Faculty of Physics and Astronomy, Astronomical Institute (AIRUB), German Centre for Cosmological Lensing (GCCL), 44780 Bochum, Germany\label{aff122}
\and
Universit\'e Paris-Saclay, CNRS/IN2P3, IJCLab, 91405 Orsay, France\label{aff123}
\and
Univ. Grenoble Alpes, CNRS, Grenoble INP, LPSC-IN2P3, 53, Avenue des Martyrs, 38000, Grenoble, France\label{aff124}
\and
Department of Physics and Astronomy, Vesilinnantie 5, 20014 University of Turku, Finland\label{aff125}
\and
Serco for European Space Agency (ESA), Camino bajo del Castillo, s/n, Urbanizacion Villafranca del Castillo, Villanueva de la Ca\~nada, 28692 Madrid, Spain\label{aff126}
\and
ARC Centre of Excellence for Dark Matter Particle Physics, Melbourne, Australia\label{aff127}
\and
Centre for Astrophysics \& Supercomputing, Swinburne University of Technology, Victoria 3122, Australia\label{aff128}
\and
W.M. Keck Observatory, 65-1120 Mamalahoa Hwy, Kamuela, HI, USA\label{aff129}
\and
Department of Physics and Astronomy, University of the Western Cape, Bellville, Cape Town, 7535, South Africa\label{aff130}
\and
Dipartimento di Fisica, Sapienza Universit\`a di Roma, Piazzale Aldo Moro 2, 00185 Roma, Italy\label{aff131}
\and
Centro de Astrof\'{\i}sica da Universidade do Porto, Rua das Estrelas, 4150-762 Porto, Portugal\label{aff132}
\and
Zentrum f\"ur Astronomie, Universit\"at Heidelberg, Philosophenweg 12, 69120 Heidelberg, Germany\label{aff133}
\and
Dipartimento di Fisica, Universit\`a di Roma Tor Vergata, Via della Ricerca Scientifica 1, Roma, Italy\label{aff134}
\and
INFN, Sezione di Roma 2, Via della Ricerca Scientifica 1, Roma, Italy\label{aff135}
\and
Department of Astrophysics, University of Zurich, Winterthurerstrasse 190, 8057 Zurich, Switzerland\label{aff136}
\and
Department of Astrophysical Sciences, Peyton Hall, Princeton University, Princeton, NJ 08544, USA\label{aff137}
\and
Niels Bohr Institute, University of Copenhagen, Jagtvej 128, 2200 Copenhagen, Denmark\label{aff138}
\and
Cosmic Dawn Center (DAWN)\label{aff139}}

%% file: tex/sec1_introduction.tex
\section{Introduction}

Weak gravitational lensing by large-scale structure is a mature cosmological tool
to measure the distribution of dark matter and study dark energy through
its evolution with redshift \citep{schneider2006,kilbinger2015,mandelbaum2018a}.
Weak lensing is particularly sensitive to modifications of the theory of gravity
and the emergence of physics beyond the concordance $\Lambda$-cold dark matter model, 
which affect the clustering of dark matter \citep{amendola2018}.

Galaxy surveys from the ground, such as the Dark Energy Survey \citep[DES;][]{abbott2018a}, 
the Kilo Degree Survey \citep[KiDS;][]{kuijken2015}, and the Hyper Suprime-Cam survey \citep[HSC;][]{aihara2017},
are now achieving constraints on the dark matter sector
(primarily in the $\Omega_\text{m}$--$\sigma_8$ parameter space
and their combination $S_8$) to a few percent \citep{abbott2018b,hildebrandt2017,hikage2019,amon2022a,secco2022,dalal2023,lix2023a}. 
After extensive consistency checks and sensitivity studies, recent lensing measurements from galaxy surveys
are shown to be broadly in agreement with each other
but in mild tension with the \planck satellite at the $2\,\sigma$ level or greater \citep{aghanim2020,joudaki2020,amon2022b,heymans2021,asgari2021,loureiro2022} -- this
includes the latest joint analysis of DES and KiDS \citep{abbott2023}.

In the coming years, galaxy surveys will enter a new regime of area, depth, and image quality.
The space-based \euclid telescope  
\citep[survey area of $14\,000\,\unit{\deg^2}$, full width at half maximum (FWHM) resolution of $\ang{;;0.2}$, depth of 24.5, $\IE$+$\YE\JE\HE$ filters, see][]{laureijs2011,cropper2016,scaramella2022,mellier2024};
the planned space-based \textit{Roman} telescope \cite[$1700\,\unit{\deg^2}$, $\ang{;;0.2}$, 26.5, YJH+F184, see][]{spergel2015,akeson2019};
and the ground-based \rubin Observatory Legacy Survey of Space and Time \citep[LSST; $18\,000\,\unit{\deg^2}$, $\ang{;;0.7}$, 27.5, $ugrizy$, see][]{ivezic2019}
will substantially increase the number of observable galaxies compared to current surveys.
The systematic observation of a billion galaxies or more across one-third of the visible sky will then be possible for the first time.
The combined effect of improved survey area and angular resolution will be an enhanced ability to probe both the large and small scales via weak lensing and galaxy clustering,
allowing us to constrain cosmological models and dark energy to percent-level precision \citep{mandelbaum2018b,blanchard2020},
or even an order of magnitude better when combined with data from \planck \citep{ilic2022}.

With the dramatic improvement in precision that will be achieved in the coming years,
experiments are now focussing on understanding the accuracy of their analyses \citep[see, e.g.,][]{paykari2020}.
Along with theory uncertainties, the cosmic shear measurement and
redshift estimation are the most challenging aspects of any large-scale weak lensing surveys.
The concern of this paper is on cosmic shear measurement,
which provides the necessary data for the weak lensing cosmological analysis \citep{kilbinger2015}.
In order to achieve an order of one-percent precision on the dark energy equation of state, a billion galaxies or more with median redshift around one need to be observed.
This observation has to be carried out consistently so the same shape measurement procedure is applied to all objects.
This measurement has to be conducted with outstanding accuracy to satisfy the stringent requirement of $2\times10^{-3}$ and $3\times10^{-4}$ on the measured multiplicative and additive shear biases that were set in the early development phase of weak lensing space telescopes \citep{massey2012,cropper2013}.

Throughout the past years, a number of shear measurement methods were developed, tested on data challenges, and applied to real data.
These can be categorised into two main classes: non-parametric and parametric.
Among the non-parametric is Kaiser-Squires-Broadhurst
\citep[KSB;][]{kaiser1995,hoekstra1998}, which is based on weighted moments of image data.
Because of its simplicity, these estimators were used for the very first attempts at measuring cosmic shear in the early 2000s.
These methods are fast and can be quickly calibrated, but they are sensitive to effects that need to be characterised.
With better precision, it later became clear that more effects needed to be factored in,
particularly a realistic PSF
and the sensitivity of bias on the PSF ellipticity, known as leakage.
Parametric methods, based on forward modelling and model fitting, soon 
appeared to be better suited to accurately incorporating such real data features building on solid statistical grounds.
In recent years, a more systematic use of model fitting techniques was 
observed across all major lensing surveys.
The Bayesian-inspired shape method \lensfit \citep{miller2007,kitching2008,miller2013}
was extremely successful, first in the Canada-France Hawaii
Telescope Lensing Survey \citep[CFHTLenS;][]{heymans2012} and
more recently in KiDS.
This method is based on forward modelling and marginalisation over galaxy nuisance parameters.
A similar method is \imthreeshape \citep{zuntz2013},
which is a maximum likelihood estimator based again on analytic forward modelling that was applied to DES.
While many real-data effects, including the PSF, are accounted for and can be directly built in parametric methods,
any in-built correction clearly introduces extra computational overhead, as the parameter probability distribution needs to be sampled accurately.

While lensing measurements became more precise over time,
accuracy also needed to be examined more carefully.
Methods were compared 
in data challenges and were run on common simulations with increasing realism,
such as in the Shear TEsting Programme \citep[STEP;][]{heymans2006,massey2007} and
the Gravitational LEnsing Accuracy Testing \citep[GREAT;][]{bridle2010,kitching2012,mandelbaum2015}.
With no method outshining in absolute terms,
and methods being better at some aspects of the measurement but worse at others,
it became evident that some form of calibration was still necessary.
More than ever before, the field has become reliant on galaxy simulations.
Sophisticated high-fidelity simulations now need to reproduce the realism of actual observations as close as possible
so all biases from detection, measurement, and selection can be fully captured \citep{fenech2017,kannawadi2019,martinet2019,maccrann2021,liss2023a,liss2023b}.
Calibration naturally raises the question about how sensitive results are to the assumptions that are made in simulations \citep{hoekstra2017},
or how large these simulations need to be to meet the desired precision \citep{pujol2019,jansen2024}.
Other methods now rely on some form of calibration that is directly built in the measurement process.
Galaxy images used in the calibration were simulated internally,
inferred from real data, or obtained through a combination of the two methods.
\metacal \citep{sheldon2017,huff2017} derives internal estimates of
the sensitivity of the ellipticity estimator to input shears
and was extremely successful on DES Year 3 \citep{gatti2021};
Bayesian Fourier Domain \citep[BFD;][]{bernstein2014,bernstein2016} estimates
the Taylor coefficients of the galaxy likelihood expanded over shear with information about moments measured from calibration fields;
a similar implementation to BFD uses forward modelling \citep{sheldon2014};
the KiDS self-calibration \citep{fenech2017} derives internal estimates of the ellipticity bias from noise-free galaxy images;
\texttt{MomentsML} relies on simulated images to train shear-predicting artificial neural networks \citep{tewes2019}.
Because many selection biases happen before the shear measurement introduces its own bias,
the field has gradually become more aware that those will probably be the limiting factor in future lensing surveys.
Further work around \metacal led to \metadet to address the issue \citep{sheldon2020}.
An application of the method to \euclid{-like} simulations showed that while selection biases may exceed requirements,
the outlook is still positive, with demonstrated success at handling detection and blending biases \citep{hoekstra2021b, hoekstra2021a,melchior2021}.
Furthermore, the recent Anacal method aims to correct measurement and selection biases via analytic differentiation \citep{lix2023b}.

The impact of neighbours to lensing measurements has also become one of the most important issues that current and future surveys will need to address.
In space, the large number density of detected galaxies of about $\qty{30}{arcmin^{-2}}$ ($\IE<24.5$) is compensated by a good image resolution, so the impact of neighbours may not be as bad as from the ground.
In fact, due to the worse resolution on the ground, the impact of neighbours is serious,
affecting $60\%$ of the sample \citep{bosch2017,arcelin2021}.
DES Year 1 results \citep{samuroff2017} showed that the total neighbour bias can be as large as $9\%$ (reaching $80\%$ at a close distance to the neighbour, if uncorrected).
Moreover, the impact of `unrecognised' blends (undetected neighbours) on the $S_8$ parameter from simulated \rubin Year 1 data can be as large as $15\%$ \citep{nourbakhsh2022}.
Cuts to the catalogue to remove those objects can reduce the total bias to $1\%$ (reaching $30\%$ at a close distance to the neighbour),
however at the cost of reducing the effective number density by $30\%$ and leaving a residual bias on $S_8$ of $2\,\sigma$.
While \metacal is extremely successful in a few idealised cases (e.g., isolated galaxies)
and \metadet in the handling of blending and detection bias,
a suite of advanced simulations were required for the tomographic calibration of DES Year 3 \citep{maccrann2021}.
These simulations show that an external calibration is still required, as
the unresolved neighbour introduces a correlation between the two galaxies at different redshift,
plus the \metacal shear responses will be biased by the presence of the neighbour itself.
Therefore, calibration simulations are necessary to correctly capture neighbour bias and the interplay between shear and redshift.
However, these simulations assume uniform random distributions of galaxies that were re-weighted to mimic clustering, which raises the question as to whether the inferred bias is likely underestimated.
The most recent simulations by KiDS have realistic clustering of $N$-body simulations, mimic a number of measurement effects, and address the shear-redshift interplay \citep{liss2023a}.
With a larger number density, it is expected that the situation may be more serious in future surveys.

In this paper, we present our advanced shear measurement method \lensmc specifically developed for \euclid
that builds on the knowledge and success of ground-based measurement at handling real data effects.
Similarly to \lensfit, it adopts a mean estimator.
Contrary to \lensfit, it does not marginalise over nuisance parameters with numerical approximations.
With full flexibility in the choice of the prior, all the marginalisation in \lensmc is performed by Markov chain Monte Carlo (MCMC) sampling
for individual galaxies or jointly across groups of neighbouring galaxies.
While \imthreeshape returns the maximum of the likelihood estimated via the Levenberg-Marquardt algorithm,
\lensmc employs a combination of large-scale and small-scale algorithms (such as conjugate gradient and simplex methods)
to estimate a suitable initial guess for the MCMC sampling, thus requiring no information about the model derivatives
and dramatically reducing the sensitivity on the initial guess (which is assumed to always be the same for all galaxies). 
The galaxy models in \lensmc are rendered directly in the Fourier space; hence only a single Fourier transform is required.
A recent profile-fitting method, \texttt{The Farmer}, has also drawn attention \citep{weaver2023}.
This method  is a maximum likelihood estimator whose initial guesses of position and shape are provided by the detection method.
It includes a decision tree based on $\chi^2$ values to classify objects on their likely type
and provides error estimates via Cramer-Rao bounds.
Preliminary results are encouraging; however, the method has not been tested on full space-based cosmic shear accuracy yet.

Accurate cosmic shear measurement requires controlling the bias from a number of sources.
Key examples include source clustering, faint objects, neighbours, PSF leakage, astrometry, image artefacts, and cosmic rays.
Additionally, any forward-modelling method is plagued by potential model bias.
One of the main sources of model bias was addressed in this work.
In summary, \lensmc employs: 
(\textit{i}) forward modelling to deal with \euclid image undersampling and convolution by a PSF with comparable size to many galaxies;
(\textit{ii}) joint measuring object groups to correctly handle bias due to neighbours;
(\textit{iii}) masking out objects belonging to different groups;
(\textit{iv}) MCMC sampling to sample the posterior in a multi-dimensional parameter space, calculate weights, 
and correctly marginalise ellipticity over nuisance parameters and other objects in the same group.
We particularly focussed on the realism of our simulations,
including clustering, stars, object detection,
the handling of neighbours due to high number density,
and the use of realistic undersampled chromatic PSF images with spatial variation across the field of view.
We did not include further real data effects such as non-linearities or cosmic rays,
as these will be addressed separately.
Also we assumed the same broadband PSF as obtained from a spectral energy distribution (SED) of an SBc-type galaxy at a redshift of one in both simulations and measurements. 

\secref{sec:method} introduces our method and the practical advantages in addressing real data problems.
Our forward-modelling method is sufficiently fast to analyse the typical data volume of Stage-IV surveys and can be applied to
the complexity of \euclid measurements, including undersampled data and a complex PSF while accounting for the full degrees of freedom in the galaxy modelling.
Additionally, it allows for the proper handling of resolved neighbours by joint measurement and masking of more distant galaxies, stars, and artefacts in the images.
\secref{sec:simulations} describes the simulations used for our intensive testing of the method.
The images are fiducial realisations of the \euclid VIS detector \citep{cropper2024}, and galaxies were rendered based on $N$-body simulations with full variability of the morphological properties \citep{potter2017,castander2024}.
All galaxies were convolved with a realistic pre-flight PSF model with full spatial variation, but the chromatic variability was ignored.
\secref{sec:results} presents the main results of this testing.
When model bias, chromaticity, and selection biases are suppressed, the obtained biases are close to the \euclid requirement.
This measurement bias is largely dominated by undetected faint galaxies in the images.
The bias was also found to be stable and mostly insensitive to the many effects in the simulations.
As the \euclid analysis will also need to correct for other artefacts in the images, the residual bias will be straightforward to calibrate through image simulations.
Once we included the model bias by allowing the full variability in the galaxy models, the overall bias became significant.
However, since the sensitivity is weak (the derivative of the bias with respect to the assumptions made in the simulations appears negligible),
it will then be straightforward to also calibrate the model bias through image simulations.
\secref{sec:conclusions} discusses the main findings and draws the conclusions of our work.

%% file: tex/sec2_method.tex
\section{Method} \label{sec:method}

The main physical quantity of interest in weak lensing is the reduced cosmic shear \citep{kilbinger2015},
\begin{equation}
g=\frac{\gamma}{1-\kappa}~,
\end{equation}
where $\kappa$ and $\gamma$ are convergence and shear (both related to the gravitational potential),
and $g\approx\gamma$ in the weak lensing regime.
The related observable in weak lensing is the ellipticity of a galaxy,
\begin{equation}\label{eq:ellipticity}
\epsilon=\frac{a-b}{a+b}\mathrm{e}^{2\mathrm{i}\varphi}~,
\end{equation}
where $a$ and $b$ are, respectively, the semi-major and semi-minor axes,\footnote{This holds true only for elliptical isophotes, but the ellipticity remains well-defined if one specifies how it is measured, that is, it becomes method dependent.}
$\varphi\in\left[0,\pi\right)$ is the orientation angle of the galaxy, and $|\epsilon|\le1$.
The effect of weak lensing is to distort the ellipticity of a source galaxy, $\epsilon_\text{s}$,
by the canonical transformation \citep{seitz1997},
\begin{equation}\label{eq:shear_transformation}
\epsilon=\frac{\epsilon_\text{s}+g}{1+\epsilon_\text{s}\,g^*}~,
\end{equation}
where all spin-2 quantities are expressed in complex notation (e.g., $\epsilon=\epsilon_1+\mathrm{i}\,\epsilon_2$, where $\epsilon_1$ quantifies the distortion along $x$ and $y$, and $\epsilon_2$ along the coordinate axes rotated by $\pi/4$).
As it is customary in weak lensing, we will refer to $\epsilon_\text{s}$ as the intrinsic ellipticity of the source galaxy, and $\epsilon$ as the lensed or observed ellipticity.
The ellipticity in Eq.\;\eqref{eq:shear_transformation} is a point 
estimate for shear in that information on the underlying cosmic shear can 
be derived in a statistical sense as a sample average,
$g=\langle\epsilon\rangle$,
which holds whenever the distribution of orientation angles is uniform, for example, when there are no astrophysical intrinsic alignments \citep{joachimi2015}
or shear dependent selection effects.

In weak lensing measurements we infer the reduced shear through sample averages. 
In this work, we use the ellipticity as a point estimator for shear and
the problem of measuring ellipticity can be formulated fully in Bayesian terms.
Suppose we have a pixel data vector, $\vec{D}$, and a model for the galaxy brightness distribution, $\vec{I}=\vec{I}(\epsilon,\theta,\phi)$, as a function of ellipticity, $\epsilon$, 
intrinsic nuisance parameters, $\theta$, and linear nuisance parameters,
$\phi$.\footnote{The parameter vectors $\theta$ and $\phi$ are not to be confused with angular coordinates.
Here $\theta$ represents non-linear parameters (such as object size and position offsets)
and $\phi$ represents linear parameters (such as component fluxes) that can be analytically marginalised out.
The modelling is linear in the $\phi$ parameters as long as the flux components are co-centred and do not depend on the positions.}
Thanks to Bayes' theorem, we can define a joint posterior as follows:
\begin{equation}\label{eq:posterior}
p(\epsilon,\theta,\phi|\vec{D}) = \frac{p(\vec{D}|\epsilon,\theta,\phi)p(\epsilon,\theta,\phi)}{p(\vec{D})}~,
\end{equation}
where $p(\vec{D}|\epsilon,\theta,\phi)$ is the likelihood, $p(\epsilon,\theta,\phi)$ is the prior on ellipticity and nuisance parameters, and $p(\vec{D})$ is the evidence or marginal likelihood,
\begin{equation}
p(\vec{D})=\int p(\vec{D}|\epsilon,\theta,\phi)p(\epsilon,\theta,\phi)\de\epsilon\de\theta\de\phi~.
\end{equation}
We can then construct the ellipticity marginal posterior:
\begin{equation}\label{eq:marginalised_posterior}
p(\epsilon|\vec{D}) = \frac{1}{p(\vec{D})} \int p(\vec{D}|\epsilon,\theta,\phi)p(\epsilon,\theta,\phi)\de\theta\de\phi~,
\end{equation}
marginalising out the nuisance parameters.
Common choices of estimators are the maximum likelihood or maximum posterior probability,
but these are usually biased if the distribution is not Gaussian.
However, the bias can be predicted in simple cases of low dimensionality or when the probability function is fully analytic \citep{cox1968,hall2017}.
Another option that was successful in ground-based surveys 
\citep{miller2013} is to set the estimator to the mean of the posterior distribution,
\begin{equation}\label{eq:mean_estimator}
\hat{\epsilon}=\int \epsilon\,p(\epsilon|\vec{D}) \de\epsilon~.
\end{equation}
We adopt this definition, as it has some useful properties:
\begin{enumerate}
\item as the nuisance parameters are marginalised out, their impact on the ellipticity estimator via their correlation is mitigated;
\item overfitting\footnote{This is the tendency of some estimators, in particular the maximum of the probability distribution function,
to interpret random fluctuations in the noise as actual signal in the data.}
is inherently reduced as we pick an average representative of all possible realisations that are statistically equivalent;
\item following on from the previous point, we expect the mean estimator to be, in general, less biased than the maximum estimators;
\item the mean of the distribution can be estimated through MCMC sampling techniques (see \secref{sec:mcmc});
such estimators satisfy the central limit theorem and therefore converge to the true mean.
\end{enumerate}
We will discuss more about those points later in this section.
Whatever choice is made, any estimator can be seen as a non-linear mapping between $\vec{D}$ and $\epsilon$.
Therefore even if $\vec{D}$ were to be Gaussian distributed, the estimator will not, hence leading to a fundamental bias in the measurement, which is commonly referred to as noise bias \citep{melchior2012,refregier2012,viola2014}.
Moreover, as the shear is estimated through a sample average over a population of galaxies with varying morphological properties and complex selections, the properties of the shear bias will be different from that of galaxy ellipticity.
Assuming shear is small, it is customary in the field to model the shear bias on each component with a linear model \citep{guzik2005,huterer2006,heymans2006},
\begin{equation}\label{eq:shear_bias}
\hat{g}_i=(1+m_i)\,g_i+c_i+n_i~,
\end{equation}
where $m_i$ and $c_i$ are the multiplicative and additive biases for the $i$-th component,
$g_i$ is the true shear, $\hat{g}_i$ is an estimate of it, and $n_i$ is the corresponding statistical noise.
The transformation in Eq.\;\eqref{eq:shear_bias} should in principle have $m_i$ replaced by a
matrix $m_{ij}$ to model any potential cross-talk between shear components.
Alternatively, it could be rewritten as a spin-2 equation \citep{kitching2022},
\begin{equation}
\hat g = (1 + m_0)\,g + m_4\,g^* +c +n~.
\end{equation}
However, generalising upon previous work, $m_0$ and $m_4$ are now spin 0 and 4 complex numbers acting on spin-2 shear fields.
Physically, this added flexibility allows for complete mode-mixing: $m_0$ models a dilation and rotation of the true shear,
whereas $m_4$ models a reflection around the axis determined by its phase.
We defer the application of this approach to future work.
Multiplicative terms can be induced by, for example, non-Gaussianities in the posterior (skewness at first order),
caused by, for example, pixel noise and a small galaxy size relative to the PSF.
Additive terms are due to anisotropies induced by, for example, the PSF and its spatial variability across the field of view.
This effect is referred to as PSF leakage and is defined by the dependence of $c_i$ on the PSF ellipticity $\epsilon_{\text{PSF},i}$
\citep[see, e.g.][and references therein]{gatti2021},
\begin{equation}
\alpha_i=\frac{\de c_i}{\de \epsilon_{\text{PSF},i}}~.
\end{equation}
We focussed primarily on the $c$ dependence on $\epsilon_{\text{PSF}}$, as the $m$ dependence is negligible as long as the PSF is stable (i.e., the variation in PSF size is within a percent level).
Earlier studies \citep{massey2012,cropper2013} set out requirements for space-based missions on $m$ and $c$ based on a top-down error breakdown from cosmology
to two-point statistics.
For \euclid, the requirement is on the statistical error on bias,
$\sigma_m < 2\times10^{-3}$ and $\sigma_c < 3\times10^{-4}$.
That is roughly equivalent to saying that a shear of $1\%$ should be measured with a fractional accuracy and precision of $0.2\%$. 
We note that the requirement is an order of magnitude more stringent than 
current ground-based experiments \citep{hildebrandt2017}.
The detailed breakdown of the total budget on $m$ and $c$ into various error terms \citep{cropper2013}
suggests we can set the required statistical error on the bias due to the measurement alone to
$\sigma_m < 5\times10^{-4}$ and $\sigma_c < 5\times10^{-5}$.
Therefore, in order to measure $|g|\approx 0.03$ with a residual post-calibration multiplicative bias smaller than $\sigma_m$, one will need at least $N=\sigma_\epsilon^2\,|g|^{-2}\sigma_m^{-2}\approx4\times10^8$ galaxies,\footnote{This assumes we measure shear with accuracy given by $\hat{g}=(1+m)\,g$. The standard deviation of the measured shear scales as $\sigma_\epsilon/\sqrt{N}$ and therefore
we required that $\sigma_\epsilon/\sqrt{N}\lesssim\sigma_m|g|$.} where $\sigma_\epsilon\approx 0.3$ is the `shape noise', that is, the standard deviation of the per-component intrinsic ellipticity distribution.
Obviously measurement noise and intrinsic scatter in the morphological properties will also need to be factored in.
A ballpark estimate for the \euclid requirement on PSF leakage that we will be using as benchmark in our analysis is
$\sigma_\alpha\lesssim\sigma_c/\left|\delta \epsilon_\text{PSF}\right|$,
where $|\delta \epsilon_\text{PSF}|\approx0.1$ is the order of magnitude (absolute) variation in PSF ellipticity across the field of view,
which yields $\sigma_\alpha<1\times10^{-3}$ if we assume a budget of $\sigma_c<1\times10^{-4}$.
This derivation may be too conservative, as a full propagation of the errors and biases through to cosmological parameters is demonstrated to be able to capture the spatial pattern imprinted by the PSF and other effects \citep{paykari2020}.
Other surveys implemented other solutions such a first-order expansion on PSF ellipticity and PSF model residuals in KiDS \citep{heymans2006,giblin2021}
or the angular correlations between PSF ellipticity and size implemented in the rho statistics in DES \citep{jarvis2021}.

In the next subsections we address the key elements of the \lensmc measurement method:
galaxy modelling, PSF convolution, likelihood, sampling, and a further discussion about handling real data effects.
We emphasise the role of joint measurement of objects to address neighbour bias, which is a concern for current and upcoming surveys,
and also our MCMC strategy to sample a multi-dimensional parameter space and marginalise each lensing target 
over all nuisance parameters and other objects.

\subsection{PSF-convolved galaxy models} \label{sec:galaxy_models}

We assumed 2D-projected galaxy models as a mixture of two circular S\'ersic profiles \citep{sersic1963}.
The disc component is
\begin{equation}\label{eq:disc_model}
I_\text{d}(r) \propto \exp\left(-\frac{r}{r_\text{e}}\right)~,
\end{equation}
and the bulge component is
\begin{equation}\label{eq:bulge_model}
I_\text{b}(r) \propto \exp\left[-a_\text{b}\left(\frac{r}{r_\text{h}}\right)^\frac{1}{n_\text{b}}\right]~,
\end{equation}
where $r$ is the distance from the centre, $r_\text{e}$ is the exponential scale length of the disc,
$r_\text{h}$ is the bulge half-light radius, $n_\text{b}=1$, and $a_\text{b}\approx2\,n_\text{b} - 0.331$ \citep{peng2002}.
The bulge S\'ersic index was fixed to $1$ based on recent multi-wavelength observations of the \hubble CANDELS/GOODS-South field (Welikala et al., in prep.),
\footnote{Their work highlights that both the dust in the disc and the 3D modelling of the galaxy influence the inferred bulge structural parameters.}
while bulge profiles with $n_\text{b}=4$ (de Vaucouleurs) are instead more typical for early-type galaxies at low redshift.
Both profiles were normalised so that their integral is 1.
In the measurement, $r_\text{e}$ plays the role of object size parameter,
and we fixed the bulge-to-disc scale length ratio to $r_\text{h}/r_\text{e}=0.15$ based on the same \hst measurements (Welikala et al., in prep.). 
Finally, we imposed a hard cut-off on the surface brightness profile at $r_{\max}/r_\text{e}=4.5$ since observations indicate that galaxies have truncated surface brightness distributions \citep{vanderkruit1981,vanderkruit1982}.
The parameters $n_\text{b}$, $r_{\max}/r_\text{e}$, and $r_\text{h}/r_\text{e}$ are assumptions made in the modelling that
can potentially lead to large biases in presence of a mismatch in the assumed S\'ersic index compared to simulations \citep{simon2017}.
We stress that our choice of fixed values are based on recent observations, and the model bias due to incorrect assumptions are often intertwined with the particular simulation setup and its complexity.
A detailed investigation of sensitivity of the calibration to bulge parameters is presented later in this paper.

The S\'{e}rsic model introduced above is an isotropic profile with zero ellipticity.
To make it anisotropic (i.e., elliptical with ellipticity $\epsilon=\epsilon_1+\mathrm{i}\,\epsilon_2$,
we used the following distortion matrix:
\begin{equation}\label{eq:shear_matrix}
\tens{S}=\frac{\bar{r}_0}{q_\epsilon\,r_0}
\begin{pmatrix}
1-\epsilon_1 & -\epsilon_2 \\
-\epsilon_2 & 1+\epsilon_1
\end{pmatrix}~,
\end{equation}
where $r_0/\bar{r}_0$ is the scale factor necessary to scale a model
of size $\bar{r}_0$ to the desired size $r_0$.
Because observed galaxy shapes are a 2D visual projection of an intrinsically 3D distribution,
we introduce the additional scale factor, $q_\epsilon=1-|\epsilon|$, to make the profile semi-major axis invariant under ellipticity transformation.\footnote{Instead, $q_\epsilon=1$ would make the profile \textit{area} invariant under an ellipticity transformation.
The choice of $q_\epsilon$ leads to different shear bias properties that can have significant impact on the final calibration \citep{fenech2017}.}
Discs and bulges typically show different intrinsic ellipticity.
As discs will be observed more elliptical if edge-on, their ellipticity is primarily driven by inclination angle.
In contrast, bulges are spheroids that are almost invariant under inclination, so they will appear more circular.
In the measurement, we still applied the same ellipticity to both components as part of our 2D modelling,
but we are aware that a positive ellipticity gradient from the intrinsically 3D distribution would induce a bias if not fully captured \citep{bernstein2012}.
Our ellipticity estimate will therefore be a proxy of the inclination angle,
especially for disc-dominated galaxies.
Any residual ellipticity gradient, if significant, will have to be addressed separately as part of the calibration.

\begin{figure}
\centering
\includegraphics[width=\columnwidth,trim=30 0 10 0,clip]{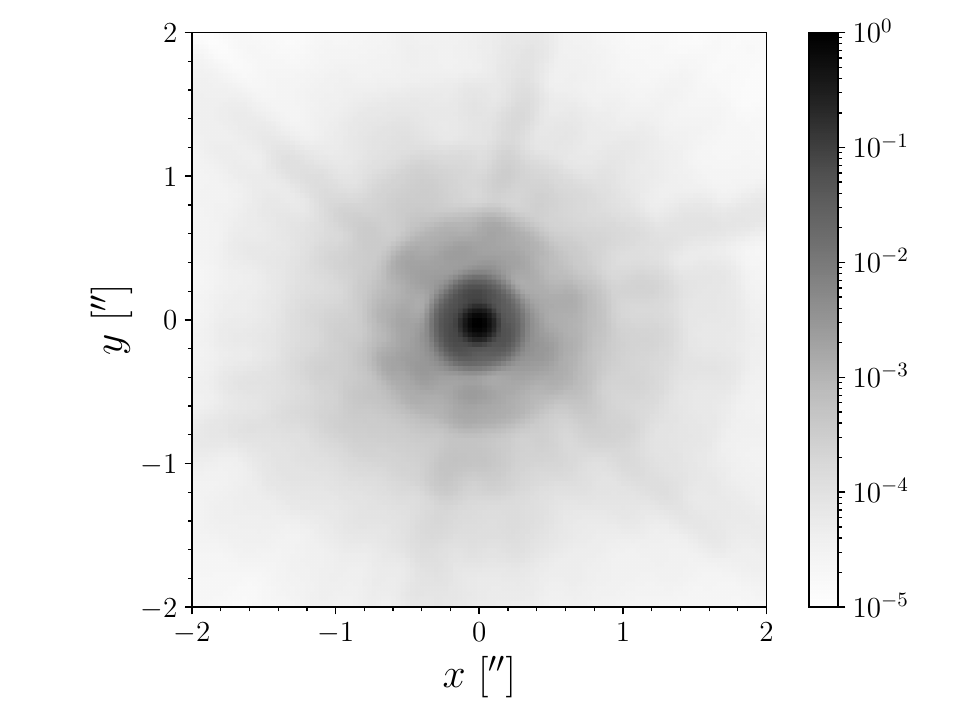}
\vspace{-10pt}
\caption{Example of a \euclid chromatic PSF for an assumed SED of a SBc-type galaxy at a redshift of one.
The flux in the image was rescaled to its maximum value
and oversampled by a factor of three with respect to the native VIS pixel size of $\ang{;;0.1}$.
Diffraction spikes are clearly visible at a significant distance from the centre despite the blurring due to the chromaticity.
The full width at half maximum, including its variation across the field of view, is $\ang{;;0.1564}^{+\ang{;;0.0040}}_{-\ang{;;0.0019}}$, comparable with the \euclid pixel size,
so images will be undersampled at the Nyquist limit.
The ellipticity is $\epsilon_{1,\text{PSF}}=0.017^{+0.038}_{-0.024}$ and $\epsilon_{2,\text{PSF}}=0.001^{+0.042}_{-0.020}$,
with the superscript and subscript denoting absolute ranges.}
\label{fig:psf}
\end{figure}

The \euclid telescope, optical elements, and detector introduce distortions of the input galaxy brightness distribution, which must be corrected.
The effect is mostly convolutive, which tends to blur the galaxy image further.
An example of a typical PSF image for a space-based telescope like \euclid is given in \figref{fig:psf}.
The PSF is:
(\textit{i}) close to being diffraction limited; (\textit{ii}) undersampled due to the half width being comparable with the pixel size;
(\textit{iii}) chromatic due to the integration over a wide range of wavelengths in the VIS filter;
(\textit{iv}) SED dependent due to integration being weighted by a combination of galaxy bulge and disc SEDs;\footnote{With galaxy bulges being, on average, redder than discs.
However, PSF images will be generated from the total galaxy SED.}
(\textit{v}) spatially variant across the field of view due to optical distortions at the exit pupil;
and (\textit{vi}) epoch variant due to varying Solar aspect angle throughout the mission inducing thermal distortion on the optics.
A comprehensive study of the modelling will be presented elsewhere (Duncan et al., in prep.). 
A smaller contribution also comes from the CCD pixel response function,
which models the response of the detector pixel as an integrated measure of the incoming flux illuminating individual pixels.
In forward modelling, we included all the convolutive effects due to the telescope PSF and CCD,
individually for bulge, disc, and for each image exposure.
Colour gradients originate from incorrectly using the total galaxy SED when generating the PSF image, 
while the bulge and disc will have naturally different SEDs \citep{semboloni2013,er2018}. 
Using individually PSF-convolved model components may help control colour gradients,
if individual SEDs were available.
However, the impact of colour dependence and gradients on our analysis was not addressed here since
we assumed the same broadband PSF as obtained from an SED of an SBc-type galaxy at a redshift of one in both simulations and measurements. 
Further non-linear distortions, such as in the case of charge transfer inefficiency \citep[CTI,][]{rhodes2010}
and the brighter-fatter effect \citep[BFE,][]{antilogus2014},
are typically corrected for at the image pre-processing stage,
but residuals could still affect the shear measurement \citep{massey2014,israel2015}, which we did not include here.

Galaxy modelling for large-volume surveys like \euclid requires fast and efficient rendering of the images.
All operations described so far are best suited to work in the Fourier space.
We adopted a similar approach to \texttt{galsim} \citep{rowe2015}.
Consider the generic profile $I(r)$, which could be either Eq.\;\eqref{eq:disc_model} or \eqref{eq:bulge_model}. 
Because of its isotropy, the 2D Fourier transform is the 1D Hankel transform,
\begin{equation}\label{eq:hankel_transform}
\tilde{I}(k)=2\pi\int_0^\infty I(r)\,\text{J}_0(kr)\,r\de r~,
\end{equation}
where $k$ is the Fourier frequency (sampled on an oversampled grid) and $\text{J}_0$ is the Bessel function of the first kind.
We call $\tilde{I}(k)$ the template model which any profile with arbitrary choice of parameter values (ellipticity, size, and position offset) can be 
derived from.\footnote{For computational purposes, the transform is calculated once and saved in a cached 1D array to minimise computing time.}
To render a galaxy profile with parameters $\epsilon_1$, $\epsilon_2$, and $r_\text{e}$,
we applied the inverse distortion matrix $\tens{S}^{-1}$ to coordinates in Fourier space, so anisotropic coordinates are now defined by $\vec{k}'=\tens{S}^{-1} \vec{k}$.
A position shift by $\vec{\delta r} =(\delta x,\delta y)$ from the centre\footnote{As it will be explained later in the section,
in reality the modelling actually includes position shifts from right ascension, $\alpha$, and declination, $\delta$.
Therefore, the position parameters will be $\Delta\alpha$ and $\Delta\delta$.} is equivalent to the phase $\vec{k}'\cdot \vec{\delta r}$.
The sheared-stretched-shifted model becomes
\begin{equation}\label{eq:hankel_affine_transformation}
\tilde{I}'(\vec{k'})=\frac{1}{\det\tens{S}}\,\mathrm{e}^{-2\pi \mathrm{i}\,\vec{k}'\cdot\,\vec{\delta r}}\,\tilde{I}(k')~,
\end{equation}
where $\tilde{I}(k')$ is the template calculated at the sheared-stretched Fourier mode $\vec{k'}$.
Since isotropy is lost through the operation above, $\tilde{I}'(\vec{k'})$ is no longer a Hankel transform but a full Fourier transform,
which is a function of the vector $\vec{k'}$.
We alleviate the problem of undersampling by calculating the PSF and galaxy model 
on coordinates with a common oversampling factor of three.
Finally, the oversampled models for PSF and galaxy is multiplied together, the
convolved model is downsampled by the same factor to the actual pixel scale,\footnote{Downsampling in real space corresponds to aliasing in Fourier space, 
that is, $n$-folding the transform and summing up.} and the downsampled convolved model is inverse fast Fourier transformed to real space.
We applied the operations of shear and stretch to the template bank described in Appendix\;\ref{app:templates} to get any object of arbitrary ellipticity and size
before the convolution with the PSF takes place as previously discussed.

The final galaxy model is a linear mixture of PSF-convolved co-centred components.
We label the profiles with a subscript `d' for disc and `b' bulge:
\begin{equation}\label{eq:galaxy_model}
I(r) = F_\text{d}\,I_\text{d}(r) + F_\text{b}\,I_\text{b}(r)~,
\end{equation}
where $F_\text{d}$ and $F_\text{b}$ are disc and bulge fluxes, and $F=F_\text{d}+F_\text{b}$ is the total flux.
If $\bt$ is the bulge fraction, then the fluxes are also defined by $F_\text{d}=F(1-\bt)$ and $F_\text{b}=F\;\bt$.

To summarise, given pre-computed template models for disc and bulge, we can generate a galaxy model with a desired ellipticity, size, and position by carrying out all the operations with simple algebra in Fourier space on oversampled coordinates, and then take one Fourier transform each time at the end.
This is sufficiently fast for an intensive, repeated calculation of the same model with varying realisations of galaxy parameters (ellipticity, size, position offset, and fluxes).
However, we kept $n_\text{b}$, $r_{\max}/r_\text{e}$, and $r_\text{h}/r_\text{e}$ fixed in our modelling as allowing too much freedom would induce strong degeneracies between parameters and complicate the measurement substantially.
We address the model bias sensitivity in \secref{sec:model_bias}.

\subsection{Likelihood} \label{sec:likelihood}

Suppose we have multi-exposure image data vectors $\vec{D}=\{\vec{D}_\text{exp}\}$.\footnote{On average 4 exposures for the Euclid Wide Survey and 64 for the Deep Survey fields.}
We wish to estimate the model, $\vec{I}=\{\vec{I}_\text{exp}\}$, that best represents the available exposures.
The model $\vec{I}=\vec{I}(\epsilon,\theta,\phi)$ is a function of ellipticity $\epsilon=(\epsilon_1,\epsilon_2)$,
nuisance parameters $\theta=(r_\text{e}, \delta x, \delta y)$,\footnote{As explained, though, 
we model positions in world coordinates.}
and linear flux nuisance parameters $\phi=(F_\text{d},F_\text{b})$.

Assuming Gaussian data,\footnote{The Gaussian approximation holds true in the limit of large counts in the image.} the likelihood can be written as
\begin{equation}\label{eq:likelihood}
\ln p(\vec{D}|\epsilon,\theta,\phi)=-\frac{1}{2}\Big[\vec{D}-\vec{I}(\epsilon,\theta,\phi)\Big]^\top \tens{C}^{-1} \Big[\vec{D}-\vec{I}(\epsilon,\theta,\phi)\Big]+\text{const}~,
\end{equation}
where $\tens{C}$ is the noise covariance matrix usually estimated from the data as a block diagonal matrix,
and the normalisation constant, $1/2\ln[(2\pi)^\lambda\det\tens{C}]$ ($\lambda$ is the dimensionality), is ignored.
The noise is intrinsically non-stationary since various noise sources (such as the Poisson noise\footnote{Poisson noise is a significant 
noise source especially for bright objects. This term is included in the simulations, but not in the measurement as it would require prior knowledge of the distribution profile that is being measured.} from the objects in the image) vary spatially.
Because the model is linear in the component fluxes, $\vec{I}=F_\text{d}\,\vec{I_\text{d}}+F_\text{b}\,\vec{I_\text{b}}$,
it is straightforward to integrate over the fluxes, $\phi=(F_\text{d},F_\text{b})$, and we have the following marginalised likelihood,
\begin{equation}\label{eq:marginalised_likelihood}
\ln p(\vec{D}|\epsilon,\theta)=\frac{1}{2}\,\mathcal{F}^{-1}_{ij}(\epsilon,\theta)\,\rho_i(\epsilon,\theta)\,\rho_j(\epsilon,\theta)+\text{const}~,
\end{equation}
where $i$ indexes the model component, $\rho_i(\epsilon,\theta)=\vec{D}^\top \tens{C}^{-1} \vec{I}_i(\epsilon,\theta)$ is a $2\times1$ vector,
$\vec{I}_i=\partial\vec{I}/\partial F_i$ ($i=\text{d},\text{b})$,
and $\mathcal{F}_{ij}$ is the $2\times2$ Fisher matrix,
\begin{equation}\label{eq:fisher}
\mathcal{F}_{ij}(\epsilon,\theta)=\vec{I}_i(\epsilon,\theta)^\top \tens{C}^{-1} \vec{I}_j(\epsilon,\theta)~.
\end{equation}
We note that because the right-hand side of Eq.\;\eqref{eq:marginalised_likelihood} is quadratic in $\rho_i$ and $\mathcal{F}_{ij}$ is positive definite,
we find $\ln p(\vec{D}|\epsilon,\theta)>0$.
A full derivation of the marginalised likelihood, including edge cases and implementation, can be found in Appendix\;\ref{app:likelihood}.
The dimensionality of the problem has now been reduced from 7 free parameters to 5: ellipticity, size, and position offsets.\footnote{Having
assumed that the two components are co-centred and the bulge size is locked to the disc size by a fixed rescaling.}

Forward modelling provides solid grounds for a further generalisation to measuring multiple objects jointly,
especially if they are observed within a short angular separation
such as for neighbours.
We label each likelihood with the index $\omega$ running through the objects being jointly measured, $\ln p_\omega(\vec{D}|\epsilon,\theta,\phi)$.
The joint likelihood is then
\begin{equation}\label{eq:joint_likelihood}
\ln p(\vec{D}|\{\epsilon,\theta,\phi\}_\omega) = \sum_\omega \ln p_\omega(\vec{D}|\epsilon,\theta,\phi)~,
\end{equation}
where $\{\epsilon,\theta,\phi\}_\omega$ is the set of all parameters for all the objects being measured.
In the above equation we assumed the independence of the individual likelihoods.
This is a fair assumption since close neighbours will very often be so due to random visual alignment.
Consequently, those galaxies will be at different redshift and will have different shear.
A much smaller fraction will include tidal interaction. In this case, the galaxies will be at the same redshift and have the same shear.
It it then expected to have some degree of correlation between the individual likelihoods.
In more extreme but much rarer cases, the galaxies will be tidally interacting, and therefore our S\`ersic modelling would break down entirely
as we did not include any extra correlation term.
Despite affecting a very small fraction of objects, dedicated simulations would be required to assess the impact on shear bias.
Also, it is worth noting that we need to be careful with the marginalisation of the individual likelihoods.
The main issue lies in the marginalised likelihood of Eq.\;\eqref{eq:marginalised_likelihood}.
This relies on calculating $\rho_i(\epsilon,\theta)$ for the various model components.
However, when multiple objects are present in the same neighbourhood, this quantity
will effectively introduce correlation between the likelihoods of the two objects.
Therefore, the statistical independence required to multiply likelihoods together will not be ensured.
We verified in testing that not marginalising individual likelihoods is indeed the correct approach to the problem.
The joint likelihood is defined in a $7\times N$-dimensional parameter space,
where $N$ is the number of objects being measured jointly, with $N=2$ being a typical number found in testing. 
For increased stability, we first optimised the likelihood for fluxes and positions offsets, and then also for ellipticity and sizes.
This proved to be very robust as opposed to iterating over individual objects after masking neighbours
to achieve a reliable initial guess \citep{drlica2018}.
One key benefit of MCMC is that it marginalises the ellipticity of an object over all remaining nuisance parameters, which include object nuisance parameters as well as the parameters of the other objects included in the joint sampling (see \secref{sec:mcmc}).

Our prior is based on enforcing hard bounds on all parameters:
$0\le|\epsilon|<1$ given that the modulus of ellipticity in Eq.\;\eqref{eq:ellipticity} cannot exceed 1;
$0\le r_\text{e} \le \ang{;;2}$, where the upper bound is based on observations made in the \hubble Deep Fields;
position offsets are restricted to $\pm\,\ang{;;0.3}$ since the accuracy to which detections are made is typically sub-pixel;
and fluxes are positive.
A more informative prior could be derived from real observations in the future.
A summary of all parameters, being free, fixed or derived, is presented in \tabref{tab:parameters}.

\begin{table*}
\caption{Summary of free, fixed, and derived parameters in the modelling.}
\label{tab:parameters}
\centering
\begin{tabular}{lccccccp{7cm}}
\hline\hline
& Free & Fixed & Derived & Initial value & Bounds & Unit & Description \\
\hline
$\epsilon_1$ & \checkmark & & & 0 & $|\epsilon|<1$ & & first component of ellipticity in the tangent plane to $(-\alpha,\delta)$ as defined in Eq.\;\eqref{eq:shear_matrix} \\
$\epsilon_2$ & \checkmark & & & 0 & $|\epsilon|<1$ & & second component of ellipticity in the tangent plane to $(-\alpha,\delta)$ as defined in Eq.\;\eqref{eq:shear_matrix} \\
$r_\text{e}$ & \checkmark & & & 0.3 & $[0,2]$ & arcsec & effective radius: disc scale length as defined in Eq.\;\eqref{eq:shear_matrix} \\
$\Delta\alpha$ & \checkmark & & & 0 & $[-0.3,0.3]$ & arcsec & offset in right ascension as a phase prefactor in Eq.\;\eqref{eq:hankel_affine_transformation} \\
$\Delta\delta$ & \checkmark & & & 0 & $[-0.3,0.3]$ & arcsec & offset in declination as a phase prefactor in Eq.\;\eqref{eq:hankel_affine_transformation} \\
$\alpha$ & & & \checkmark & & & degree & right ascension \\
$\delta$ & & & \checkmark & & & degree & declination \\
$F_\text{d}$ & \checkmark & & & & $F_\text{d}\ge0$ & & disc flux: marginalised over or free as defined in Eq.\;\eqref{eq:galaxy_model} \\
$F_\text{b}$ & \checkmark & & & & $F_\text{b}\ge0$ & & bulge flux: marginalised over or free as defined in Eq.\;\eqref{eq:galaxy_model} \\
$F$ & & & \checkmark & & $F\ge0$ & & total flux \\
$\bt$ & & & \checkmark & & $[0,1]$ & & bulge fraction as defined in text after Eq.\;\eqref{eq:galaxy_model} \\
$\snr$ & & & \checkmark & & $\snr\ge0$ & & signal-to-noise ratio \\
$\IE$ & & & \checkmark & & & & magnitude: depends on the assumed \zeropoint, exposure gain and integration time \\
$n_\text{b}$ & & \checkmark & & 1 & & & S\'ersic index of the bulge as defined in Eq.\;\eqref{eq:bulge_model} \\
$a_\text{b}$ & & \checkmark & & depends on $n_\text{b}$ & & & S\'ersic coefficient of the bulge as defined in Eq.\;\eqref{eq:bulge_model}, see \citet{peng2002} \\
$r_\text{h}/r_\text{e}$ & & \checkmark & & 0.15 & & & bulge half-light radius to effective radius ratio as defined in text after Eq.\;\eqref{eq:bulge_model}, see Welikala et al., in prep. \\
$r_\text{max}/r_\text{e}$ & & \checkmark & & 4.5 & & & truncation radius to effective radius ratio as defined in text after Eq.\;\eqref{eq:bulge_model} \\
\hline\hline
\end{tabular}
\tablefoot{The value should be interpreted as an initial guess or constant depending on whether the parameter was allowed to vary or remain constant.}
\end{table*}

\subsection{Massive Markov chain Monte Carlo sampling} \label{sec:mcmc}

Shear measurement poses serious difficulties in identifying the best strategy to sample
the posterior probability distribution of Eq.\;\eqref{eq:posterior}, assuming the likelihood of Eq.\;\eqref{eq:marginalised_likelihood} or \eqref{eq:joint_likelihood}.
\begin{enumerate}
\item Since the lensing sample is very broad in morphological properties, it will contain both low and high $\snr$ objects,
whose posterior probability distribution can be either very broad or very narrow;
hence any sampling strategy must be robust to this variability.
\item If $N$ is the number of objects being measured jointly to mitigate the neighbour bias, the dimensionality of the problem is $7\times N$ (with $N$ being typically 2 and rarely 3 or 4); sampling must then be resilient to the large dimensionality, and provide marginalisation and error estimations
with minimum overheads.
\item The shape of the distribution is a strong function of object parameters,
such as ellipticity and size, and therefore it varies significantly across the sample;
without prior knowledge of the physical properties of each galaxy,
any sampling method must run in a consistent, robust way.
\item Given the large sample size of order $10^9$ galaxies, sampling the posterior is a computational challenge,
so a trade-off between method complexity, runtime, and access to computing resources must be identified.
\item The sampling must be completely automated, without human supervision, and no fine tuning of sampling parameters and options is allowed.
\end{enumerate}
Considering all these challenges, our priority must be the average convergence property of the sampler.
The best strategy identified is MCMC, which allowed us to sample the posterior generated from the marginalised likelihood of Eq.\;\eqref{eq:marginalised_likelihood}
for an individual object, or the joint likelihood of Eq.\;\eqref{eq:joint_likelihood} for a group of objects
in an efficient and consistent way,
for all $10^9$ objects in the sample (hence the adjective `massive').
More importantly, MCMC seems to be the best choice to tackle neighbours, particularly as an estimate can be found
in a high dimensional parameter space.
It is worth noting that another key benefit of MCMC sampling is that
it is both a maximisation and sampling procedure.
The maximisation happens during the burn-in phase where the sampler tries to reach the region of higher probability.
The actual sampling happens in the later stage of the chain after the burn-in phase.
The marginalisation over nuisance and error estimation are then natural by-products with no extra overhead.
This implies that not only can ellipticity estimates be marginalised over object nuisance parameters, but also
over other object parameters in the joint group,
hence minimising the impact of neighbours in the final ellipticity estimate.

When searching for such an algorithm that could potentially suit our needs, we considered a number of potential candidates
that are widely used in cosmology and other fields \citep{mackay2002}.
The development of various sampling methods is primarily
driven by the quest to achieve lower auto-correlation and higher acceptance rate \citep{hastings1970,swendsen1986,skilling2006,goodman2010,
foreman2013,betancourt2017,karamanis2021,lemos2022}.
Although appealing, all these methods do suffer from increased complexity,
which is the limiting factor in large-scale applications, where `large' in this context implies runs repeated over a billion times.
Even for the most sophisticated methods, it is often realised that a good initial guess is the key for good sampling of the posterior.

For shear measurement on the scale of large galaxy surveys, there is not much room for sampling complexity.
The method has to be light enough and yet robust to all the posteriors that need to be sampled.
Furthermore, the likelihood runtime limits the maximum number of samples that can be drawn for each galaxy
without having an overall impact on the survey analysis runtime.
The likelihood runtime is mostly dominated by the model component generation.
The measurement is dominated by sampling the posterior, plus some additional pre-processing,
therefore to limit the galaxy runtime to around a few seconds per galaxy, the MCMC sampling must be sparse.
This is considered acceptable as we are interested in accurate shear estimates, which are found by averaging over ellipticity measurements.
We chose an improved version of the Metropolis-Hastings algorithm, which was modified in two ways:
(\textit{i}) it incorporates some of the ideas of parallel tempering \citep{swendsen1986,sambridge2013}, so the parameter space can be sampled on a larger scale initially,
and then on the smaller scale;
(\textit{ii}) it updates the proposal distribution during the burn-in phase, automatically tuning it to find a good match with the target distribution.
Let $\vartheta$ be the generic vector of all parameters.
As explained in \secref{sec:likelihood}, this is $\vartheta=(\epsilon,\theta)$ for individual galaxy measurement or $\vartheta=(\epsilon,\theta,\phi)$ when jointly measuring groups of galaxies.
Ignoring the evidence since our method is invariant to it, the Bayes' theorem gives us
\begin{equation}
p(\vartheta|\vec{D}) \propto p(\vec{D}|\vartheta)\,p(\vartheta)~.
\end{equation}
The parameter vector at the current iteration step is denoted with $\vartheta_t$, 
where $t=0,\ldots$ is the MCMC sample index.
We also define a tempering function, $T_t$, as a function of $t$.
This acts as a Boltzmann temperature and its expression will be defined later in the text.
When the temperature is high, $T_t\gg1$, sampling from the posterior is equivalent to sampling globally from the prior.
When the temperature is gradually reduced, as in annealing, $T_t\rightarrow1$, we begin sampling directly from the posterior.
We define such a tempering function for application during the burn-in phase only, and make sure $T_t=1$ for the final part of the chain where we will take sample from.
The method goes as follows.
\begin{enumerate}
\item At $t$, draw a new sample $\vartheta'_t$ from the proposal distribution $q(\vartheta'_t|\vartheta_t)$;
here we assume a symmetric Gaussian proposal with mean $\vartheta_t$ and a pre-defined diagonal covariance of $10^{-4}$ on all parameters (in units of arcsec for size and position offsets).
\item Calculate the logarithm of the acceptance ratio
\begin{equation}
\ln A=\frac{\ln p(\vec{D}|\vartheta'_t) - \ln p(\vec{D}|\vartheta_t)}{T_t} + \ln p(\vartheta'_t) - \ln p(\vartheta_t)~.
\end{equation}
\item Accept or reject $\vartheta'_t$ with probability $A$, that is, draw $u$ from the uniform distribution on $[0,1]$
and accept $\vartheta'_t$ if $u<\min(1, A)$;
to speed things up and avoid calculating the likelihood outside the prior, we immediately reject $\vartheta'_n$ if $p(\vartheta'_n)=0$.
\end{enumerate}
For consistency, all posteriors are sampled from an initial guess that is a circular galaxy of mean size, $r_\text{e}=\ang{;;0.3}$, 
and zero offset from the nominal position in the sky.
If we were to run any MCMC method from this point onward we would end up with varying autocorrelation time depending on how far the actual galaxy is from the initial guess; therefore, we would need to wait longer for very elliptical, small, or large galaxies.
To improve the convergence of the chains within a smaller number of iterations, we achieved a better initial guess by running an initial maximisation of the posterior before the actual MCMC.
We ran the conjugate-gradient search method \citep{powell1964} restricted to only 100 function evaluations,
and then a downhill simplex search \citep{nelder1965}.
The burn-in phase of the MCMC starts right afterwards.
During this phase the temperature is gradually lowered
to one. We adopted the following power law cooling scheme \citep{cornish2007},
\begin{equation}
T_t=
\begin{cases}
10^{\,f_\text{heat}(1-t/t_\text{cool})}, & \text{if}~t<t_\text{cool} \\
1, & \text{otherwise}
\end{cases}~,
\end{equation}
where $f_\text{heat}=10$ is the heat factor and $t_\text{cool}=100$ is cooling-down sample index.
The parameter $t_\text{cool}$ represents the number of samples it takes for the tempering function to become one.
We note that $T_0=10^{\,f_{\text{heat}}}$ and $T_{t_\text{cool}}=1$.
We begin with a diagonal Gaussian proposal of width 0.01 on all parameters,
which is then recalculated from the chains every 100 samples
and rescaled by the factor $2.4\,{\lambda}^{-1/2}$, with $\lambda$ being the number of parameters \citep{dunkley2005}.
The burn-in phase lasts for a total of 500 samples,
which is long enough for the tempering function to become 1,
the proposal covariance to be recalculated 5 times,
and the chain to stabilise and reach the high probability region
(well before we start accumulating the final chain samples).
The final phase lasts for an additional $N_\sfont{MC}=200$ samples. 
Again, this is plenty to estimate both the mean and covariance of the chains with enough precision, 
but we recognise that sampling noise may still be non-negligible.

A good quantitative way to test the convergence of the chains is to investigate their auto-correlation.
We do so for a variety of galaxies and results are shown in Appendix\;\ref{app:mcmc}.
A less quantitative way is to verify that the acceptance rate is within the expected range.
We also increased the final 200 samples up by a factor of five, without noticing any significant difference in the shear results.
For further verification, we compared the method with our implementation of affine invariant \citep{goodman2010} and parallel tempering \citep{swendsen1986,sambridge2013}.\footnote{These alternative methods can be controlled by specific parameters in the code.}
While these methods produce better ellipticity chains, they did not show any significant advantage in recovering shear,
but increased complexity and therefore runtime overhead.

Once the samples are drawn from the distribution function,
calculating the mean and width of the marginalised distribution becomes straightforward.
Our point estimate for the ellipticity component marginalised over nuisance is the mean of the chain after the burn-in phase,
\begin{equation}
\hat{\epsilon}_i = \frac{1}{N_\sfont{MC}}\sum_t \epsilon_{i,t}~,
\end{equation}
where $i=1,2$, and $\epsilon_{1,t}$ and $\epsilon_{2,t}$ are the two ellipticity chains.
The marginalised ellipticity covariance matrix is
\begin{equation}
C_{ij} = \frac{1}{N_\sfont{MC}-1}\sum_t (\epsilon_{i,t}-\hat{\epsilon}_{i})\,(\epsilon_{j,t}-\hat{\epsilon}_{j})~.
\end{equation}
We calculated the averaged per-component variance (ignoring negligible covariance between components),
\begin{equation}
C_\epsilon=\frac{1}{2}(C_{11}+C_{22})~,
\end{equation}
and chose to define the galaxy shear weight by
\begin{equation}\label{eq:weight}
w=\frac{1}{C_\epsilon+\sigma_\epsilon^2}~,
\end{equation}
where $\sigma_\epsilon$ is the assumed shape noise (i.e., the width of the 1D intrinsic ellipticity distribution),
ideally estimated in tomographic bins from deeper measurements.
In this modelling, as $C_\epsilon$ quantifies the noise level in the data, faint galaxies will be automatically downweighted as opposed to very bright galaxies that will always have a maximum finite weight.
Additionally, we note the negligible sensitivity to the choice of the $1/2$ factor in $C_\epsilon$.\footnote{In fact, 
ignoring the $1/2$ factor would lead to a redefinition of the weight,
$w=1/[(C_{11}+C_{22})/2+\sigma_\epsilon^2]\propto1/[(C_{11}+C_{22})+\sigma_\epsilon'^2]$ with $\sigma_\epsilon'=\sqrt{2}\,\sigma_\epsilon^2$,
but results show weak sensitivity to the value assumed for $\sigma_\epsilon$,
as it will be demonstrated at the end of \secref{sec:shear_bias}.}
The MCMC provides a convenient and efficient way to calculate both the mean and width of the ellipticity posterior
at no extra computational cost.
The weights can then be used to define sample averages, such as in one-point estimates:
\begin{equation}
\hat{g}_i=\frac{1}{\sum_k w_k}\sum_k w_k\,\hat{e}_{i,k}~,
\end{equation}
where $k$ indexes the galaxies in the lensing catalogue. The generalisation to two-point estimates is straightforward.
Please note that any choice of weight leads to shear bias due to correlation with the measured ellipticity,
and this is tested in \secref{sec:shear_bias}.

We also implemented the self-calibration of ellipticity proposed by \citet{fenech2017}\footnote{The same correction can also be proved to map,
within some approximations, to other studies \citep{cox1968,refregier2012,hall2017}}
via importance sampling and likelihood ratio while checking the quality of the sampling weights \citep{wraith2009}
without finding strong evidence of improvement.
As results will be dominated by other larger effects, we leave out further discussion from this paper.

\subsection{Handling real data} \label{sec:real_data}

Handling real data requires being careful with additional aspects of the measurement.
For instance, throughout our discussion we proposed that our sampling strategy is best suited to handle
the presence of neighbours, that is, resolved objects\footnote{In this context `resolved' implies that the object has been detected and at least partially deblended so that our measurement can be applied to all reported object positions.} close to the lensing targets.
However, the situation is complicated by the fact that there is more variety in real data as the brightness distribution of an object can be contaminated in different ways depending on the nature of the nearby objects:
\begin{enumerate}
\item neighbours (resolved galaxies or stars);
\item blends (unresolved galaxies or stars);
\item any other contamination (cosmic rays, transients, or ghosts).
\end{enumerate}
Each case leads to a particular type of bias, and therefore
we deal with close objects in two ways.
First, neighbours are grouped with a friend-of-friend algorithm 
to a maximum angular separation of $r_\text{friend}=\ang{;;1}$.
If the separation is too small, the objects will be mostly measured in isolation;
therefore, they will still be affected by the neighbours due to improper masking.
If the separation is too large, the groups will begin to be unmanageable in size, 
but the benefit in controlling the neighbour bias will be negligible.
We found that $\ang{;;1}$ is a good trade-off between measuring $N$ close neighbours jointly 
within a default postage stamp size of $\ang{;;38.4}$,
and the non-negligible overhead in sampling a $7\times N$ dimensional posterior.
The joint analysis of object groups also gave us a good control of neighbour bias, leading to a correction of $m\approx-7\times10^{-4}$ as it will be shown
later in the paper.\footnote{We do not attempt to optimise our choice of $r_\text{friend}$ due to a number of other effects being more substantial than this.}
Second, the segmentation maps and masks that are usually provided with the data \citep{bertin2020,kuemmel2020}
were combined in a binary map and passed on to the likelihood to mask out objects in different groups.
Detector artefacts or cosmic rays were also masked out in a similar way.
In this case, to be even more conservative, we further dilated the masks by one pixel
so most of contamination bias should be taken care of.
But masking also helps partially with blends because objects that are false negatives according to the detection strategy may sometimes be true positives according to the masking procedure and would therefore be masked out.
Blending with faint galaxies were demonstrated to be relevant when trying to calibrate methods that are particularly sensitive to the problem \citep{martinet2019}.
We demonstrate that, to some extent, this is also the case in our simulations where we measured objects deeper than the \euclid nominal survey depth, as we will show in the next section. 

Real images have a background level that needs to be subtracted.
\lensmc uses the background estimates and noise maps that the \euclid processing provides,
but residual local background variations are subtracted at the individual object group level.
This is implemented by post-masking median subtraction.
Likewise, the standard deviation of the background noise is estimated after masking.
We measured galaxies and stars with the same model described earlier in this section.
We found that good star-galaxy separation is based on selecting objects whose measured size is greater than the PSF size.
This method fits well with the joint measurement of groups, however at the price of rejecting faint galaxies that would nevertheless
have a negligible weight or be hard to distinguish from faint stars.
More details will be given in \secref{sec:results}.

The measurement was made in sky coordinates using the supplied world coordinate system (WCS) solution, which includes both linear and non-linear distortions \citep{calabretta2002}.
We assumed the default coordinate system $(-\alpha,\delta)$, where $\alpha$ is the right ascension and $\delta$ is the declination.
We measured position offsets from the provided nominal position in arcsec.
The resulting $\alpha$ and $\delta$ were reported in degrees, and $r_\text{e}$ in arcsec.
Likewise, the measured ellipticity is defined in the tangent plane to the $(-\alpha,\delta)$ coordinate system centred at the object position.
We used the WCS to estimate a local linear approximation of the mapping from sky coordinates to tangent plane coordinates at the nominal position.
We defined 9 points in a square grid of size $\ang{;;0.3}$ in pixel coordinates centred at the nominal position, 
mapped them to sky coordinates,
and finally mapped the sky coordinates back to the undistorted tangent plane. 
The Jacobian matrix, which models the local linear approximation of the mapping, is the least-square solution to the mapping from sky coordinates to tangent coordinates.  
As part of this procedure, we also calculated the astrometric offsets due to the exposures being dithered differently.
The brightness model was then correctly generated taking into account both the local distortion and astrometric offsets so
all the observables were measured uniquely in tangent plane to sky coordinates.

When reporting our measurement we always compute $\chi^2=-2\ln p/\nu$,
where $p$ is the likelihood of Eq.\;\eqref{eq:marginalised_likelihood} or \eqref{eq:joint_likelihood} calculated at the mean estimate
and $\nu$ is the number of degrees of freedom.
The $\chi^2$ will not in general follow the theoretical distribution for a number of reasons.
The noise is only approximately Gaussian and non-Gaussianities will always be present in the data.
For instance, key examples are the Poisson noise from the background and the object, digitalisation noise, non-linear artefacts, modelling mismatches, or failures in the sampling.
Nonetheless, the $\chi^2$ metric defined in this way is still a good statistical measure of the quality of the measurement.
We also compared the $\chi^2$ calculated above with the same quantity, which we call $\chi^2_\text{bkg}$, after having masked out all the objects, 
which is expected to be just noise. Objects will be flagged up if the $\chi^2$ is not consistent with the background.
Following an F-test procedure, we calculated the test statistic $(\chi^2 / \nu)\,(\chi^2_\text{bkg} / \nu_\text{bkg})^{-1}$
and rejected the null hypothesis (the measured $\chi^2$ is consistent with the background) if the p-value was less than 0.01.
Nonetheless we found that the impact of flagged objects is negligible, so we usually included them in our results.
However, that may not be true for real data where the contamination from data artefacts will be more important.

The measurement includes estimation of the object magnitude based on the supplied \zeropoint, gain, and exposure time.
Each exposure may come with its own values, as these vary both spatially and temporally. Therefore it is important to normalise the data to common units.
As the data is measured in analogue-to-digital units (ADU), we multiply each exposure by $G\,10^{-(\IEzp - \bar{I}_\sfont{E,\,0})/2.5}/\tau$,
where $G$ is the gain in $\unit{e^-/ADU}$, $\IEzp$ is the magnitude \zeropoint, $\bar{I}_\sfont{E,\,0}$ is the average magnitude \zeropoint across the exposures, $\tau$ is the exposure time,
and the data is now in normalised photoelectron count rate of $\unit{e^-/s}$.
The flux, F, is then measured in the same units, and we can estimate the magnitude as follows:
\begin{equation}\label{eq:magnitude_flux_relation}
\IE= -2.5\logten\frac{F}{\unit{e^-/s}}+\bar{I}_\sfont{E,\,0}~.
\end{equation}
The specific values for \zeropoint, gain and exposure time assumed in our simulations will be provided in \secref{sec:simulations}.

Analysing a volume of about 1.5 billion galaxies for \euclid will be a massive computational challenge,
especially if employing MCMC to sample the posterior.
Our measurement on highly realistic images ran at about 5 seconds per galaxy per exposure per computing core,
including joint measurement of groups.\footnote{The overhead of the joint measurement is about half of the quoted total.}
We discussed the benefits of using a fast, efficient implementation of MCMC
in the previous section. Here we want to highlight the fact that all the pre- and post-processing described above adds very little overhead to
the measurement.
We found that the maximum posterior does suffer from a large bias
of $m\approx-1\times10^{-2}$,
which is about twice the bias obtained when using the mean of the MCMC samples.
Since the bias tends to increase with magnitude, we interpret it as the maximum posterior estimate of the ellipticity being more prone to noise bias.
This is further evidence that the MCMC can mitigate noise bias by consistent sampling and marginalisation of a multi-dimensional posterior,
in particular when jointly measuring groups of objects,
with the full complexity of real data and at the modest cost of slightly more overhead.\footnote{Compared to the maximum estimate, the MCMC adds only $40\%$ to the total runtime.}

%% file: tex/sec3_simulations.tex
\section{Simulations} \label{sec:simulations}

In order to validate our measurement method in a realistic setup, we designed a suite of simulations
that incorporate most of the real data effects that future lensing surveys like \euclid will need to account for.
It is essential then to bring in as much realism as possible.
One problem that all shear methods have to deal with is clustering that leads to close neighbours,
which is a concern for \euclid, Rubin, and present surveys as well.
Because the inferred bias depends on the details about the realism of clustering of faint galaxies,
this has to be incorporated in simulations particularly for calibration purposes \citep{kannawadi2019,martinet2019}.
To make our custom simulations realistic and bring in all those effects we are most concerned about,
we utilised the exquisite, state-of-the-art \flagship simulation mock galaxy catalogue \citep{potter2017,castander2024},\footnote{We ingested 
the catalogue version 2.1.10 retrieved from the official website \href{https://cosmohub.pic.es}{cosmohub.pic.es} \citep{carretero2017,tallada2020}.} developed specifically for \euclid.
The same \flagship simulation is also used for the Euclid Science Ground Segment simulations \citep{serrano2024}.
\flagship provides, in particular, a realistic distribution of galaxy morphologies, 
and clustering of galaxies obtained through a full $N$-body dark matter simulation.

\begin{figure}
\centering
\includegraphics[width=\columnwidth]{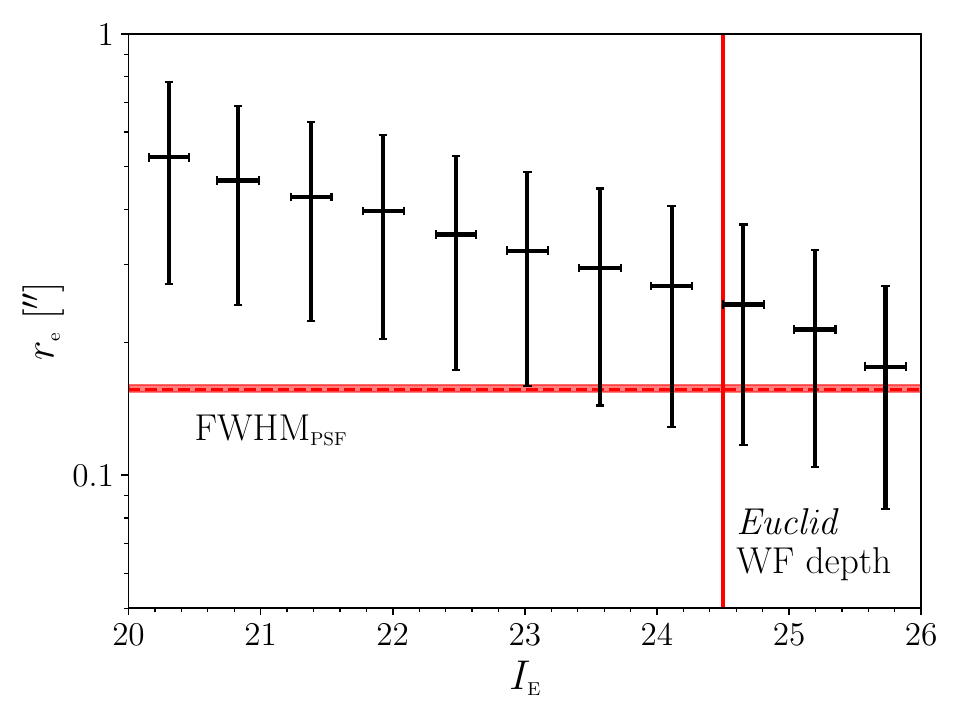}
\vspace{-10pt}
\caption{Input magnitude-size distribution of galaxies.
The data points are the mean $r_\text{e}$ as a function of $\IE$.
Also shown are the standard deviation of $r_\text{e}$ and $\IE$ in each bin.
The horizontal band denotes the PSF full width at half maximum and its variation across the field of view.
A significant fraction of the galaxies have intrinsic effective radii similar to the PSF, especially at the Euclid wide field (WF) depth.}
\label{fig:input_size_vs_mag_distribution}
\end{figure}

\begin{figure}
\centering
\includegraphics[width=\columnwidth]{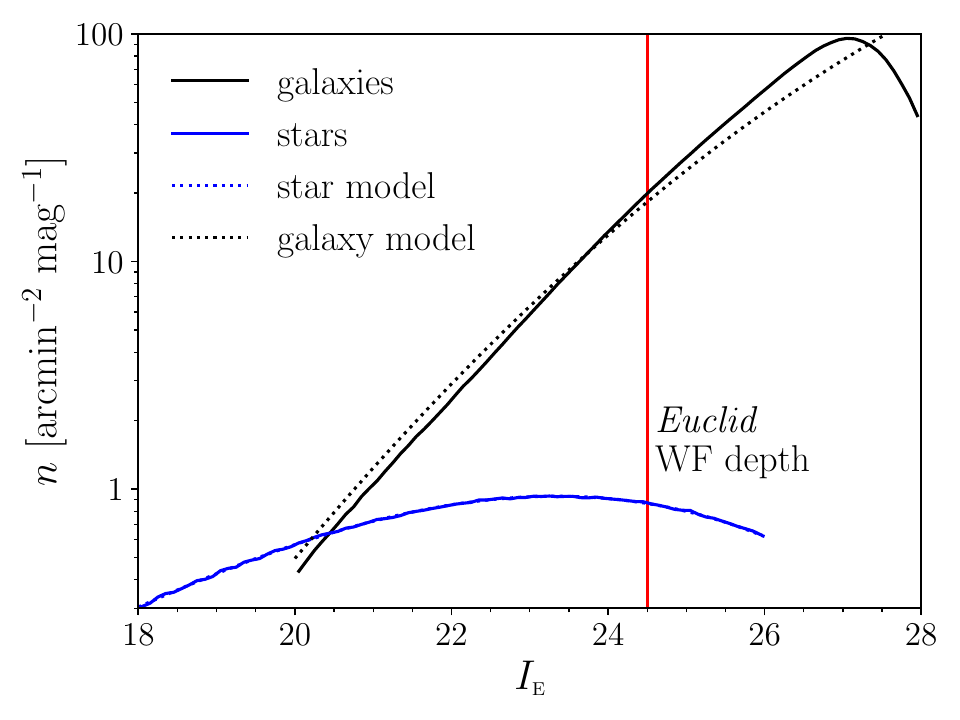}
\vspace{-10pt}
\caption{Input differential number count for galaxies and stars. The solid line is the measured galaxy count from \flagship.
The dashed line is a polynomial model of VIS-corrected magnitudes in the GOODS South ($\IE<26$) and Ultra Deep Field ($\IE>26$).
Stars were drawn from a polynomial model of $i$ magnitudes generated with the Besançon model in the north ecliptic pole.
The cumulative number counts are $\qty{250}{arcmin^{-2}}$ ($\IE<29.5$) for galaxies and $\qty{6}{arcmin^{-2}}$ ($\IE<26$) for stars.}
\label{fig:number_count}
\end{figure}

The morphological parameters and spatial distributions are provided over an octant of the full sky,
which is just less than 40\% of the Euclid Wide Survey.
Here we used values for the provided disc ellipticity and orientation angle, disc scale length, VIS flux, bulge fraction,
and position over a region defined by $150^\circ<\alpha<230^\circ$ and $15^\circ<\delta<85^\circ$.
We also selected all galaxies that are classified in the catalogue as being either central or satellite in the halo, and excluded quasars or high-redshift galaxies.
Figure\;\ref{fig:input_size_vs_mag_distribution} shows the joint size-magnitude distribution of galaxies.
A significant fraction of the galaxies have intrinsic effective radii similar to the PSF, which has a full width at half maximum of $\ang{;;0.1564}^{+\ang{;;0.0040}}_{-\ang{;;0.0019}}$,
and therefore appear only marginally resolved in the PSF-convolved images.
It is worth highlighting that because of the very faint magnitude limit ($\IE<29.5$, but complete to $\IE<27$) a significant fraction of the objects will be too faint
to be detected,
but these will be still included in the background noise.
In addition to galaxies provided by \flagship, we also simulated a uniform spatial distribution of stars up to $\IE<26$.
Figure\;\ref{fig:number_count} shows the number count of galaxies and stars.
The galaxy count was obtained from \flagship and compared against a polynomial model to VIS-corrected magnitudes in the GOODS-South field up to 26 \citep{giavalisco2004} and Ultra Deep Field beyond 26 \citep{beckwith2006}.
Stars were drawn from a polynomial model of $i$ magnitudes generated with the Besan\c{c}on model \citep{czekaj2014} in an area of $\qty{10}{deg^2}$ 
around the north ecliptic pole.
Overall, we obtained a number density of $\qty{250}{arcmin^{-2}}$ ($\IE<29.5$) for galaxies and $\qty{6}{arcmin^{-2}}$ ($\IE<26$) for stars.

We defined sky patches of size $\ang{;;380}$ (about $\qty{40}{arcmin^2}$), broadly corresponding to the size of a single \euclid CCD,
but also included an adjacent area of extra 10\% (called buffer/guard region) all around the patch to draw objects in the image that will not be part of the measurement.
When selecting the morphological properties from the \flagship catalogue
(ellipticity, disc scale length, bulge fraction, position, and flux),
their AB flux was first converted to AB magnitude, then further converted to VIS photoelectron count rate via the magnitude-flux relation of Eq.\;\eqref{eq:magnitude_flux_relation}.
We assumed a constant magnitude \zeropoint of $\IEzp=25.719$,
which was calculated using \euclid as-designed system throughput data.
Star positions are drawn from the count model uniformly in each patch.

All galaxies in each patch had ellipticity assigned by \flagship.
In principle we could use the cosmic shear from \flagship.
However, in this work we applied the same constant shear to all galaxies in a patch,
with the shear varying from one patch to another according to a uniform distribution on a circle of radius $|g|=0.02$ and random orientation.
This choice mimics the typical shear expected for a real survey and 
also minimises the error on multiplicative bias.
We assumed an (infinitely thin) annular distribution as opposed to a disc distribution or an even more realistic log-normal distribution
because we wanted to minimise the statistical error on multiplicative bias, and obviously be as cosmology agnostic as possible.
On the other hand, a variable shear field might in principle introduce an additional correlation with neighbour bias,
particularly if neighbours at different redshifts are considered \citep{maccrann2021}.
However, capturing realistic clustering is the most important aspect of the simulations, which is what we focussed on in this work.
Similarly to previous work \citep{bridle2010,kannawadi2019}, we applied shape noise mitigation by making, in total, 4 clones of each patch 
with all ellipticities rotated by $45^\circ$ while maintaining the same overall constant shear,
which gave us significant leverage on the required simulation volume.
It is worth noting, though, that a varying shear could also be dealt with in a shear response approach,
leading to an increased effective sample size and reduce simulation volume in calibration \citep{pujol2019,jansen2024}.

We set up a suite of simulations for each of 9 realisations of the PSF image drawn at different positions in the field of view.
While varying the PSF image, we kept the objects at the same positions.
We assumed a \euclid PSF model for a fixed SBc-type galaxy SED at a redshift of one,
the median of the distribution.
The mean ellipticity and its variation across the field of view was:
$\epsilon_{1,\text{PSF}}=0.017^{+0.038}_{-0.024}$ and $\epsilon_{2,\text{PSF}}=0.001^{+0.042}_{-0.020}$,
with the superscript and subscript denoting absolute ranges. 
We note that this variation, if not included in the modelling, would be responsible for an error in the shear measurement that would far exceed science requirements. 
We did not include PSF mismodelling in our simulations as the current \euclid requirement on PSF ellipticity error is already quite stringent, but will be addressed elsewhere.
The Euclid Wide Survey is designed to take 4 dithered exposures (pointings), plus two extra short exposures, of the same sky area.
Most often these will be taken in the same observation.
Hence the PSF model is not expected to vary too much across the exposures, but the images will be different because taken at different positions in the field of view.

We generated \euclid detector images containing galaxies rendered with the brightness model of \secref{sec:galaxy_models} with varying ellipticity, 
$r_\text{e}$, position, and fluxes.
In our initial tests, we made our results insensitive to model bias by construction, and therefore we fixed $n_\text{b}$, $r_\text{h}/r_\text{e}$ and $r_{\max}/r_\text{e}$.
Later on, we address model bias sensitivity by allowing $n_\text{b}$, and $r_\text{h}/r_\text{e}$ to vary.
For stars, we used a restricted model with zero ellipticity and $r_\text{e}$, so we effectively rendered point-like PSF images.
For the measurement, we used the same galaxy profile (with fixed bulge parameters) for all detected objects.

\begin{figure}
\centering
\includegraphics[width=\columnwidth,trim=40 0 60 0,clip]{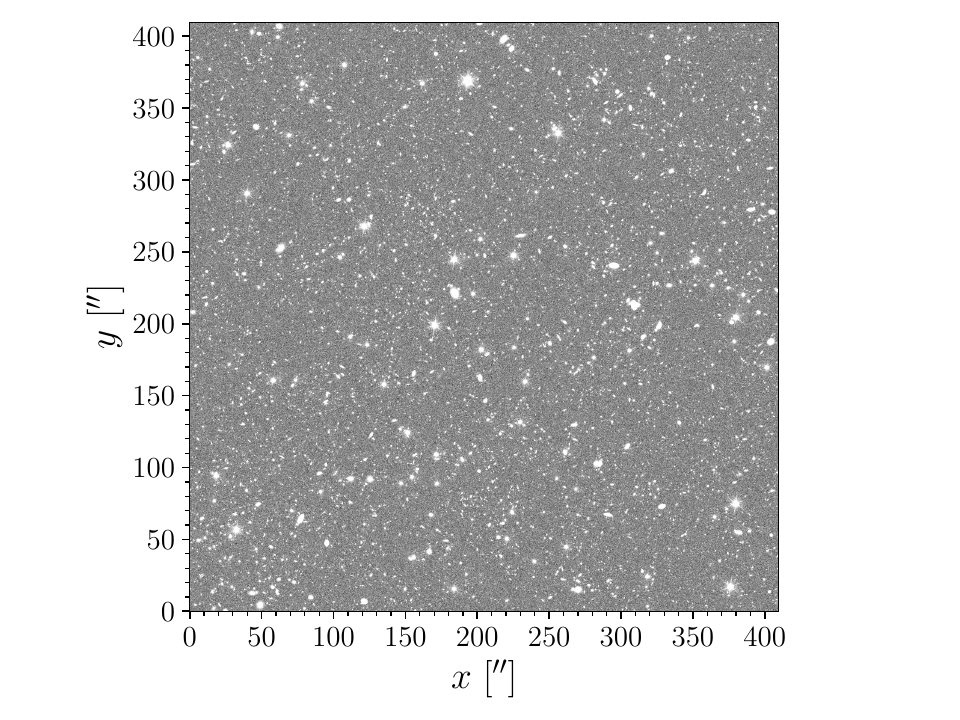}
\caption{Example of \lensmc-\flagship image.
The input galaxy distribution was provided by \flagship and
stars were drawn from a model.
We emulated the VIS detector by including realistic image properties and noise, but we did not include non-linearities, CTI, BFE, or cosmic rays.
To aid the visualisation of the faint objects, the image was clipped between the tenth and ninetieth percentiles,
illustrating the sheer number of objects and their clustering.
The image size is 4096 pixels.}
\label{fig:image}
\end{figure}

The pixel photoelectron noise is given by
\begin{equation}
\sigma_\text{px}^2(x,y)=\left(R_\text{bkg}+R_\text{dark}\right)\tau+\lambda_\text{obj}(x,y)+\sigma_\text{read}^2~.
\end{equation}
The first term is Poisson noise from a constant zodiacal light background and dark current,
with rates $R_\text{bkg}=\qty{0.225}{e^-/s}$ and $R_\text{dark}=\qty{0.001}{e^-/s}$,
and exposure time $\tau=\qty{565}{s}$.
The second term, $\lambda_\text{obj}(x,y)$, is spatially varying Poisson noise from all the objects in the image,
which is non-negligible in the \euclid VIS images.
The third term is Gaussian noise from the CCD readout with a constant $\sigma_\text{read}=\qty{4.5}{e^-}$.
For the purpose of this work, we assumed that all noise sources are uncorrelated.\footnote{Pixel noise is also uncorrelated with good approximation in real data especially when working with individual exposures, except for residual detector artefacts.}
In generating the images, we also applied a bias of $\qty{2000}{e^-}$ (about as expected for \euclid),
and finally digitised the data.
Digitalisation corresponds to dividing the image by a gain of $\qty{3.1}{e^-/ADU}$ and floor truncating it to nearest integer,
which itself adds uniform noise of variance $1/12\,\unit{ADU}$.\footnote{The in-orbit detectors will have slightly larger gain, likely around $\qty{3.4}{e^-/ADU}$, but this is not expected to change any of our results.}
We set a tangent projection as our WCS at the centre of the patch, drew 4 undithered exposures and stacked them up by taking their average.
These images were used by the object detection for the main results presented here,
but we will also include a discussion about the dithering.
An example of stacked CCD image is shown in \figref{fig:image}.


%% file: tex/sec4_results.tex
\section{Results} \label{sec:results}

\begin{figure}
\centering
\setlength{\tabcolsep}{0pt}
\begin{tabular}{lccc}
& Data & Model(s) & Residuals \\
\rotatebox[origin=l]{90}{\hspace{10pt}One galaxy} &
\includegraphics[width=0.3\columnwidth,trim={3cm 1cm 3cm 1cm},clip]{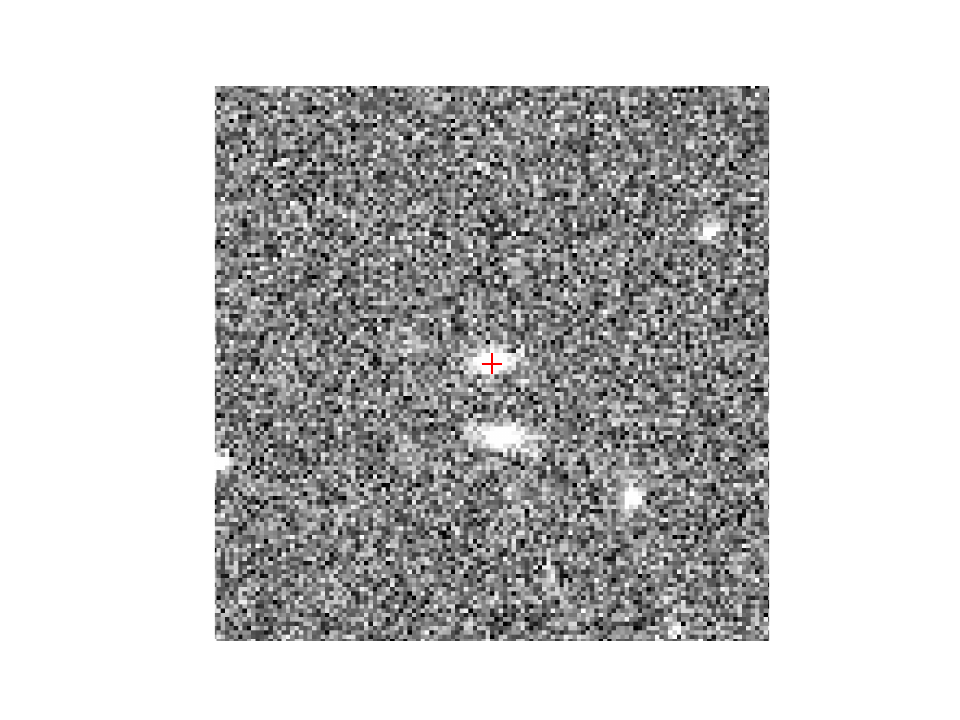} &
\includegraphics[width=0.3\columnwidth,trim={3cm 1cm 3cm 1cm},clip]{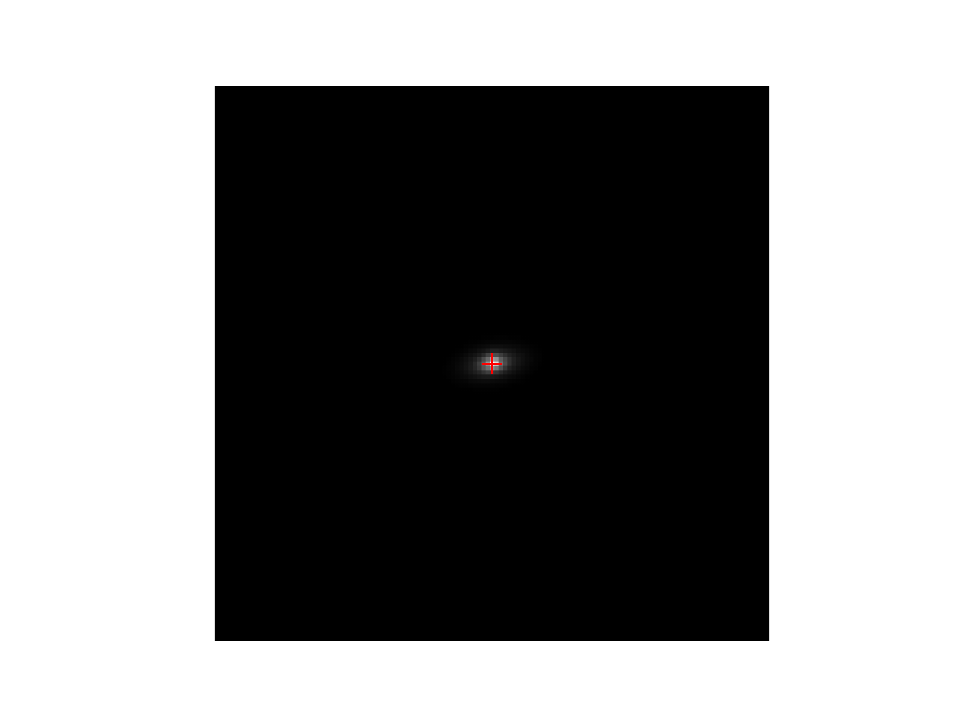} &
\includegraphics[width=0.3\columnwidth,trim={3cm 1cm 3cm 1cm},clip]{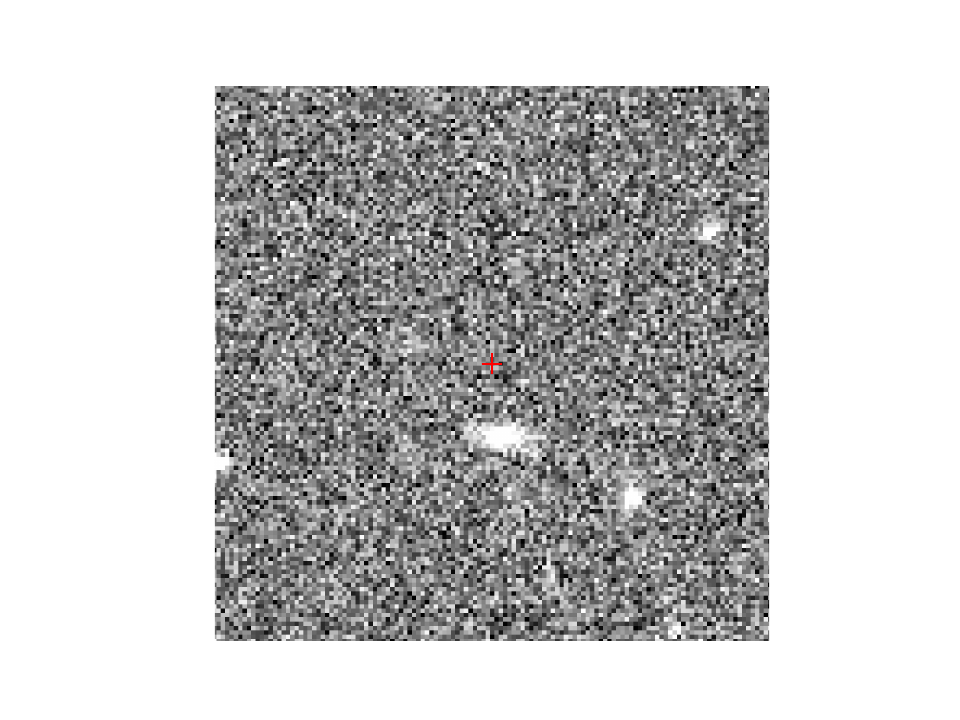} \\
\rotatebox[origin=l]{90}{\hspace{10pt}One galaxy} &
\includegraphics[width=0.3\columnwidth,trim={3cm 1cm 3cm 1cm},clip]{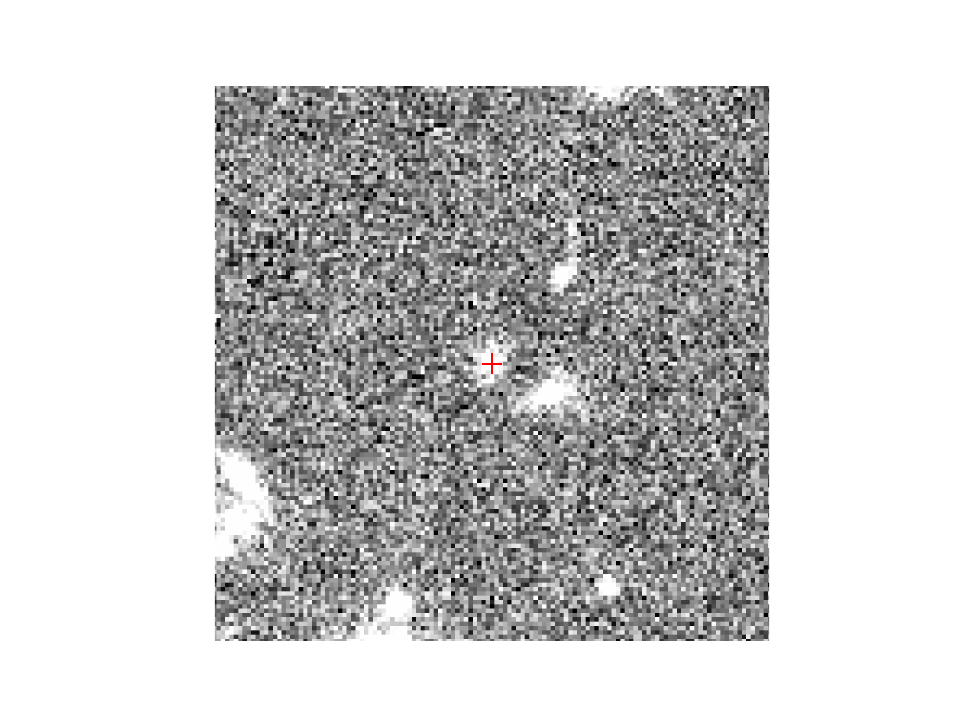} &
\includegraphics[width=0.3\columnwidth,trim={3cm 1cm 3cm 1cm},clip]{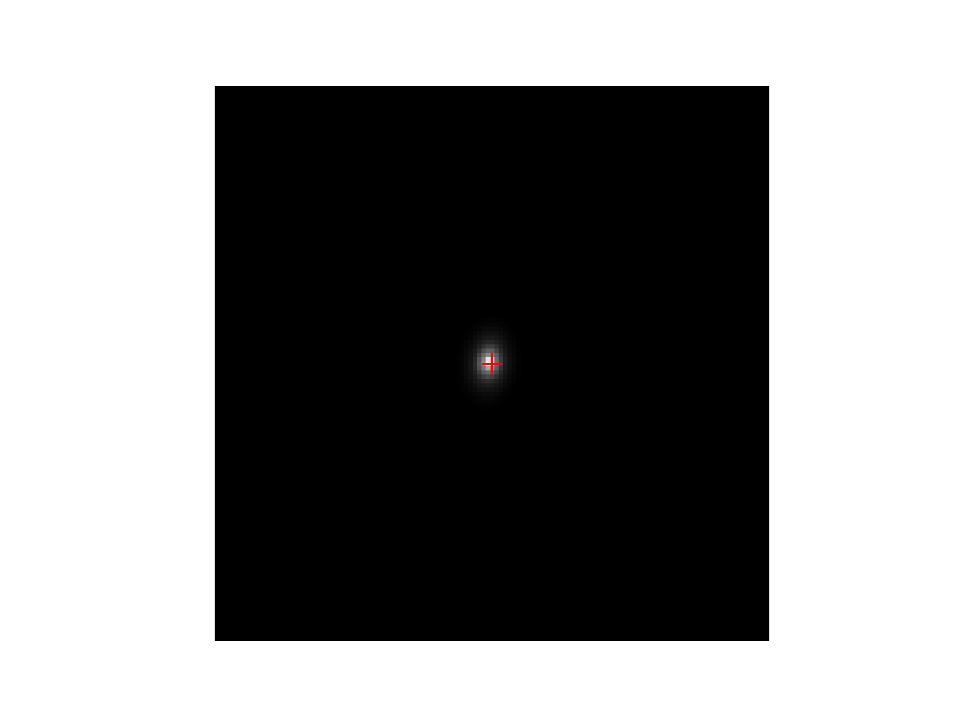} &
\includegraphics[width=0.3\columnwidth,trim={3cm 1cm 3cm 1cm},clip]{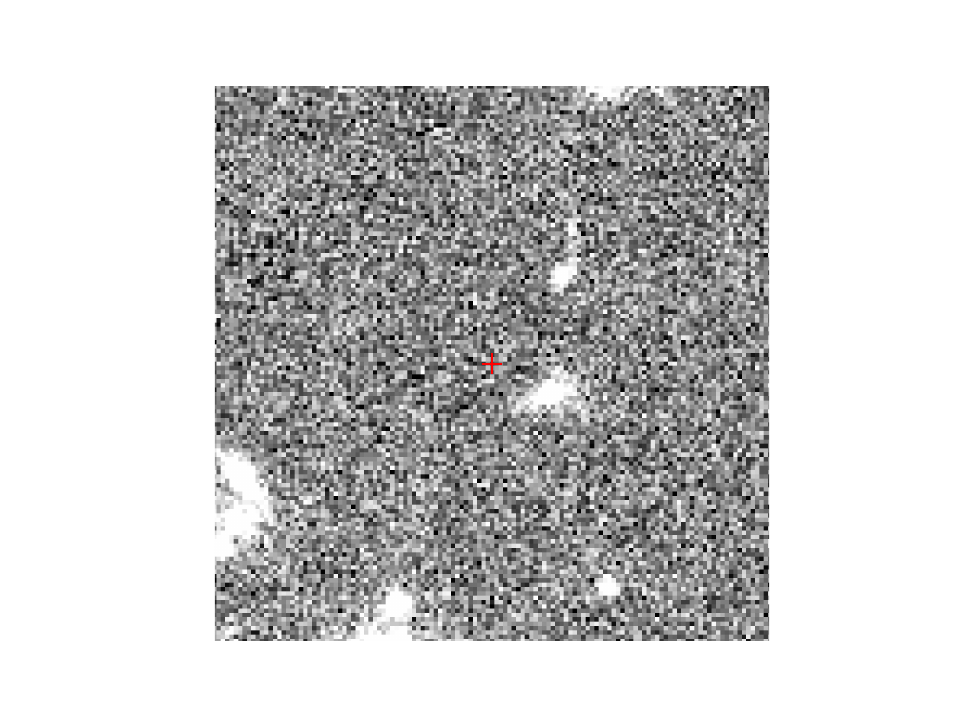} \\
\rotatebox[origin=l]{90}{\hspace{8pt}Two galaxies} &
\includegraphics[width=0.3\columnwidth,trim={3cm 1cm 3cm 1cm},clip]{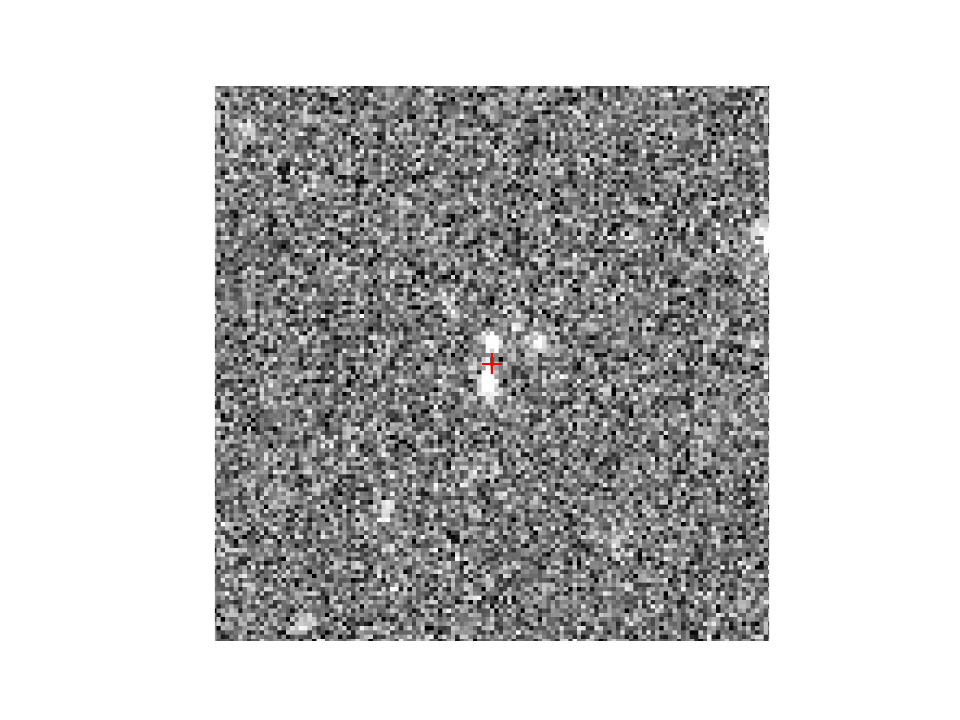} &
\includegraphics[width=0.3\columnwidth,trim={3cm 1cm 3cm 1cm},clip]{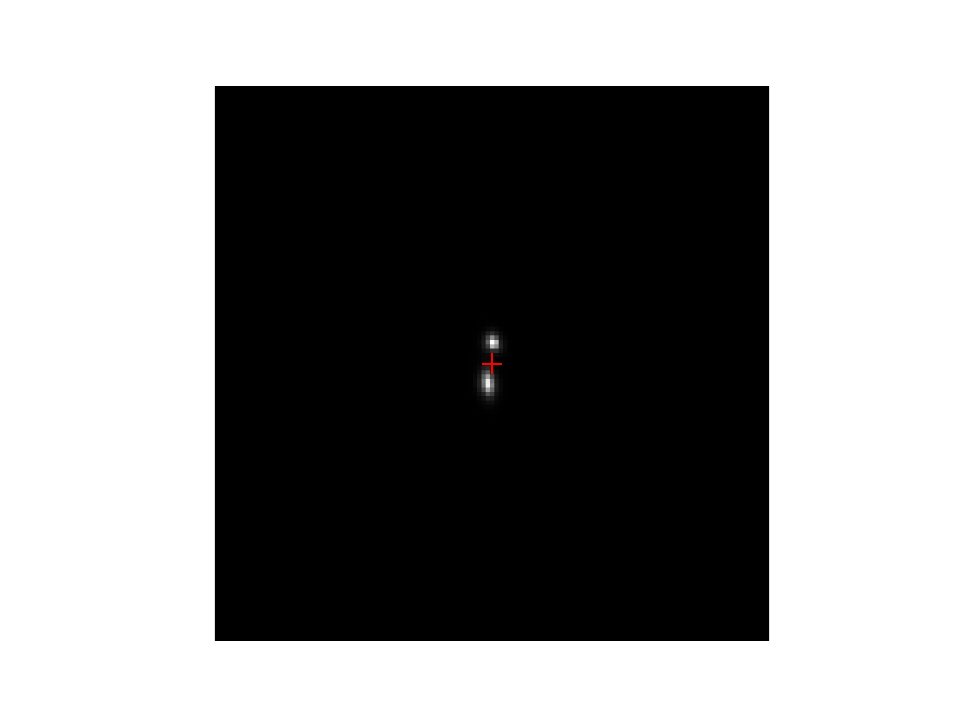} &
\includegraphics[width=0.3\columnwidth,trim={3cm 1cm 3cm 1cm},clip]{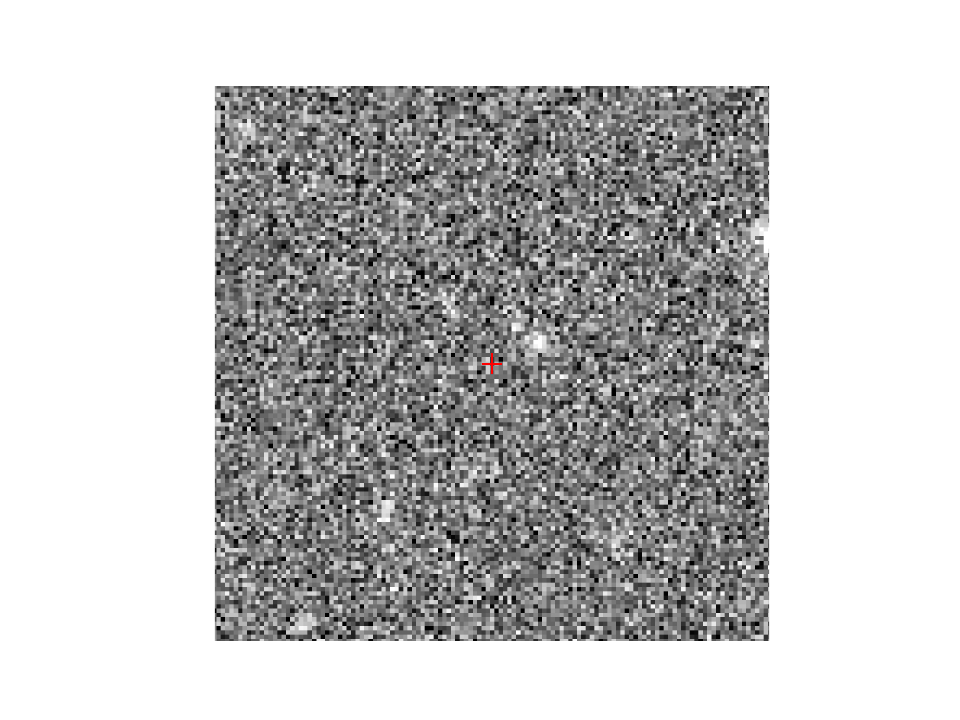} \\
\rotatebox[origin=l]{90}{\hspace{8pt}Two galaxies} &
\includegraphics[width=0.3\columnwidth,trim={3cm 1cm 3cm 1cm},clip]{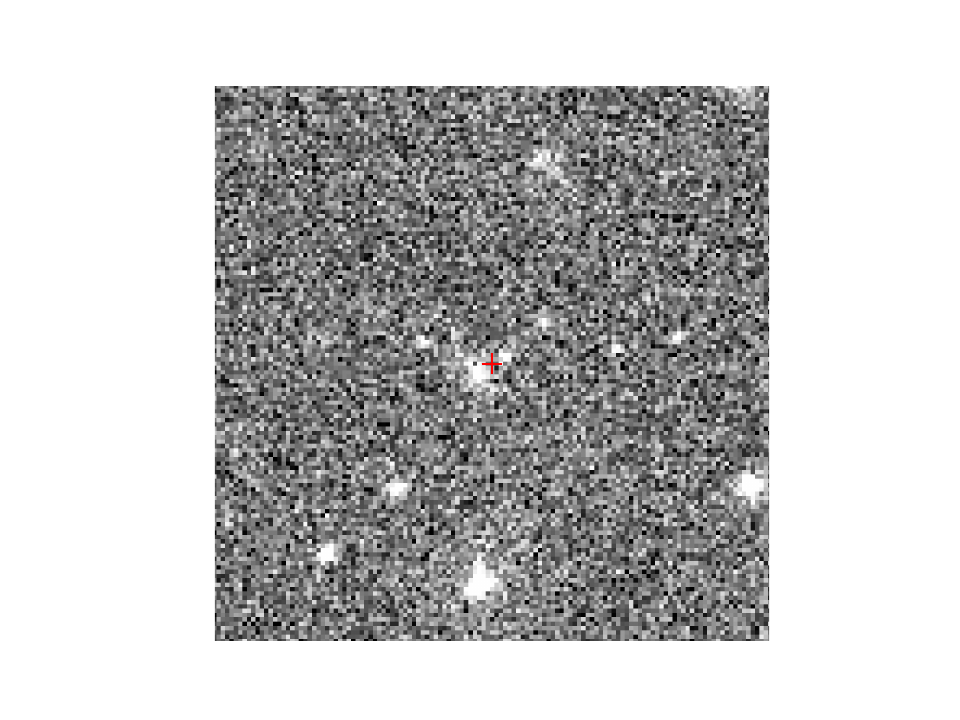} &
\includegraphics[width=0.3\columnwidth,trim={3cm 1cm 3cm 1cm},clip]{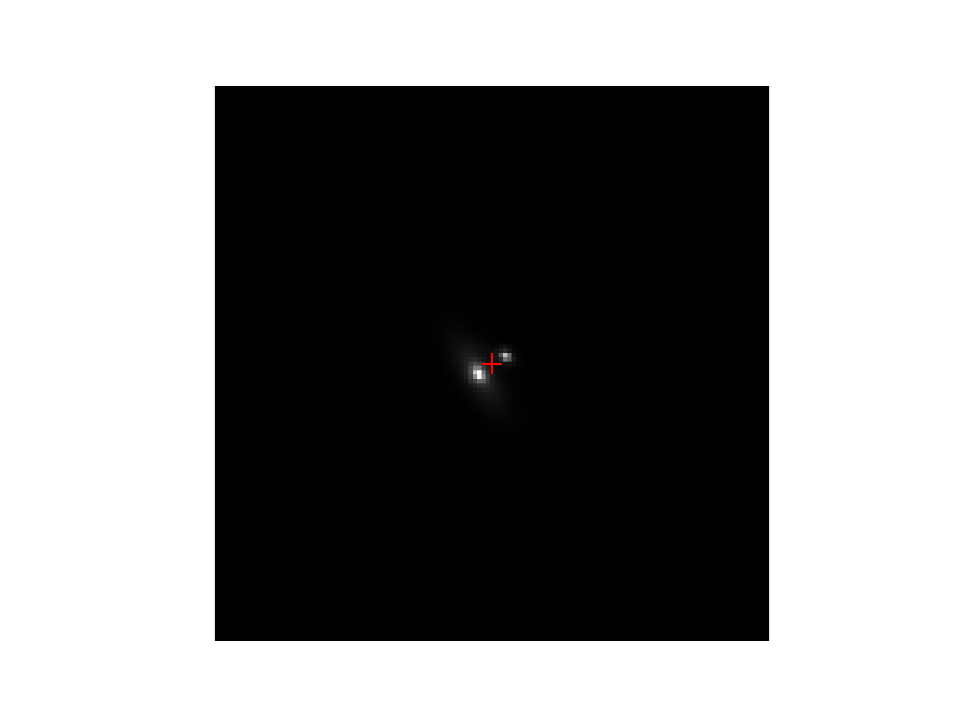} &
\includegraphics[width=0.3\columnwidth,trim={3cm 1cm 3cm 1cm},clip]{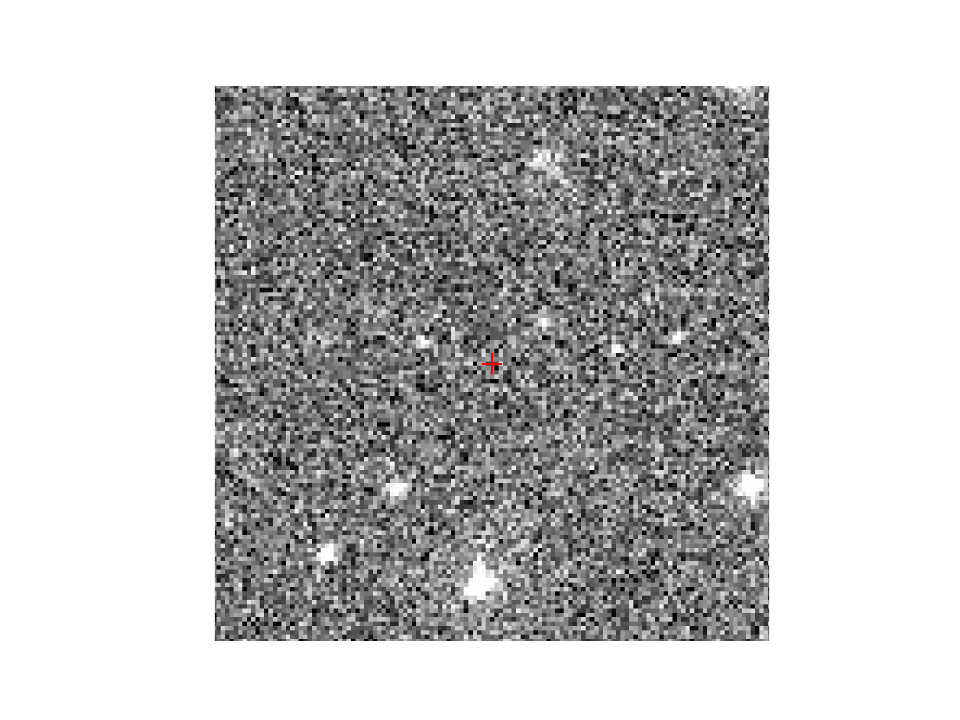} \\
\rotatebox[origin=l]{90}{\hspace{6pt}Three galaxies} &
\includegraphics[width=0.3\columnwidth,trim={3cm 1cm 3cm 1cm},clip]{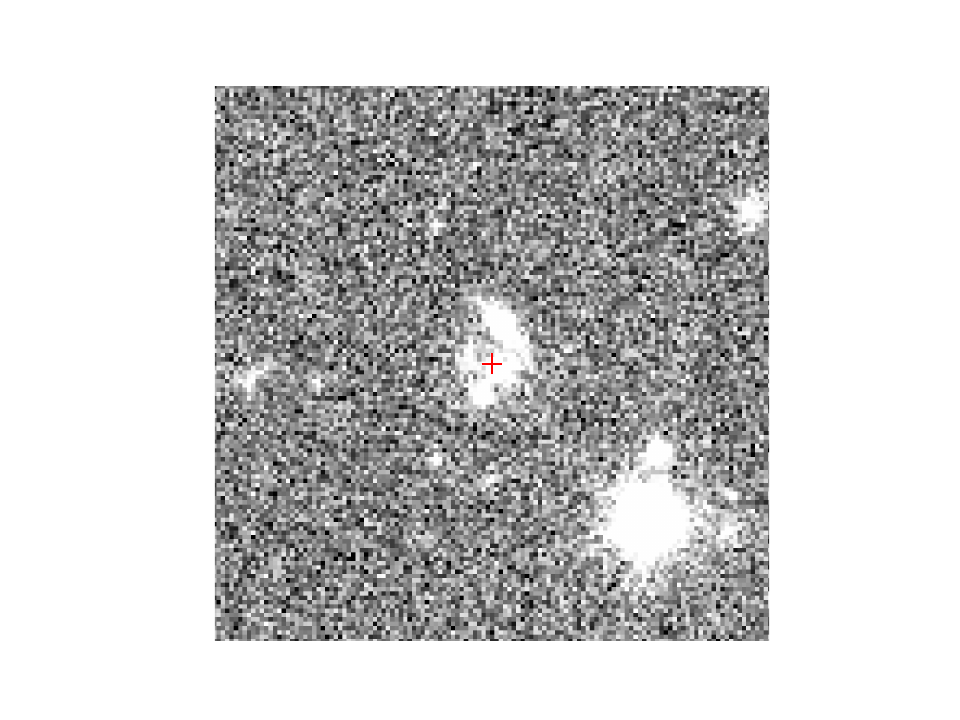} &
\includegraphics[width=0.3\columnwidth,trim={3cm 1cm 3cm 1cm},clip]{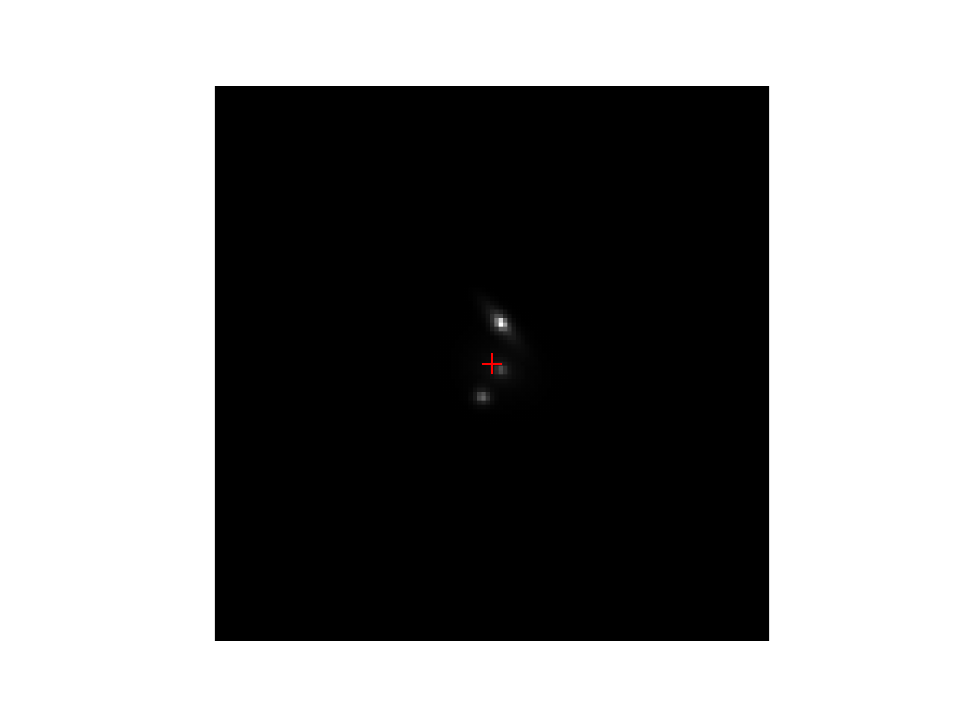} &
\includegraphics[width=0.3\columnwidth,trim={3cm 1cm 3cm 1cm},clip]{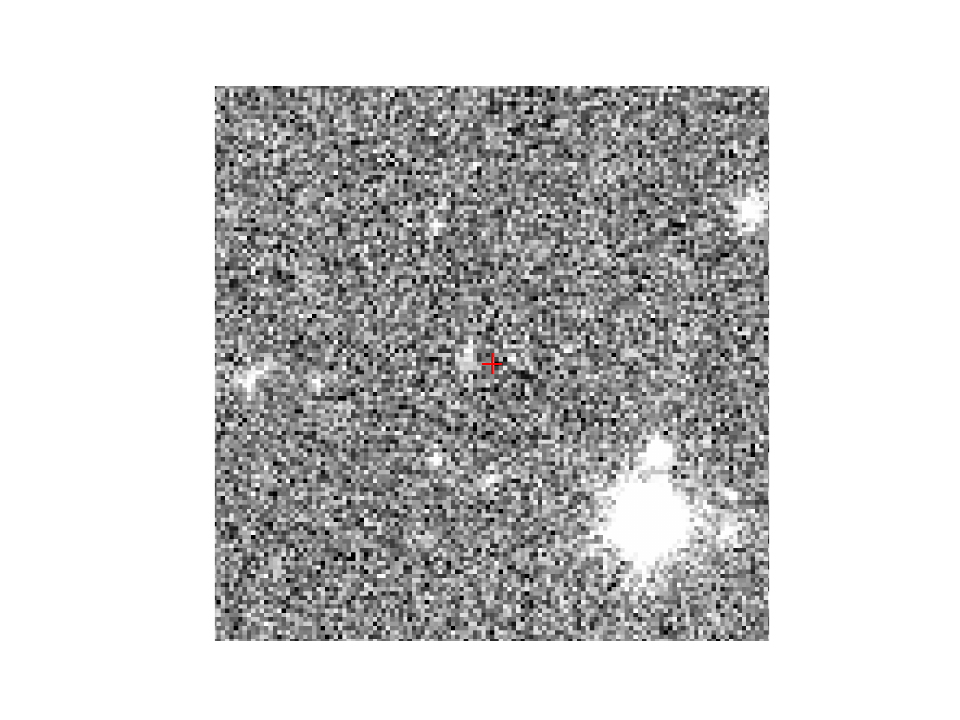} \\
\end{tabular}
\caption{Examples of measurement performance. The target galaxies are denoted with a cross.
The image residuals look consistent with noise, for galaxies measured individually or jointly in groups, despite the presence of neighbours.
All images have a size of 128 pixels.}
\label{fig:residuals}
\end{figure}

To quantify the performance of \lensmc on our realistic \lensmc-\flagship simulations,
we ran the measurement on about $45\,000$ random patches,
which is equivalent to an area of about $\qty{500}{deg^2}$,
with mean number density, according to \figref{fig:number_count},
of $\qty{250}{arcmin^{-2}}$ ($\IE<29.5$) for galaxies and $\qty{6}{arcmin^{-2}}$ ($\IE<26$) for stars.
We measured the same area (with the objects at the same positions) 9 times for varying noise realisations
and PSF across the field of view,
totalling an equivalent, effective area of $4500\,\unit{deg^2}$.
We ran all our simulations across the \gridpp UK network
\citep{faulkner2005,britton2009}.\footnote{Testing took 6 months, with our final run averaging 15\,000 cores/day for two weeks.}
A qualitative test of the measurement performance is shown in \figref{fig:residuals}.
After the galaxy models was subtracted from the image data, the residuals looked consistent with noise, for galaxies measured individually or jointly in groups, despite the presence of neighbours.
More quantitative tests will be discussed as part of the validation presented in Appendix\;\ref{app:validation}.

\subsection{Selection} \label{sec:selection}

\begin{figure}
\centering
\hspace{-10pt} \includegraphics[width=\columnwidth]{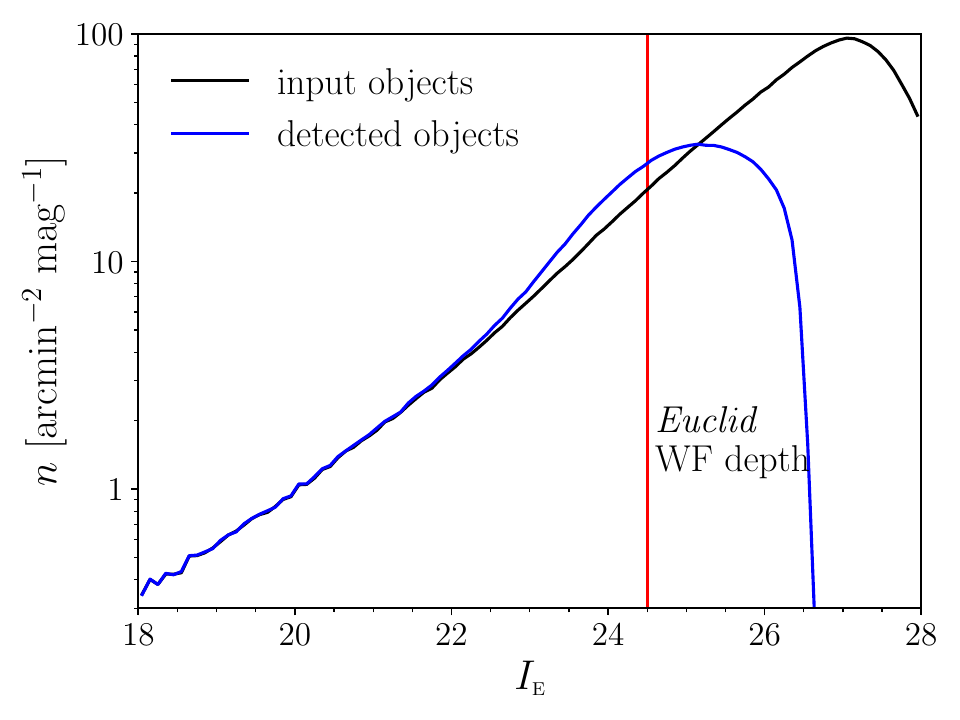}
\caption{Differential number count for all objects in the simulation and after the detection by \sepp.
The detection catalogue is mostly complete to $\IE<24.5$,
apart from false positives of about $\qty{6}{arcmin^{-2}}$ ($\IE<24.5$).
The distribution starts rolling off at 26, so a large fraction of faint objects is not detected.}
\label{fig:selection}
\end{figure}

For our baseline test, we ran \sepp \citep{bertin2020,kuemmel2020}\footnote{Version 0.19.2 with default settings as used in \euclid.
We do not test the sensitivity to changes in these settings and that will be the focus of future work.}
to detect galaxies and stars in each of the undithered stacked images.
The code attempts to deblend detected objects
and produces a detection catalogue with a total number density of $\qty{88}{arcmin^{-2}}$ ($\IE<26.5$), and $\qty{34}{arcmin^{-2}}$ ($\IE<24.5$).
Figure\;\ref{fig:selection} contains the galaxy magnitude selection applied by \sepp. 
This shows the number count of the objects in the simulation
and after the detection by \sepp.
The detection catalogue is complete to the magnitudes of interest,
apart from false positives of about $\qty{6}{arcmin^{-2}}$ ($\IE<24.5$),
probably due to a combination of sub-optimal detection and mismatching with the true input catalogue in presence of neighbours at those magnitudes.

The detections were grouped (with $r_\text{friend}=\ang{;;1}$) according to their reported \sepp positions.
\lensmc went through each object group and measured the object parameters starting from an initial guess at the provided \sepp positions.
If the size of the group is 1, \lensmc will measure the object in isolation and masks out neighbours through the supplied segmentation maps.
Instead, if it is greater than 1, \lensmc will measure the objects jointly, while masking out neighbours belonging to other groups.
We matched the input catalogue with the measurement catalogue and within a maximum angular distance of $\ang{;;0.3}$ from the measured object
(which also corresponds to the \lensmc maximum search region around the detected object).
The few measured objects that did not get a useful match to within that distance were then flagged up and removed from the analysis.
We tested the sensitivity to the maximum match distance without noticing any appreciable change to the bias.

A key selection applied to the measurement catalogue is the star-galaxy separation.
As found in applications to real data, the object size is an excellent statistic to discriminate between galaxies and stars \citep{sevilla2018}.
Therefore we classified objects according to measured $r_\text{e}>r_\text{s/g}$ where $r_\text{s/g}=\ang{;;0.15}$, which is slightly larger than the pixel size and image resolution.
We note that we applied our star-galaxy separation to broadband data simulated with a fixed choice of SED representative of a typical galaxy at a redshift of one.
However, this does not test how well the star-galaxy separation works with a broad range of galaxy SEDs and with a clear distinction between galaxy and star SEDs.
We quantified the performance of our separation by calculating the following:
\textit{i}) $N_\text{g}$, the number of true positives -- galaxies correctly identified as such;
\textit{ii}) $N_\text{s}$, the number of true negatives -- stars correctly identified as such;
\textit{iii}) $N_{\neg{\text{g}}}$, the number of false positives -- stars wrongly identified as galaxies;
\textit{iv}) $N_{\neg{\text{s}}}$, the number of false negatives -- galaxies wrongly identified as stars.
The above numbers are always defined in the measurement catalogue.
The true positive rate (TPR) and false positive rate (FPR) are
\begin{align}
\text{TPR}&=\frac{N_\text{g}}{N_\text{g}+N_{\neg{\text{s}}}}~, \\
\text{FPR}&=\frac{N_{\neg{\text{g}}}}{N_{\neg{\text{g}}}+N_\text{s}}~.
\end{align}
Realistic values of $\text{FPR}>0$ and $\text{TPR}<1$ are always linked to type I and II errors in the shear analysis.
Type I is the inclusion of stars in the lensing sample, hence leading to potentially large negative multiplicative bias.
Type II is the omission of galaxies (with potentially large shear signal) from the lensing sample which introduces selection bias and also a dilation in statistical error.

For the sample of detected objects to the detection limit ($\IE<26.5$) we found $\text{TPR}=93.3\%$, $\text{FPR}=4.6\%$, purity of 99.8\%,\footnote{Astronomical completeness coincides with TPR, but $\text{purity} = N_\text{g}/(N_\text{g}+N_{\neg{\text{g}}})$.}
and a star fraction of 6.6\%.\footnote{$\text{Star fraction}=N_\text{s}/(N_\text{g}+N_\text{s})$.}
The TPR gives us the frequentist probability of a positive being a galaxy, so $\text{TPR}=p(+|\text{g})$.
Similarly, $\text{FPR}=p(+|\text{s})$.
Bayesian posterior probabilities provide a more meaningful interpretation of those numbers.
The prior probability of an object being a galaxy is $p(\text{g})$ and a star is $p(\text{s}) = 1 - p(\text{g})$ (i.e., the star fraction).
Applying Bayes' theorem, we get the probability of a galaxy given a positive detection,
\begin{equation}
p(\text{g}|+) = \frac{p(+|\text{g})\,p(\text{g})}{p(+|\text{g})\,p(\text{g}) + p(+|\text{s})\,p(\text{s})}~,
\end{equation}
and similarly for $p(\text{s}|+)$.
With the numbers above, we obtained $p(\text{g}|+)=99.7\%$ and $p(\text{s}|+)=0.3\%$ for all objects in the detection catalogue.
A more relaxed FPR of about $20\%$ would still give us
$p(\text{g}|+)=99\%$ and $p(\text{s}|+)=1\%$,
given the strong imbalance between the galaxy and star samples.
These numbers give us reassurance that once an object is classified as a galaxy,
there will be an average $3\,\sigma$ confidence that it will indeed be a galaxy for the entire sample up to the detection limit ($\IE<26.5$).


%
%
%

\begin{figure}
\centering
\hspace{-10pt} \includegraphics[width=\columnwidth]{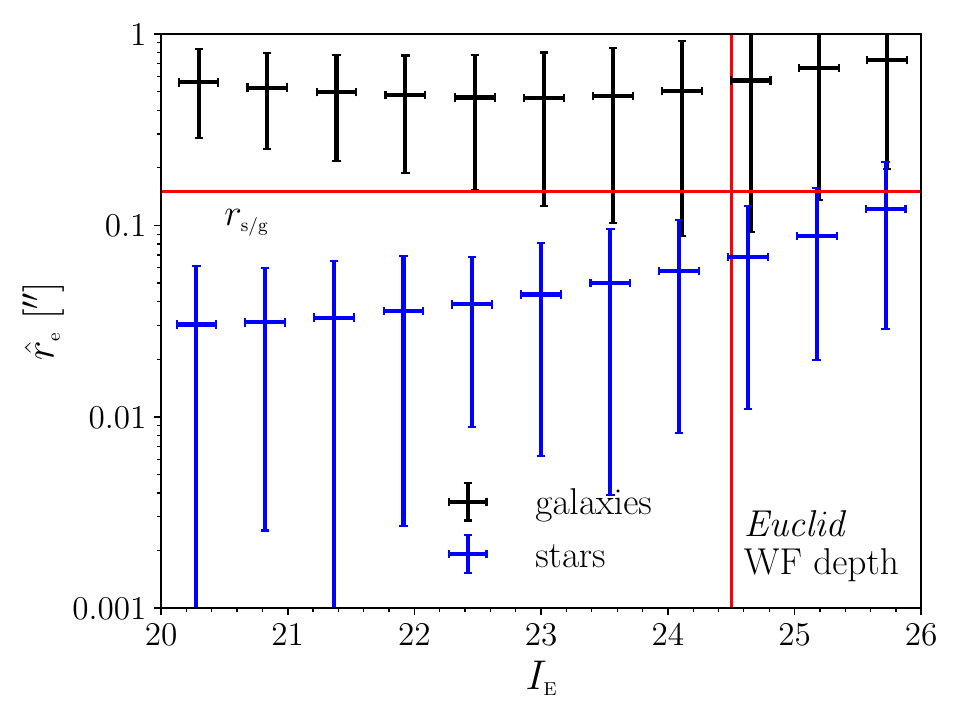} \\
\includegraphics[width=\columnwidth]{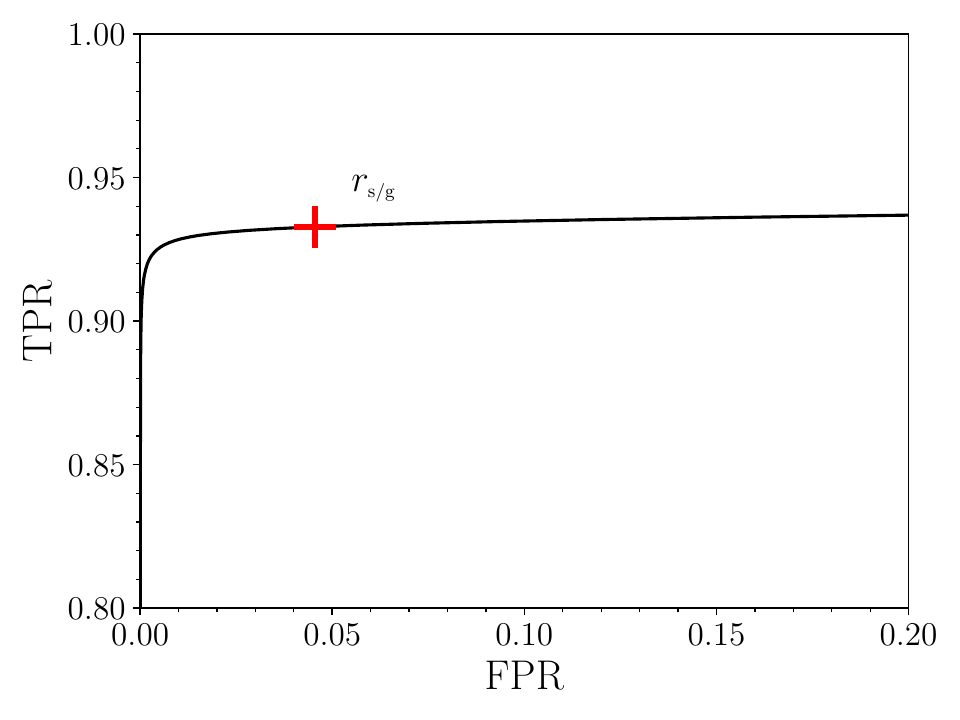}
\vspace{-10pt}
\caption{Star-galaxy separation of detected objects.
(Top) Observed size-magnitude distribution of true galaxies and stars.
The data points are the mean of $r_\text{e}$ as a function of $\IE$.
Also shown are the standard deviation of $r_\text{e}$ and $\IE$ in each bin.
$r_\text{s/g}=\ang{;;0.15}$ provides a good separation threshold working up to $\IE<24$.
(Bottom) Operating curve showing where our threshold (indicated with a cross) sits in terms of true positives (92.9\%) and false positives (4.3\%).
The area under the operating curve is large, and the curve itself is reasonably flat for a wide range of false positive rate suggesting excellent discrimination and weak sensitivity on the threshold (the shear bias does not appreciably change).
In real measurements this could be further optimised through access to external data or simulations.}
\label{fig:sg_separation}
\end{figure}

Figure\;\ref{fig:sg_separation} shows the size distribution of true galaxies and true stars
and the operating curve (TPR versus FPR for varying threshold) of our classification
with either a horizontal line or a cross to denote the default threshold, $r_\text{s/g}=\ang{;;0.15}$.
Both plots provide solid justification for our choice of $r_\text{s/g}$,
but confusion is evident around $\IE=24$, which might explain most of the false positives.
The area under the operating curve at the bottom of \figref{fig:sg_separation} is large, and the curve itself is reasonably flat for a wide range of FPRs, suggesting excellent discrimination and weak sensitivity on the threshold
(in that the shear bias does not appreciably change for a wide range of threshold values around the nominal one).
However, in real measurements, an optimal value could be inferred from external data or simulations,
hence allowing for a dramatic reduction of the false positives at the expense of a modest reduction in the true positives.
Our TPR, FPR, and operating curve of \figref{fig:sg_separation} are consistent or better than the best estimators presented in \citet{sevilla2018},
although a key caveat in our work is likely to be that we did not include a full colour variation of galaxy and star SEDs,
which would lead to variable PSFs and potentially harder separation.
Moreover, we did not investigate any effect due to star density variation, which might well change by a factor of two or three going from the high to the low latitudes.

\begin{figure}\label{fig:weight}
\centering
\hspace{-15pt} \includegraphics[width=\columnwidth]{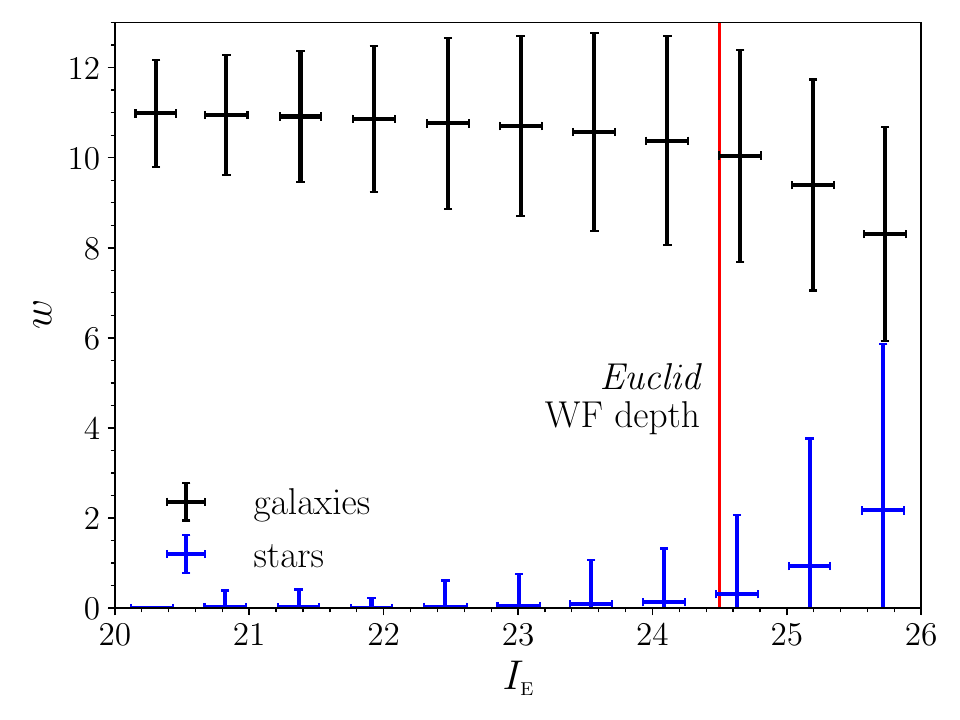}
\caption{Shear weight after star-galaxy separation of detected objects as a function of magnitude separately for true galaxies and stars.
As the star weight is systematically lower than the galaxy weight,
this drastically reduces the impact of those residual stars in the catalogue up to the faint magnitudes.}
\label{fig:weight_vs_mag}
\end{figure}

We defined our final shear weight by multiplying Eq.\;\eqref{eq:weight} by the step function, $\mathrm{H}(r_\text{e}-r_\text{s/g})$ and show this as a function of magnitude for detected true galaxies and stars in \figref{fig:weight_vs_mag}.
As the star weight is systematically lower than the galaxy weight,
this drastically reduces the impact of those residual stars (false positives) in the catalogue up to the faint magnitudes.

The quality of our star-galaxy separation can only be tested by fully propagating results through shear bias.
We calculated the shear bias for perfect star-galaxy separation
(where we enforce knowledge about the truth; that is, we do not use our classification but exclude true stars from the galaxy catalogue),
and compared it with our nominal analysis.
We did not see any statistically significant difference in shear bias between the two cases.
Additionally, we varied the value of $r_\text{s/g}$ and again found that the bias does not change appreciably, but this may not hold true for more realistic SED variation.

\subsection{Shear bias} \label{sec:shear_bias}

As a preliminary validation, Appendix\;\ref{app:validation} contains a few distributions and correlations.
We tested the bias as a function of input true magnitude to avoid
large selection biases due to binning by observed quantities, such as $\snr$ or $r_\text{e}$,
which strongly correlate with shear \citep{fenech2017}.
We defined 12 bins in $20<\IE<24.5$
and, in each bin, we regressed the measured ellipticity against input true shear via weighted least square as described
in detail in Appendix\;\ref{app:shear_bias}.
We also wanted to clearly separate measurement (shear measurement method)
from purely selection (detection, catalogue matching, weights, and star-galaxy separation) effects.
We let $g_i$ be the input true shear, $\hat{\epsilon}_i$ the measured ellipticity, and
$\epsilon_{i,\text{sel}}$ the input true sheared ellipticity on the same selection.
Similarly to Eq.\;\eqref{eq:shear_bias}, where we regress estimates of shear with input true shear,
we can define the same regression of our estimate of shear (i.e., ellipticity),\footnote{It is worth noting that a least-square regression of ellipticity is a linear operation that corresponds to calculating the mean ellipticity, that is, estimating shear.}
\begin{align}\label{eq:ellipticity_vs_shear}
\hat{\epsilon}_i &= (1+m_i)\,g_i+c_i+n_i~, \\
\epsilon_{i,\text{sel}} &= (1+m_{i,\text{sel}})\,g_i+c_{i,\text{sel}}+n_{i,\text{sel}}~,
\end{align}
where $i=1,2$ indexes the ellipticity or shear component,
$n_i$ and $n_{i,\text{sel}}$ are the statistical noise components for measurement and selection.
The first equation estimates the total of measurement and selection bias, whereas the second one estimates the selection bias.
Therefore, if we take the difference between the two and regress $\hat{\epsilon}_i-\epsilon_{i,\text{sel}}$ with $g_i$,
\begin{equation}
\hat{\epsilon}_i-\epsilon_{i,\text{sel}} = m_{i,\text{meas}}\,g_i+c_{i,\text{meas}}+n_i-n_{i,\text{sel}}~,
\end{equation}
we can then directly estimate the measurement bias, having coherently subtracted the selection bias,
and reduce the statistical noise thanks to the correlation between $n_i$ and $n_{i,\text{sel}}$.
The measurement-only bias is then defined as $m_{i,\text{meas}} = m_i-m_{i,\text{sel}}$ and $c_{i,\text{meas}} = c_i-c_{i,\text{sel}}$.

\begin{figure}
\centering
\includegraphics[width=\columnwidth]{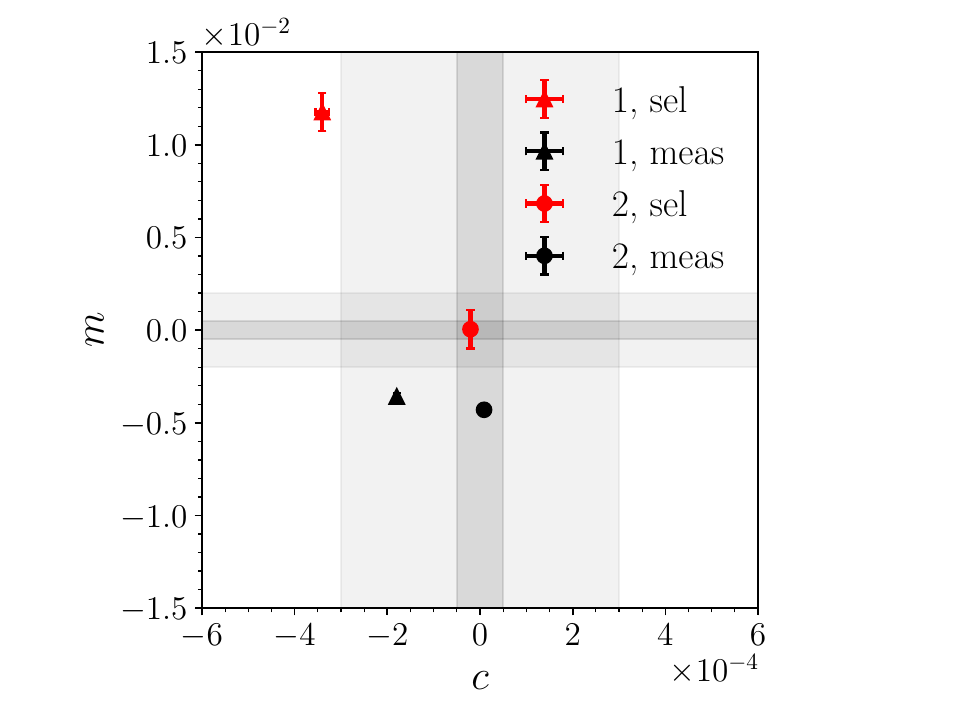}
\vspace{-10pt}
\caption{Multiplicative and additive biases
averaged over the magnitude selection $20<\IE<24.5$ and PSF variation across the field of view
for the measurement and selection.
The triangle and the circle denote each of the two components.
The light shaded area is the \euclid requirement on knowledge of $m$ and $c$,
respectively $\sigma_m<2\times10^{-3}$ and $\sigma_c<3\times10^{-4}$.
The dark shaded area is the ideal target for measurement alone, 
respectively $\sigma_m<5\times10^{-4}$ and $\sigma_c<5\times10^{-5}$.
Reference values can be found in \tabref{tab:m&c}.}
\label{fig:m&c}
\end{figure}

The main performance metric that we present here is the total bias computed as a weighted average over magnitude bins 
and PSF variation across the field of view as shown in \figref{fig:m&c}.
The performance metric is defined in an $m$-$c$ plane for each of the two components fitted independently.
We indicate the \euclid requirements $\sigma_m$ and $\sigma_c$ in the shaded areas,
both the total one and the desired for measurement alone.
From looking at the summary figures of \tabref{tab:m&c},
selection effects are dominating the error budget,
with a pronounced asymmetry between the two components.
We tested that this is not due to the star-galaxy separation, weights, or the particular PSF used here
by varying each parameter and checking that results remain consistent with the default analysis.
To investigate if the origin of this asymmetry could be due to the input distributions,
we estimated the bias of the input ellipticity (before detection) in exactly the same way as we did for our measurements, finding no bias up until the detection is run.
We defer the investigation of sensitivity to the \sepp configuration to future work.
For the time being, we highlight that the multiplicative bias owing to measurement alone (i.e., the shear measurement method) 
is about $-4\times10^{-3}$ with an uncertainty of $2\times10^{-4}$,
and a small residual asymmetry in additive bias.
All statistical errors for selection and measurement were estimated following the modelling of Eqs.\;\eqref{eq:ellipticity_vs_shear}
and the analytical solution presented in Appendix\;\ref{app:shear_bias}.
The modelling effectively uses the high correlation between measurement and selection estimates to reduce the error on bias.
Once selection and measurement biases are calculated, the total bias is just the sum of the two individual values.
The total error is then the sum in quadrature, given that the individual values have had their correlation removed.

\begin{table}
\caption{Multiplicative and additive biases.}
\label{tab:m&c}
\centering
\begin{tabular}{D{/}{\,/\,}{2} D{,}{\,\pm\,}{4} D{,}{\,\pm\,}{4}}
\hline\hline
& \multicolumn{1}{c}{\hspace{10pt}Measurement} & \multicolumn{1}{c}{Selection} \\
\hline
m_1/10^{-3} & -3.58, 0.18 & 11.8, 1.0 \\
c_1/10^{-4} & -1.797, 0.025 & -3.40, 0.15 \\
m_2/10^{-3} & -4.30, 0.18 & 0.0, 1.0 \\
c_2/10^{-4} & 0.088, 0.025 & -0.21, 0.15 \\
\hline\hline
\end{tabular}
\tablefoot{Biases averaged over the magnitude selection $20<\IE<24.5$ and PSF variation across the field of view,
for measurement and selection (detection, catalogue matching, star-galaxy separation, and weights).}
\end{table}

\begin{figure}
\centering
\hspace{-10pt} \includegraphics[width=\columnwidth]{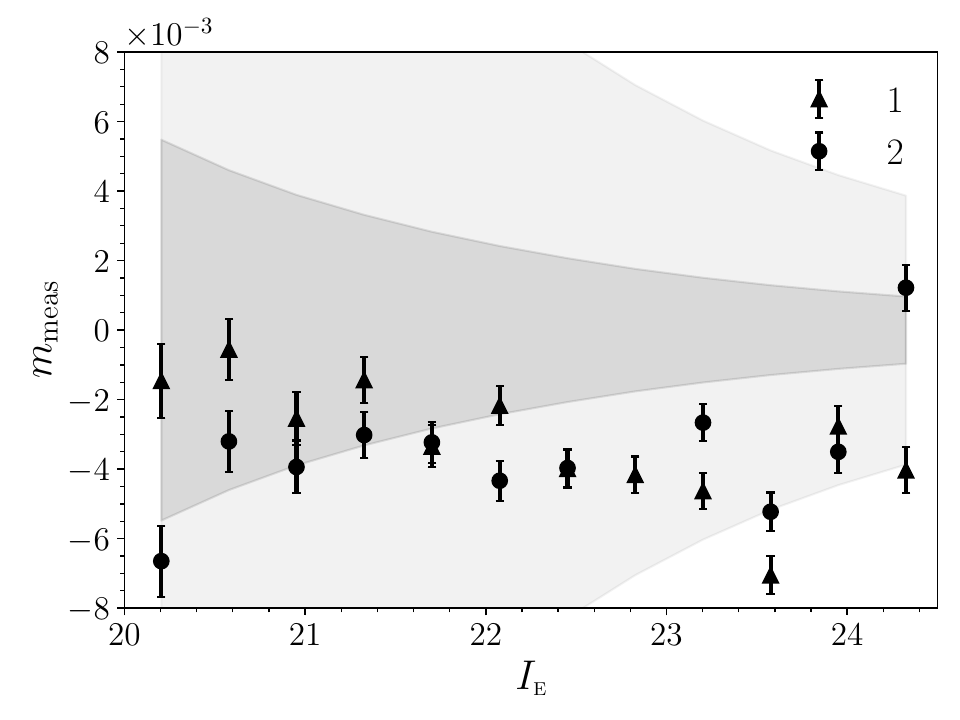} \\
\hspace{-10pt} \includegraphics[width=\columnwidth]{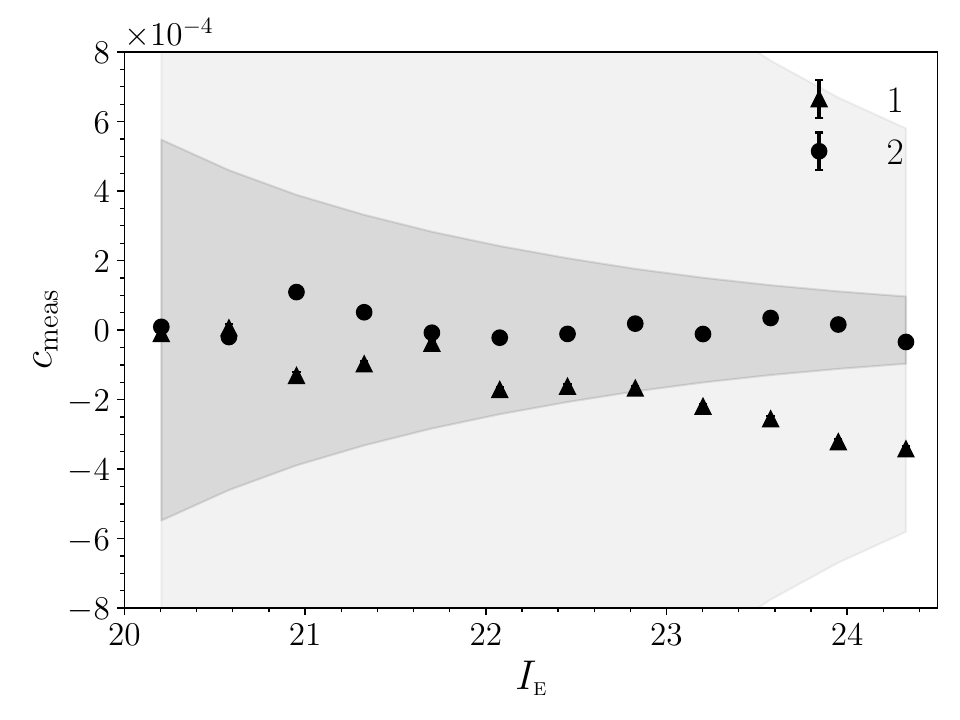}
\caption{Measurement multiplicative and additive biases
in bins of $\IE$ and averaged over the PSF variations across the field of view.
The triangles and circles denote each of the two components.
Shaded areas are the \euclid requirements relaxed by the increase in variance in each bin.
Except for $c_{1,\textrm{meas}}$, which is fully consistent with requirements,
all other biases show a slight trend in the faintest bins.}
\label{fig:m&c_vs_mag}
\end{figure}

A more in-depth investigation can be carried out when looking at bias as a function of true input magnitude as shown in \figref{fig:m&c_vs_mag}.
As already mentioned, not only is the correlation between magnitude (or flux) with shear negligible 
(apart from magnification, not included here),
but defining true input bins is also essential to minimise the impact of selection bias and not to misinterpret results.
Curves were averaged over the PSF variations across the field of view.
We note that $m$ and $c$ show a negative trend at the faint magnitudes.
This suggests that the total bias shown in \figref{fig:m&c} is mostly dominated by those bins,
which happen to have the largest relative weight due to the number count increasing with magnitude.
The requirement, $\sigma_m$ and $\sigma_c$, on each bin was derived from the total requirement by increasing 
the per-bin variance by the decrease in number count in each bin.
The error bars are consistent with this.


\begin{figure}
\centering
\hspace{-10pt} \includegraphics[width=\columnwidth]{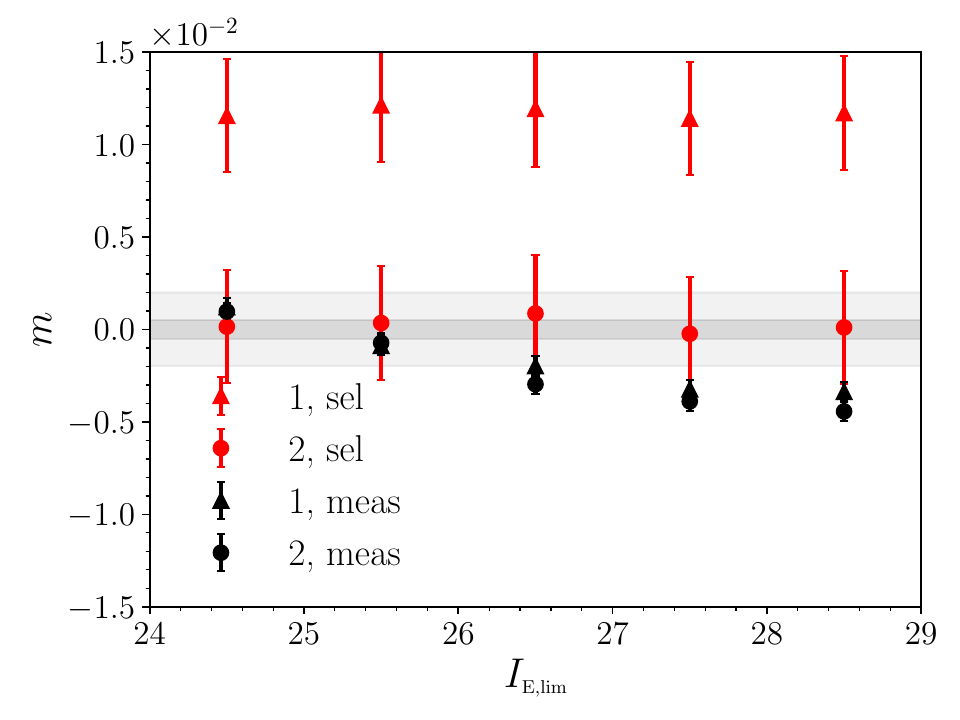} \\
\hspace{-10pt} \includegraphics[width=\columnwidth]{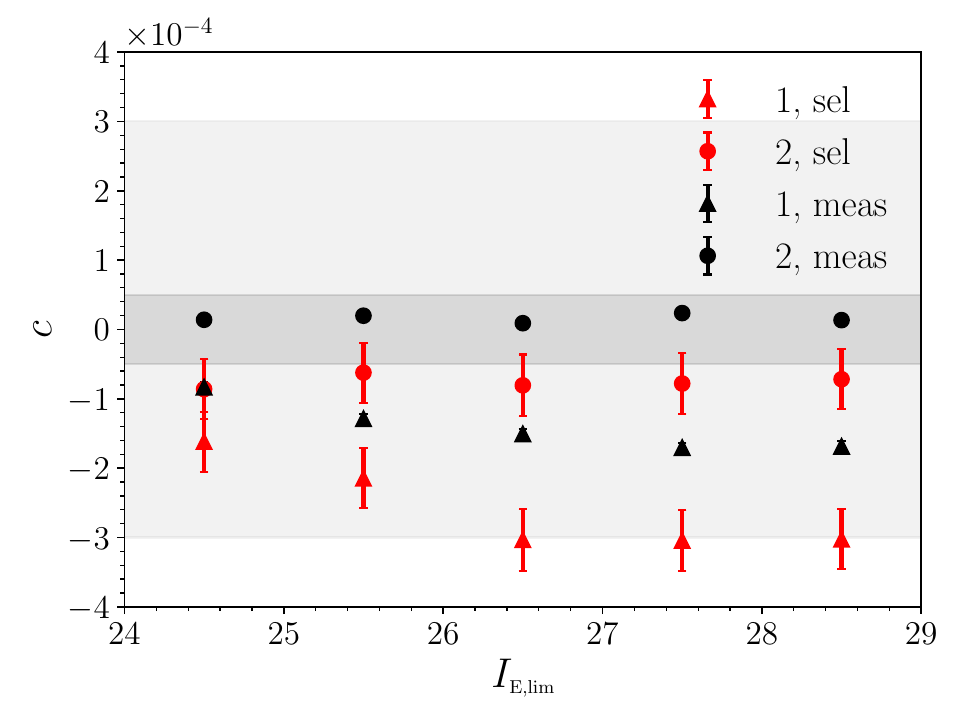}
\caption{Multiplicative and additive biases for measurement and selection
as a function of the intrinsic limiting magnitude in the simulations for a PSF chosen at the centre of the field of view.
The triangles and circles denote each of the two components.
Shaded areas are the \euclid requirements.
The measurement bias shows varying trends with the limiting magnitude
and asymmetries between components in some cases.
(See text for discussion.)}
\label{fig:m&c_vs_depth}
\end{figure}

We tested the sensitivity to the faint undetected objects 
by calculating the total bias as a function of the intrinsic limiting magnitude in the simulations.
For this we repeatedly rendered images excluding the faintest objects, with a varying magnitude limit.
We expect that the brighter the magnitude limit in the simulation, the weaker the impact of faint objects on the measurement (as their relative fraction becomes small).
Figure\;\ref{fig:m&c_vs_depth} shows different trends in bias with the magnitude limit.
For multiplicative bias, the selection bias seems to be insensitive to the magnitude limit;
instead, the measurement bias seems to be symmetric (components are consistent with one another)
and shows a trend with the magnitude limit,
with a slight hint of flattening out at the faintest end.
This effect on the measurement bias indicates a circularisation bias due to the faint objects.
Overall, we estimated $m_\text{faint}\approx-5\times10^{-3}$ just due to the presence of faint undetected objects.
This is fully consistent with earlier predictions of a few $10^{-3}$ up to $10^{-2}$ depending on the clustering of the faint objects \citep{martinet2019}.
We think the flattening of the measurement curves with the magnitude limit
might be due to faint galaxies having less impact the fainter they are or perhaps a lack of an ultra-faint population.
However, this indicates that any calibration strategy relying on external images should render galaxies with a magnitude limit of at least 27.5,
which is 3 deeper than the Euclid Wide Survey \citep[in agreement with][]{hoekstra2017},
and the sensitivity to that limit should be investigated as well.

We estimated the bias due to close detected neighbours
by comparing results from running the measurement in two modes:
$r_\text{friend}=0$ or $r_\text{friend}=\ang{;;1}$.
As discussed in \secref{sec:real_data}, the first mode corresponds
to no grouping of detections at all, so \lensmc measures
all the objects individually, hence relying on the supplied maps to implement the masking of neighbours.
In fact, masking provide limited help when the objects are too close to each other:
the final ellipticity estimate of the target object 
will be slightly biased towards the neighbour.
Because of the random orientation around the target object,
the net effect is a circularisation of the average ellipticity if not corrected for.
Instead, the second mode makes groups of objects that are measured jointly.
In total, almost 96\% of the objects were still measured in isolation, 4\% in pairs, and 0.2\% in groups that include triplets, quadruplets, and some rare quintuplets.
The measurement of the groups increases the robustness to neighbours and mitigates the reliance on the accuracy of the maps at short angular separations between objects.
We found a differential multiplicative bias of $m_\text{neighbour}\approx-7\times10^{-4}$ due to the masking of close neighbours (4.2\% of the sample)
when measuring objects individually ($r_\text{friend}=0$),
and when joint measuring them in groups ($r_\text{friend}=\ang{;;1}$).
As the neighbour bias predominantly effects the short separation between objects and the fraction of multiplets compared to the pairs is very small, we expect that increasing $r_\text{friend}$ further would provide little benefit to the bias, but at a much increased computational expense.

We also checked the dependence of bias on the weight definition
of Eq.\;\eqref{eq:weight} or the one employed by KiDS \citep{miller2013},
\begin{equation}
w_\sfont{KiDS}=\left(\frac{C_\epsilon\,\epsilon^2_{\max}}{\epsilon^2_{\max}-2C_\epsilon} + \sigma^2_\epsilon\right)^{-1},
\end{equation}
as a function of the assumed shape noise, $\sigma_\epsilon$.
We estimated the first-order sensitivity as a linear regression to the bias for varying $\sigma_\epsilon$.
For either definition, we found weak sensitivity: $\text{d}m/\text{d}\sigma_\epsilon\approx4\times10^{-4}$ and $\text{d}c/\text{d}\sigma_\epsilon\lesssim4\times10^{-6}$.
The implication is such that a change in the assumed $\sigma_\epsilon$ of, for example, 20\% would be responsible for an additional bias of order of $10^{-5}$.

Finally, we carried out a test on exposures dithered randomly between $[-0.05,0.05]\,\unit{\arcsecond}$.
In reality, the dithering could be as large as a few arcminutes
but sampling is always affected by the random sub-pixel shifts.
In a real survey, any stacking procedure, even if applied to nominally undithered exposures, will be affected by the random shifts in the telescope pointing at the sub-pixel level.
The combination of such exposures will inevitably introduce pixel correlations in the stacked images and PSFs,
a problem that would be exacerbated by combining exposures at different epochs.
The aim of this test is to verify that the results between dithered exposures and our perfectly undithered exposures (as presented throughout the text) are fully consistent, so to demonstrate that the method will perform well on dithered exposures on real data. 
However, we did not quantify the impact of stacking in our tests, nor directly assess any benefit from analysing dithered exposures over stacked exposures.
For dithered exposures, we resample-coadded the images with \swarp \citep{bertin2002}, ran \sepp, and remapped the segmentation maps back to individual exposures with \swarp.
This data was then passed on to \lensmc as usual, with the only difference that it now carried out the measurement jointly across exposures.
We ran the measurement on one of the PSF images at the centre of the field of view and processed the catalogues as usual.
We found no statistically significant difference in the estimated bias between the two cases of random dithering and perfectly undithered exposures (as used in the main simulations of this paper).
This gives us reassurance that the joint measurement of individual exposures would perform well on real data, and also better than stacking because data interpolation would be avoided.
However, it also suggests that any undersampling bias was probably below the statistical uncertainty to be seen in these simulations.

\subsection{PSF leakage}

\begin{figure}
\centering
\hspace{-10pt} \includegraphics[width=\columnwidth]{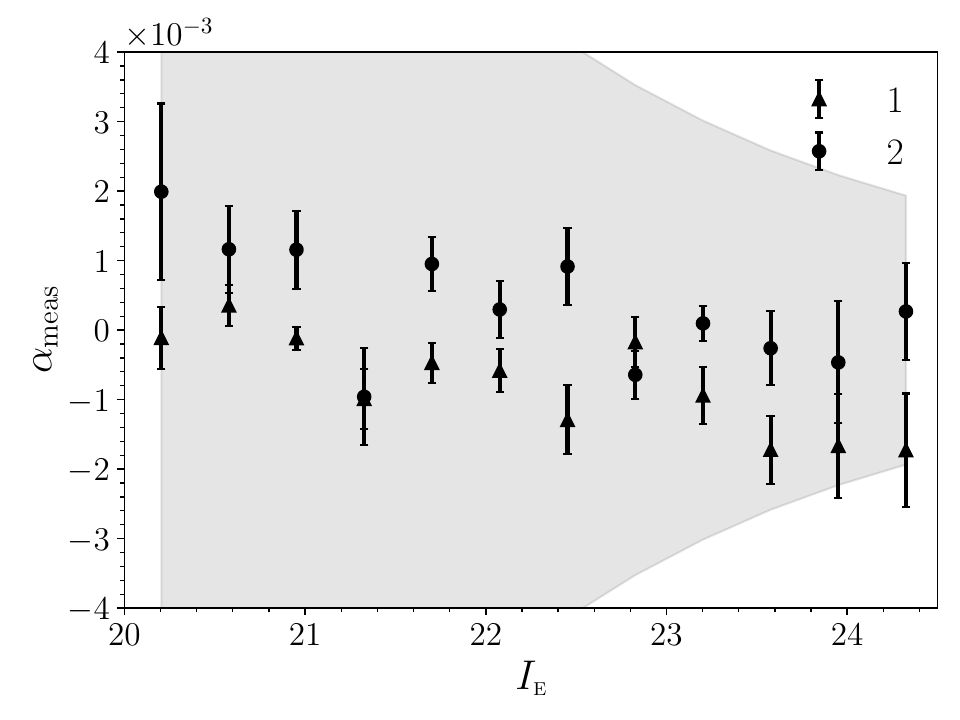}
\caption{Measurement PSF leakage in bins of $\IE$.
The triangles and circles denote each of the two components.
Shaded areas are the projected \euclid requirements forward propagated by the corresponding ones on $c$,
relaxed by the increase in variance in each bin.
The measurement terms are consistent with requirements, except for the first component in the faintest bins.}
\label{fig:alpha_vs_mag}
\end{figure}

Having multiple realisations for the spatially varying PSF model allows us to investigate the dependence of bias on the PSF.
Figure\;\ref{fig:alpha_vs_mag} shows the calculated leakage terms, $\alpha_1$ and $\alpha_2$,
obtained from a linear fit to measured $c$ against input PSF ellipticity.
We found $\alpha_{1,\text{sel}}=(-2\,\pm\,3)\times10^{-3}$ and
$\alpha_{2,\text{sel}}=(-8\,\pm\,3)\times10^{-3}$ for selection and 
$\alpha_{1,\text{meas}}=(-9\,\pm\,3)\times10^{-4}$ and
$\alpha_{2,\text{meas}}=(2\,\pm\,3)\times10^{-4}$ for measurement, 
averaged over all magnitude bins.
We note that the measurement leakage is within the empirical requirement derived in \secref{sec:method},
but $\alpha_{2,\text{sel}}$ is clearly not.
Combined with the results from the previous section, in particular \tabref{tab:m&c}, we can observe that
the asymmetry in $c$ is most likely due to the PSF having a factor of ten larger first component of ellipticity in detector frame
(which in our setup is anti-aligned with world coordinates along right ascension).
It is worth adding that the negligible measurement leakage ensures that this residual additive bias is constant across the field of view and hence potentially straightforward to calibrate.
However, the same statement would not entirely apply to the selection leakage where a residual term would still complicate the calibration.

Finally, we tested the consistency between the measurement curves
when assuming perfect star-galaxy separation or not.
In fact, the more elliptical the PSF, the larger the residual additive bias, and the harder separating galaxies from stars will be.
Due to the intertwining of chromaticity and variation across the field of view of the PSF,
we would expect some leakage due to imperfect star-galaxy separation.
Luckily, we did not see that the star-galaxy separation could be appreciably impacting our results in our simulations,
but we realise this may not be the case with the full chromaticity of real data.

\subsection{Model bias calibration} \label{sec:model_bias}

Having the same galaxy brightness model in both simulations and measurements 
allows us to isolate the various contributions to the total shear bias, while separating them from the issue of model bias.
However, this raises the reasonable concern about whether model bias could be the dominant source of error.
Bearing in mind that model bias is usually addressed as part of shear calibration (which is outside of the scope of this paper),
we still want to provide reassurance by providing results after we relax some key assumptions made in the previous sections.
Then the question of calibratability ties in closely with the sensitivity to the assumptions that are made in the calibration simulations.
To address this, the KiDS calibration relies on applying the measurement method to reference simulations 
and new realisations of the same after having scaled key parameter distributions up and down, with residual biases present in some cases \citep{liss2023a,liss2023b}.

Here we chose a similar approach by relaxing some key assumptions in the simulations and then investigated the sensitivity of the calibration.
By looking at the list of model parameters in \tabref{tab:parameters}, we can immediately recognise that the parameters that were fixed and matched in measurement and simulations take priority.
Key parameters are those of the bulges, namely $n_\text{b}$ (hence $a_\text{b}$) and 
$r_\text{h}$ (hence $r_\text{h}/r_\text{e}$).\footnote{We leave out $r_{\max}/r_\text{e}$ from this work for now as we are more concerned about the impact of the bulges on the calibration of the model bias and the uncertainty in the knowledge of their distribution.}
Investigating the impact of a distribution of variable bulges to shear bias also makes sense as these are more compact than the discs and not fully captured by \lensmc,
which always assumes a fixed $r_\text{h}/r_\text{e}$ in the measurement which could lead to bias.
In these tests, we allowed for a broad variation of bulge parameters, shifted their distributions up and down as in the KiDS calibration,
and saw what the impact was on the calibrated shear bias.
As part of this tests, we also made some changes to the reference dataset. We selected the patches randomly within the available simulation area,
simulated objects within a large model array size,
and included bulge-only galaxies (accounting for an additional 1\% of the galaxies).
The last change is required because when allowing the full realism of the bulges, this sub-population of bulge-only galaxies could play a role in model bias. 
We therefore have the following datasets:
\begin{enumerate}
\item a revised reference dataset where the bulge parameters were still fixed and bulge-only galaxies were artificially rendered as two-component galaxies;
\item a new dataset where the bulge parameters were now allowed to vary following a broad distribution of $n_\text{b}$ and $r_\text{h}$;
\item calibration datasets where the bulge parameters were scaled up and down.
\end{enumerate}
In all cases, during the measurement, \lensmc was applied with the same assumptions described earlier in this paper,
which allowed us to study the differential bias from any assumptions made in the simulations.

While the revised dataset in 1 contains slightly more galaxies, the measurement bias is consistent with results presented in the previous sections.
In particular, in this case we found $m_{1,\text{meas}}\approx m_{2,\text{meas}}\approx-3\times10^{-3}$,
$c_{1,\text{meas}}\approx-1\times10^{-4}$, and $c_{2,\text{meas}}\approx-2\times10^{-5}$ (cf. \tabref{tab:m&c}).
However, the selection bias now shows a less pronounced asymmetry between the two components: $m_{1,\text{sel}}\approx7\times10^{-3}$, and $m_{2,\text{sel}}\approx1\times10^{-2}$, suggesting a possible sensitivity to the changes introduces in the revised dataset in 1.
We hypothesise that one of the three changes included in the revised dataset 
(random patches, larger model array size, and included sub-population) could be playing a role,
but we defer the study to a future work since the property of the selection bias seems to depend on the details of the image simulation. 
After running the measurement on dataset 2 and by comparing it with 1, we saw the emergence of a bias of $\approx-8\times10^{-3}$
due to the sub-population of bulge-only galaxies
and the full variability of bulges ($n_\text{b}$ and $r_\text{h}$), now included in the simulations
but not captured in the measurement.\footnote{However, we did not see any degradation in runtime due to model bias, suggesting that the measurement is robust.}

\begin{figure}
\centering
\includegraphics[width=\columnwidth]{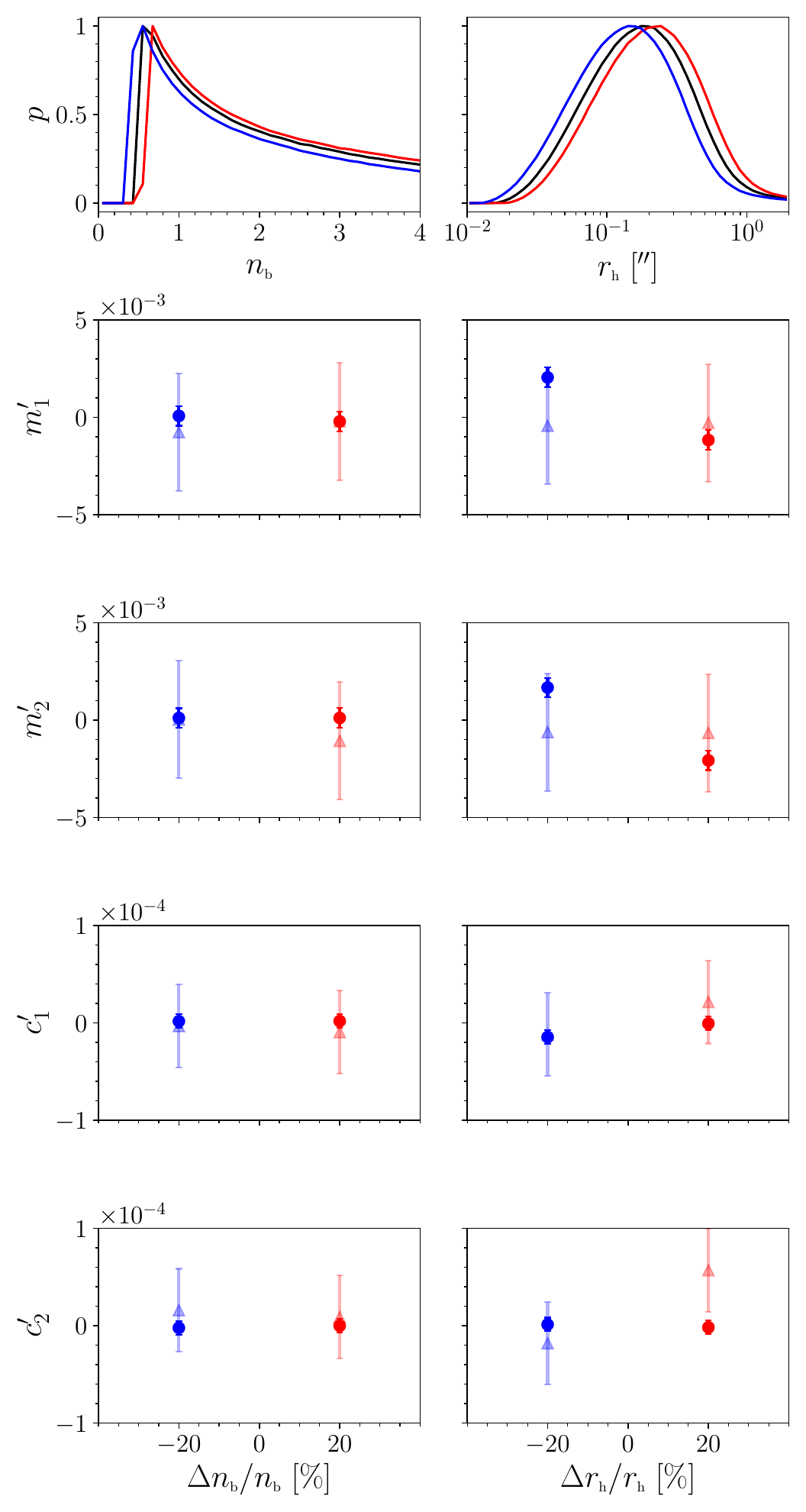}
\caption{Bias calibration sensitivity. (Top row) Distributions of bulge parameters scaled up and down by $\pm20\%$ relative to the nominal one at the centre.
(Following rows) Calibrated bias using the simulations with scaled bulge parameters.
Semi-transparent triangles are selection biases; circles are measurement biases.}
\label{fig:model_bias_calibration}
\end{figure}

This additional bias would be corrected through direct calibration via realistic image simulations.
However, it remains to be seen how sensitive the calibration would be to choices made about the bulges
in the calibration datasets.
This is the main idea behind setting up the new simulations in 3, with distributions that were scaled up and down to quantify the sensitivity.
The top row of \figref{fig:model_bias_calibration} shows the distributions of simulations in 2 and scaled distributions of $n_\text{b}$ and $r_\text{h}$ in 3.
As clarified above, the bulges were always allowed to vary in the simulations, but were not fully captured by the measurement (since \lensmc fixes both parameters to their nominal values as in \tabref{tab:parameters}).
These distributions were scaled up and down by $\pm20\%$, which is about $30\,\sigma$ away from the mean, comfortably outside the statistical uncertainty of the mean.
The following rows of \figref{fig:model_bias_calibration} show the sensitivity of the corrected multiplicative and additive biases to the variation in the scaled calibration dataset.
These are shown separately for selection and measurement biases.
While the evidence of sensitivity of selection bias is weak due to the large error bars, we do see some sensitivity particularly of measurement multiplicative bias on $r_\text{h}$.
This is important as knowing the distribution of bulge sizes is essential for an accurate calibration of the bias.
We estimated:
$\text{d}m_{1,\text{meas}}/\text{d}[\Delta n_\text{b}/n_\text{b}]\approx-7\times10^{-4}$,
$\text{d}m_{2,\text{meas}}/\text{d}[\Delta n_\text{b}/n_\text{b}]\approx4\times10^{-6}$,
$\text{d}m_{1,\text{meas}}/\text{d}[\Delta r_\text{h}/r_\text{h}]\approx-8\times10^{-3}$,
$\text{d}m_{2,\text{meas}}/\text{d}[\Delta r_\text{h}/r_\text{h}]\approx-9\times10^{-3}$.
For instance, that would imply a level of bias of $\approx-1\times10^{-4}$ for every (positive) percent variation on $r_{h}$ assumed in the calibration dataset.
Conversely, a bias requirement of, for example, $1\times10^{-3}$ would imply a calibration requirement on $r_{h}$ of 10\%. 

While the results presented in this section provides some reassurance that the bias is stable and the sensitivity of the calibration is under control,
it is worth mentioning that the variability in bulge distributions included here did not fully capture the variability observed in real data,
with realistic morphologies of faint galaxies potentially playing a major role in the \euclid analysis \citep{csizi2024}.

%% file: tex/sec5_conclusions.tex
\section{Summary and discussion} \label{sec:conclusions}

\lensmc is our advanced cosmic shear measurement method based on galaxy forward modelling and MCMC sampling 
that is being developed for \euclid and Stage-IV surveys.
We discussed the key components of the measurement and how to handle real data problems robustly.
We demonstrated its performance on a suite of suitably complex images, our \lensmc-\flagship simulations,
which take the \flagship catalogue as input to produce full \euclid-like detector images.
These images include realistic galaxy properties and clustering to $\IE<29.5$,
and stars to $\IE<26$.
We made emulations of the VIS images to include realistic pixel noise and
a broadband chromatic PSF model with spatial variation across the field of view.
\sepp was run to detect objects down to $\IE<26$,
and \lensmc used segmentation maps to mask out objects,
if not belonging to the same group,
or jointly measured all objects within the same group.

The bias can be broken down into measurement (from running the method on the detected objects) and selection (detection, catalogue matching, star-galaxy separation, and weights).
From \tabref{tab:m&c}, the selection accounts for a bias of
$m_{1,\text{sel}}=(12\pm1)\times10^{-3}$, $m_{2,\text{sel}}=(0\pm1)\times10^{-3}$,
$c_{1,\text{sel}}=(-3.4\pm0.2)\times10^{-4}$, and $c_{2,\text{sel}}=(-0.2\pm0.2)\times10^{-4}$.
The measurement instead accounts for a bias of $m_{1,\text{meas}}=(-3.6\pm0.2)\times10^{-3}$, $m_{2,\text{meas}}=(-4.3\pm0.2)\times10^{-3}$,
$c_{1,\text{meas}}=(-1.80\pm0.03)\times10^{-4}$, and $c_{2,\text{meas}}=(0.09\pm0.03)\times10^{-4}$.
We tested that the measurement bias would be larger by an 
additional multiplicative bias of $-7\times10^{-4}$ if detected objects were not measured jointly.
This alleviates the need of having extremely accurate deblended 
segmentation maps that are usually needed for masking out detected neighbours at a close angular separation.
Undetected faint objects remain buried in the noise,
and we estimated their multiplicative bias to $-5\times10^{-3}$.
Therefore, the measurement bias is dominated by the faint objects and to some extent by the measurement method.
We tested the sensitivity to other effects, including the star-galaxy separation or the weight definition,
but we did not find any statistical significance.
The leakage due to the PSF variation across the field of view was found to be limited by selection,
with the measurement contribution mostly consistent with requirements.
Our total $m$ and $c$ biases are to a large extent limited by selection and secondarily by the presence of faint objects.
The detection bias might first be improved with an optimisation of detection parameters,
for example, by a better choice of the \sepp convolution filter that should match the PSF size as closely as possible.
Second, as the faint objects account for the other largest source of bias after selection,
we aim to study the sensitivity on the choice of that distribution by adding an ultra-faint population.
Since the bias can be measured by varying the magnitude limit in the input simulations,
the differential bias between a shallow and deep limit yielded the calibration coefficient required for the correction of the effect due to the faint objects,
which is $m_\text{faint}=(-5.0\pm0.2)\times10^{-3}$.
Once the simulation complexity was increased after the inclusion of full variability in bulge parameters and bulge-only galaxies,
we observed the emergence of a bias of $\approx-8\times10^{-3}$.
As this is not captured in the measurement, it is usually corrected as part of the shear bias calibration.
In this case, we studied the sensitivity of our calibration on assumptions made about the bulges in the calibration datasets,
finding that it was modest overall (with a bias of $\approx1\times10^{-4}$ for every percent variation in bulge sizes assumed in the calibration dataset). 
As part of this investigation, we revisited the datasets and found a change in the selection bias with a less pronounced asymmetry and smaller magnitude, suggesting that the details of the simulation setup may play a key role.
We defer further study of the selection bias to future work following a further upgrade of the simulations.
We found that breaking down the bias into leading effects as shown in this work proves itself as a useful tool when deriving the calibration corrections required for real data applications.

We recognise that our simulations will need further elements of realism. 
First, we showed initial results for nominal simulation settings and studied the sensitivity to some key parameters.
This proved essential in order to break down the bias in its main contributions, but it did not give an exhaustive
answer to the full calibratability of our method.
While we tested the sensitivity to some key effects,
we will need to carry out a more thorough study of the sensitivity to choices made in \sepp (e.g., detection thresholds and convolution filter)
and the distribution of faint galaxies.
Furthermore, we showed results for models that match those in the simulations for three parameters (see fixed ones in \tabref{tab:parameters}) and then also after relaxing assumptions made for bulges.
The residual bias is still under control, and its sensitivity is weak, but more parameters will need to be varied in a follow-up work.
Additionally, we assumed the same shear magnitude for all galaxies, bright and faint, in our simulations.
In reality, bright galaxies (affected by model bias) have a small shear, while faint galaxies have a large shear. The implication is that we might have overestimated the model bias while underestimating faint bias.
We defer investigation into this effect with more realistic shear fields to a follow-up publication.
Furthermore, we did not include the study of bias due to galaxy substructure.
However, as this effect is observed for bright galaxies at relatively lower redshift and with smaller shear compared to faint galaxies, we anticipate that the substructure bias will not only be calibrated with \hubble or \euclid deep fields but will also be easily distinguishable from faint bias.
Having ignored any redshift dependence, we used a constant SED for all the objects, resulting in no SED variation in the PSF model and perfect knowledge of the resulting PSF.
There are a number of implications:
(\textit{i}) Stars are bluer than galaxies, and therefore their PSF appears more compact, which leads to a more difficult star-galaxy separation in real data. 
(\textit{ii}) Any SED mismodelling (e.g., in the slope of the spectrum) will lead to multiplicative bias due to the differing PSF size.
(\textit{iii}) The measured galaxy SEDs are based on ground-based photometry and are therefore noisy, which causes shear bias once averaged over many galaxies.
(\textit{iv}) The quantification of the shear bias due to the SED estimation depends on the particular implementation details of the SED estimation method, which is currently being finalised for the \euclid analysis.
An accurate investigation of SED variation will be included in a follow-up paper, but the impact of SED mismodelling depends on wider aspects of the \euclid pipeline, which are outside the scope of this paper.
Another point of future work is the inclusion of
cosmic rays, which are identified as one of the main causes of concern for space-based lensing surveys.
In fact, while these are usually masked out at the detector level,
residuals and undetected cosmic rays may still impact the shear measurement significantly.
Similarly, detector non-linearities, CTI, and BFE were not included, and while again there are already strategies to correct for those at the detector level,
residuals should still be fully propagated through.
A complete study of redshift-dependent biases is a further essential step,
as it will also be necessary to account for the colour dependence (leading to an effective redshift dependence) of the PSF modelling.
As the PSF modelling is a strong function of both colour and redshift, future realistic simulations that include tomographic binning will
also have to include the full spatial and colour variability of the galaxies. 
Closely related galaxy colour gradients\footnote{Bulges and discs can have different colors, and in such cases, the two components 
are convolved with a slightly wrong PSF in the shear measurement.} might lead to additional redshift-dependent biases \citep{semboloni2013,er2018}.

Furthermore, our work concentrated on weak lensing shear bias for the cosmic shear using $|g|=0.02$.
\euclid will also provide lensing measurements for galaxy clusters, where very massive systems feature reduced shears $|g|>0.1$ even outside the core region \citep[e.g.,][]{schrabback2018}.
In this regime, non-linear shear responses and increased blending can affect shear calibrations at the percent level \citep[e.g.,][]{hernandez2020}. This must be accounted for in order to reach the accuracy requirements of next-generation cluster cosmology analyses \citep[e.g.,][]{grandis2019}.
We therefore additionally plan to conduct dedicated analyses of \lensmc using cluster field image simulations,
anticipating that ongoing preliminary tests already demonstrate linearity of the method up to $|g|=0.1$.

While our simulations are already up to the standard of the most recent shear measurement simulations,
the goal of this paper was to set up suitably complex simulations that can prove the robustness and performance of \lensmc in real-data scenarios,
so we are confident about its application to real \euclid data.
The added benefit of our simulations is the relative flexibility in the possibility to incorporate and study new effects individually
and for the shear bias to be broken down into individual effects.
This is something that has not been possible with the fully fledged simulations implemented in the \euclid science ground segment.

We showed results that were derived without resorting to external calibration and will demonstrate the full calibratability
in a separate paper investigating further realism and sensitivity in much more detail.
To summarise, the key points of our measurement and why we think this is best for real data are as follows: 
(\textit{i}) forward modelling to deal with \euclid image undersampling and convolution by a PSF with comparable size to the many galaxies;
(\textit{ii}) joint measuring object groups to correctly handle neighbour bias;
(\textit{iii}) masking out objects belonging to different groups;
(\textit{iv}) MCMC sampling of the posterior in a multi-dimensional parameter space, which provides shear weights 
and correct marginalisation of ellipticity over nuisance parameters and other objects in the same group.

The main findings and takeaways can be summarised as follows.
When model bias, chromaticity, and selection biases are suppressed, the obtained biases are close to the \euclid requirement.
This measurement bias is largely dominated by undetected faint galaxies in the images.
The bias was also found to be stable and mostly insensitive to the many effects in the simulations, which we explored in detail.
As the \euclid analysis will also need to correct for other artefacts in the images, because of its stability,
the residual bias will be straightforward to calibrate through image simulations.
Once we included the model bias in the simulations, the overall bias was found to be significant.
However, since sensitivity of model bias on galaxy morphological parameters is weak,
it will be straightforward to also calibrate it through the same image simulations.

%% file: tex/acknowledgements.tex

GC acknowledges support provided by the United Kingdom Space Agency with grant numbers ST/X00189X/1, ST/W002655/1, ST/V001701/1, and ST/N001761/1.
GC thanks the University of Oxford for support, where this work was started.
GC thanks D. Bauer, S. Fayer (Imperial College London), P. Clarke, R. Currie (University of Edinburgh), and the \gridpp Collaboration for support on 
getting access to and use of \gridpp where the final simulations were run.
GC acknowledges the use of the \euclid clusters in Edinburgh,
and thanks J. Patterson for support on the Astrophysics cluster in Oxford where the initial simulations were run.
TS acknowledges support provided by the Austrian Research Promotion Agency (FFG) and the Federal Ministry of the Republic of Austria for Climate Action, Environment, Energy, Mobility, Innovation and Technology (BMK) via the Austrian Space Applications Programme with grant numbers 899537 and 900565.
MT acknowledges support from the German Federal Ministry for Economic Affairs and Climate Action (BMWK) provided by DLR under projects 50QE1103, 50QE2002, and 50QE2302.

\AckEC

\texttt{Python} \citep[][and also \href{https://www.python.org/}{python.org}]{vanrossum2009};
\texttt{astropy} \citep{price2022}, \texttt{cython} \citep{behnel2011}, \texttt{numpy} \citep{harris2020},
\texttt{pyfftw} \citep[][see also \href{https://github.com/pyFFTW/pyFFTW}{github.com/pyFFTW/pyFFTW}]{frigo2005},
and \texttt{scipy} \citep{virtanen2020} for the core measurement code;
\texttt{DIRAC} \citep{bauer2015} for the submission to \gridpp;
\texttt{dask} \citep{rocklin2015} and \texttt{h5py} \citep[see also \href{https://github.com/h5py/h5py}{github.com/h5py/h5py}]{collette2013}
for the final analysis.

For the purpose of open access, the authors have applied
a Creative Commons Attribution (CC BY) licence to any
Author Accepted Manuscript version arising from this submission.

%% file: tex/app1_templates.tex
\section{Template bank} \label{app:templates}

The Fourier model of Eq\;\eqref{eq:hankel_transform}, $\tilde{I}(k)$, was precomputed and stored as a template bank in a cached array.
Subsequently, these templates were interpolated at the new coordinates, $\vec{k}'$.
There is a residual interpolation error in the convolved model at a large radius from the centre,
but is around the same level of the precision used to store the model itself.
It is worth noting that, contrarily to the analytic approach adopted by \texttt{galsim},
our template models are numerical arrays obtained by a Fourier transform of the isotropic model arrays.
The main reason for this approach is that the theoretical definition of Hankel transform of Eq.\;\eqref{eq:hankel_transform}
is invalidated by the finite limit of integration and the oversampling of the model,
which make our models a bit more realistic.

We further optimised the calculation of the models for speed by drawing on 
square images of size proportional to the galaxy size being rendered,
so larger galaxies require larger arrays.
Given a cut-off at $r_{\max}=4.5\, r_\text{e}$ from the centre, a minimum image half-size of $2\,r_{\max}$
was required to avoid aliasing from the Fourier transform.
Therefore the minimum image size will always be larger than $18\, r_\text{e}$.
As galaxies are expected to have mean size of $\ang{;;0.3}$, but spanning the whole range from just below resolution,
$\ang{;;0.1}$, to the largest (although rare) galaxies, $\ang{;;1}$,
we defined a template bank of pre-calculated Hankel transforms of circular objects of different scale lengths $\bar{r}_0=\{0.2,0.4,0.8,1.6,3.2\}\,\unit{\arcsec}$
and model arrays of different sizes $\{3.2,4.8,6.4,9.6,12.8,19.2,25.6,38.4\}\,\unit{\arcsec}$.
To avoid aliasing from the interpolation, the template scale length $\bar{r}_0$
is required to be slightly larger than the galaxy size being rendered;
we required $\bar{r}_0>\sqrt{2}\,r_\text{e}$.

%% file: tex/app2_likelihood.tex
\section{Likelihood} \label{app:likelihood}

We wish to linearly marginalise the likelihood in Eq.\;\eqref{eq:likelihood} over flux parameters, $\phi_i$.
Taking the derivative and equating to zero, we have
\begin{equation}
\frac{\partial}{\partial\phi_i}\ln p(\vec{D}|\epsilon,\theta,\phi)=\vec{I}_i(\epsilon,\theta)^\top \tens{C}^{-1}\Big[\vec{D}-\phi_j \vec{I}_j(\epsilon,\theta)\Big]=0~,
\end{equation}
where $\vec{I}_i=\partial\vec{I}/\partial\phi_i$.
For a linear model, marginalising is equivalent to solving the equation in a least-square sense.
The Fisher matrix of the problem is defined by the derivatives of the likelihood
and is given by Eq.\;\eqref{eq:fisher}.
The least-square solution is therefore
\begin{equation}\label{eq:linear_solution}
\hat{\phi}_i(\epsilon,\theta)=\mathcal{F}^{-1}_{ij}(\epsilon,\theta)\,\rho_j(\epsilon,\theta)~.
\end{equation}
where $\rho_j(\epsilon,\theta)=\vec{D}^\top \tens{C}^{-1} \vec{I}_j(\epsilon,\theta)$.
The marginal likelihood is found by plugging the solution above into the original likelihood of Eq.\;\eqref{eq:likelihood}.
This partial marginalisation technique is customary in many fields
including cosmology \citep{taylor2010}, and gravitational wave analysis \citep{congedo2015}.

The problem is invertible whenever the Fisher matrix is full rank, $|\mathcal{F}_{ij}(\epsilon,\theta)|>0$.
The following two conditions must be satisfied:
(\textit{i}) bulge and disc components must not be the same to avoid degeneracy built in the modelling;
(\textit{ii}) both components must not go to zero.
If we were to naively apply the linear solution of Eq.\;\eqref{eq:linear_solution} to very faint galaxies,
we would often get negative fluxes.
This is undesirable; hence we adopted the non-negative least-square implementation of \citet{lawson1987}.
Effectively this is equivalent to the standard least square while enforcing a hard constraint 
on fluxes, $\phi_i>0$.
Whenever one of the two fluxes is zero, we make the likelihood collapse to single component modelling.
The Fisher matrix of the $i$-th component is now a scalar,
\begin{equation}
\mathcal{F}_{ii}(\epsilon,\theta)=\vec{I}_i(\epsilon,\theta)^\top \tens{C}^{-1} \vec{I}_i(\epsilon,\theta)~.
\end{equation}
The linear solution is
\begin{equation}
\hat{\phi}_i(\epsilon,\theta)=\frac{\rho_i(\epsilon,\theta)}{\mathcal{F}_{ii}(\epsilon,\theta)}~,
\end{equation}
and the marginalised likelihood is given by
\begin{equation}
\ln p(\vec{D}|\epsilon,\theta)=\frac{1}{2}\frac{\rho_i^2(\epsilon,\theta)}{\mathcal{F}_{ii}(\epsilon,\theta)} + \text{const}~.
\end{equation}
We note that this is just a particular case of the more general Eq.\;\eqref{eq:marginalised_likelihood},
which is valid for more than one model component.

%% file: tex/app3_mcmc.tex
\section{MCMC convergence} \label{app:mcmc}

\begin{figure}
\centering
\includegraphics[width=\columnwidth]{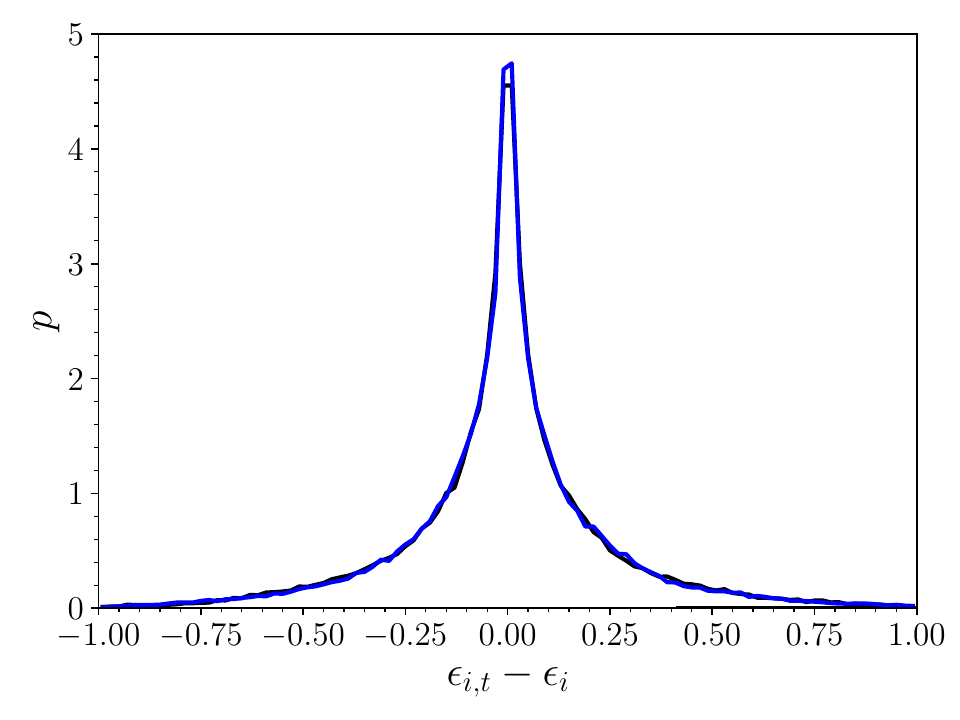}
\caption{Distribution of shifts of ellipticity components from the truth marginalised over nuisance for a variety of galaxies.
The $\epsilon_{i,t}$ is a chain for an ellipticity component with $i=1,2$ and MCMC sample $t$;
$\epsilon_i$ is the corresponding true value.
The distribution peaks at zero but shows large random values that are usually washed out by taking averages over large samples.}
\label{fig:mcmc_chains}
\end{figure}

We quantified the average convergence property by calculating the shifts of every MCMC sample from the truth.
Figure\;\ref{fig:mcmc_chains} shows the distribution of ellipticity component shifts, marginalised over nuisance, for a variety of galaxies of broad morphological properties as described in Sect.\;\ref{sec:simulations}.
This distribution peaks at zero but also shows some large random shifts from the truth. Any such large shift is usually washed out by taking averages over large samples, as it is the case for cosmic shear analyses.

\begin{figure}
\centering
\includegraphics[width=\columnwidth]{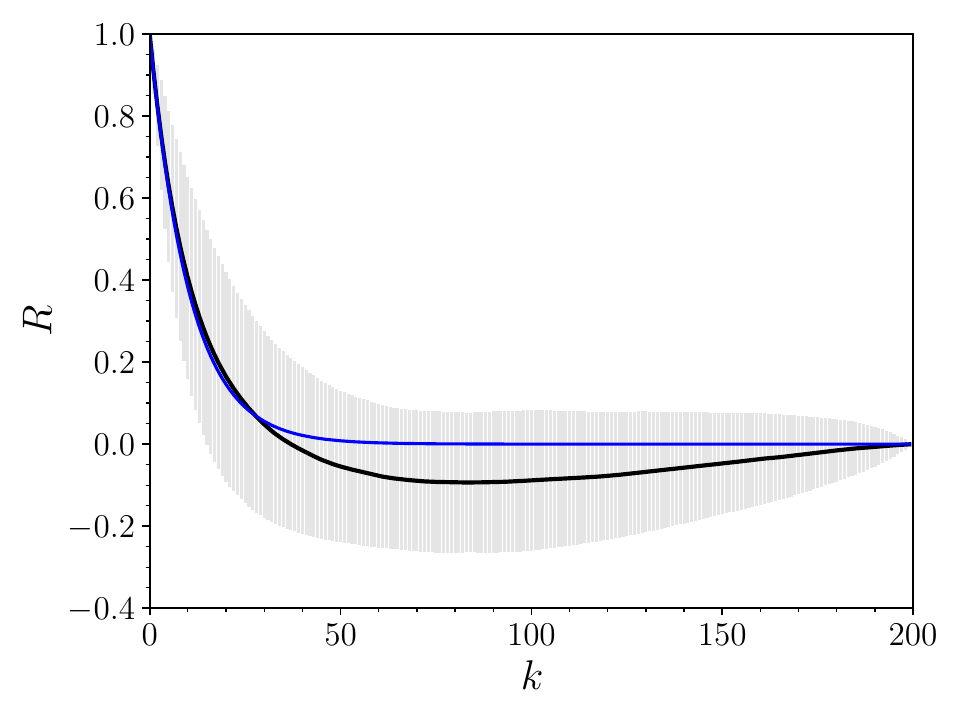}

\includegraphics[width=\columnwidth]{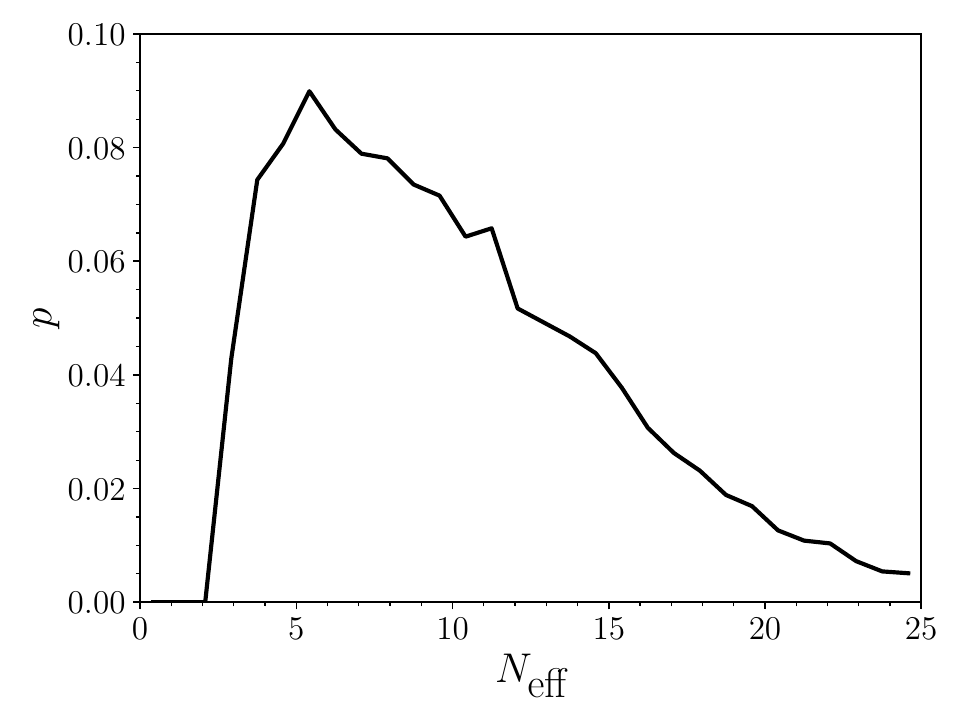}
\caption{(Top) Sample autocorrelation function of the same ellipticity chains as shown in \figref{fig:mcmc_chains}.
The function dies off rapidly approximately as an exponential decay (shown as the analytic curve without error bands) with the same autocorrelation time.
A small negative correlation at large lags is an indicator of fast convergence.
(Bottom) Distribution of effective sample size for the same chains, whose mean is at about 14 (compare with $N_\sfont{MC}=200$).}
\label{fig:mcmc_autocorr}
\end{figure}

A complementary test of convergence is through the analysis of the autocorrelation function.\footnote{Alternatively, for longer chains, the power spectrum may be better suited \citep{dunkley2005}.}
Suppose we have a chain for a generic parameter, $\vartheta_t$, with mean $\bar{\vartheta}$ and standard deviation $\hat{\sigma}_\vartheta$.
The sample autocorrelation function is defined as
\begin{equation}
R_k=\frac{1}{\sigma_\vartheta^2\,(N_\sfont{MC}-k)}\sum_t (\vartheta_{t+k}-\bar{\vartheta})\,(\vartheta_t-\bar{\vartheta})~,
\end{equation}
where $t=0,\dots,N_\sfont{MC}-k-1$ and $N_\sfont{MC}$ is the chain length.
The autocorrelation time quantifies how long the samples will be correlated for, and is given by
\begin{equation}\label{eq:tau}
\tau=1+2\sum_{k}R_k~,
\end{equation}
for $k=1,\dots,M$ and $M$ is a cut-off point set by $R_{k-1}+R_k<0$ \citep{geyer1992}.
This truncation is usually required to avoid the inclusion of too much noise at large lags.
The top of \figref{fig:mcmc_autocorr} shows the mean autocorrelation function of the same ellipticity chains.
The 1-$\sigma$ error bars are due to the random variation in the sample.
At small lags, the mean autocorrelation function approximately follows an $\exp(-2\,k/\tau)$ trend, and becomes slightly negative at large lags.
Positive autocorrelation at large lags is an unwanted feature of any MCMC method as this suggests poor convergence.
In contrast, a small negative autocorrelation as shown here suggests faster convergence.
This can be seen from, for example, the estimator of Eq.\;\eqref{eq:tau}: if $R_k<0$ consistently for some $k>0$,
the final estimate of $\tau$ would be reduced compared to the case of positive autocorrelation.
However, as discussed above, at large lags, the impact of noise on the estimator will be important.
To account for autocorrelation in the chain, the sample variance needs to be rescaled by $\tau$, so it is worth introducing the effective sample size as follows,
\begin{equation}
N_\text{eff}=\frac{N_\sfont{MC}}{\tau}~.
\end{equation}
The distribution of this quantity is shown at the bottom of \figref{fig:mcmc_autocorr}, and its mean is at about 14 (compare with $N_\sfont{MC}=200$) with a positive skewness towards the larger values.
In this context, a large effective sample size is a good indicator of the quality of the chains.
In fact, since the MC noise scales as $N_\text{eff}^{-1/2}$,
a galaxy with typical measurement ellipticity noise of 0.3 will be affected, on average, by a sampling noise of 0.08.
Both the intrinsic ellipticity dispersion and sampling noise are diluted by the average over large samples.
As the averaged shear will include both noise components, the sampling noise will always be a factor of four smaller than the intrinsic dispersion. 
However, in general, one would expect that many more sources of noise may be present in the measurement from image detection to shear,
so the sampling term would be expected to be significantly smaller than the total noise.

%% file: tex/app4_validation.tex
\section{Validation} \label{app:validation}

\begin{figure}
\centering
\includegraphics[width=\columnwidth]{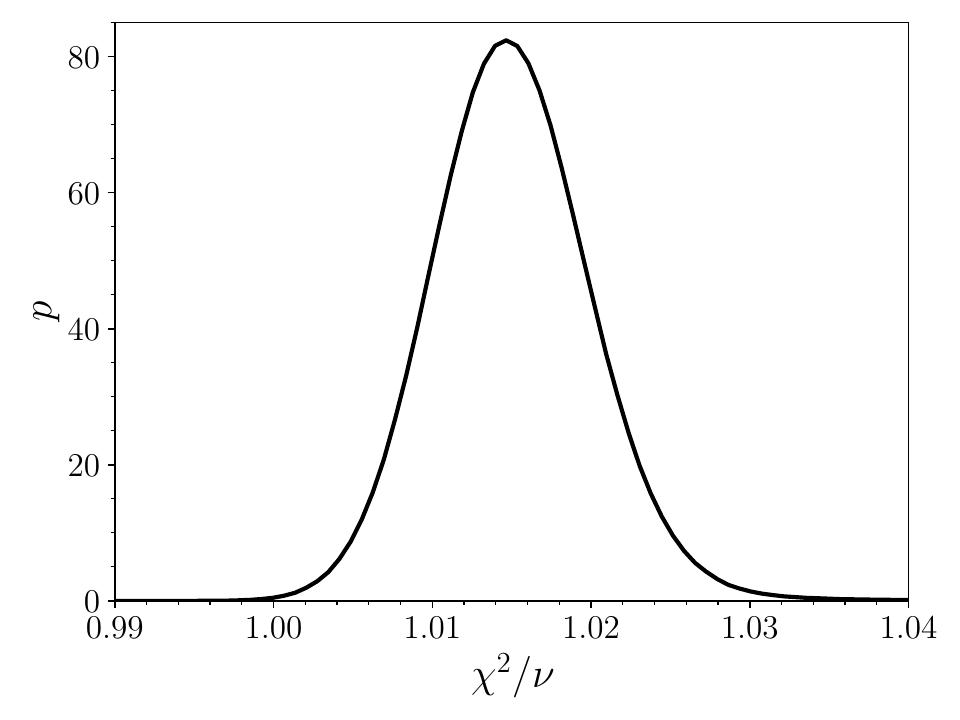}
\caption{Distribution of $\chi^2/\nu$ values.
A shift of about 0.15 as well as a slightly positive skewness can be seen in the distribution.}
\label{fig:chi2}
\end{figure}

Figure\;\ref{fig:chi2} shows the distribution of $\chi^2/\nu$ for all galaxies in the measurement,
where $\chi^2=-2\ln p$, $p$ is the likelihood calculated at the mean estimate, and $\nu$ is the number of degrees of freedom.
This distribution is broadly consistent with 1, except for a small shift of about 0.15 and a slightly positive skewness.
There are likely two reasons for this small shift. The first one could be due to residual features in the data due to inaccurate masking and the presence of undetected faint objects.
The second one, more general, is due to the fact that we calculate the $\chi^2$ at the mean, not the maximum, and therefore a positive shift should always be expected.
In our simulations, we found $\Delta\chi^2/\nu\approx0.015$.

\begin{figure}
\centering
\includegraphics[width=\columnwidth]{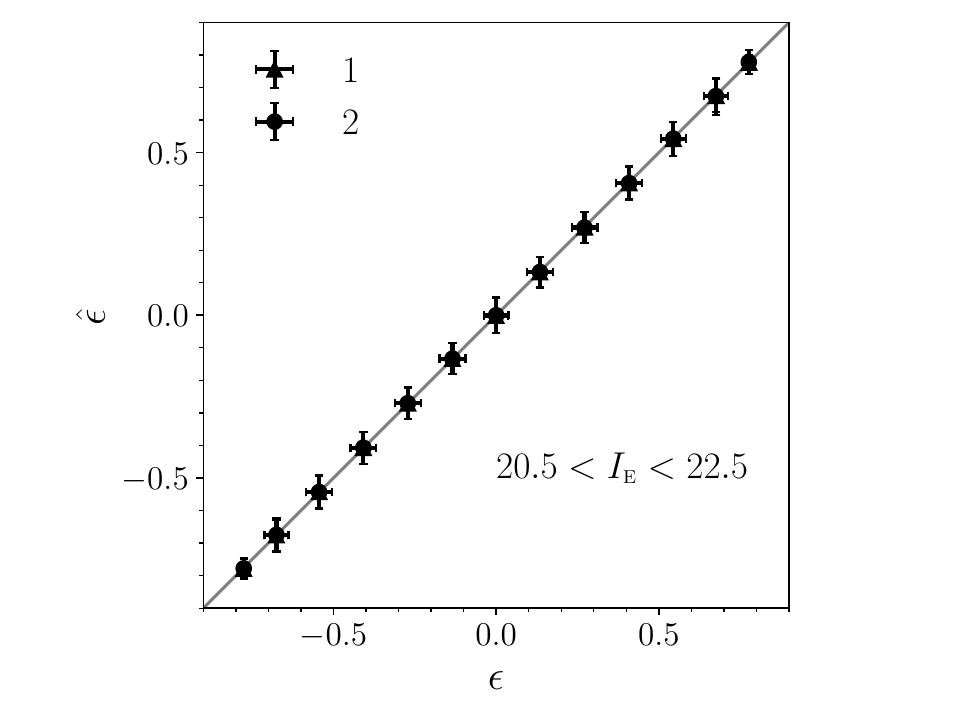}

\includegraphics[width=\columnwidth]{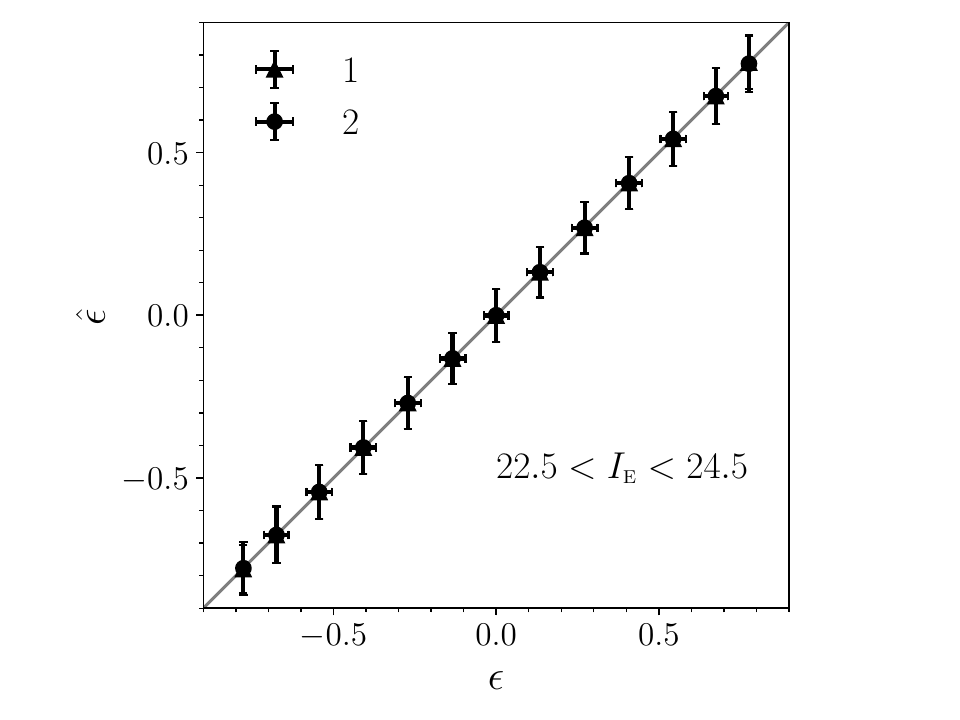}

\includegraphics[width=\columnwidth]{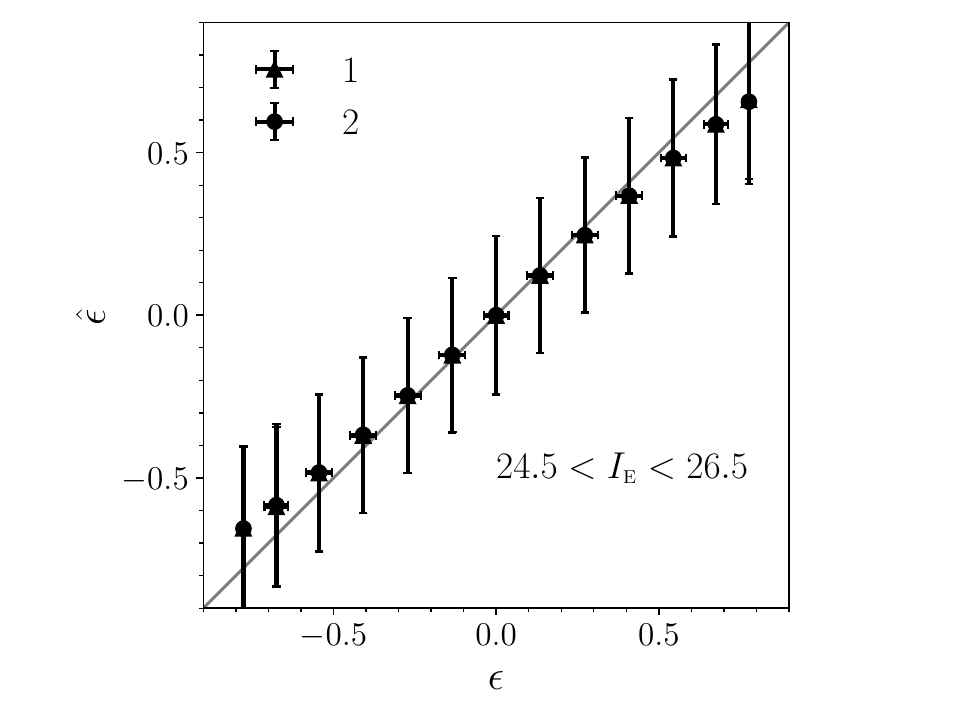}
\caption{Input-output ellipticity correlation.
The correlation has been calculated for the selected sample of galaxies for a relatively bright (top),
faint (middle), and very faint (bottom) magnitude bins.
The measured ellipticity shows increased noise and a negative bias at the faint end,
highlighted as the deviation from the perfect correlation line.}
\label{fig:ellipticity_corr}
\end{figure}

Figure\;\ref{fig:ellipticity_corr} shows the input-output ellipticity correlation for all the selected galaxies in the catalogue (see discussion about selection in Sect.\;\ref{sec:selection}),
split in three magnitude bins: relatively bright, faint, and very faint.
The measured ellipticity correlates very accurately with the true input value.
However, as the galaxies become fainter, we observe an increase of noise and a negative bias
(which is a reflection of what is noted about shear in Sect.\;\ref{sec:shear_bias}).
We quantified this bias as the deviation from the perfect correlation line in each of the three magnitude bins:
$(1\pm 7)\times 10^{-4}$, $(-2\pm 1)\times 10^{-3}$, and $(-13.0\pm 0.6)\times 10^{-2}$.

\begin{figure}
\centering
\includegraphics[width=\columnwidth]{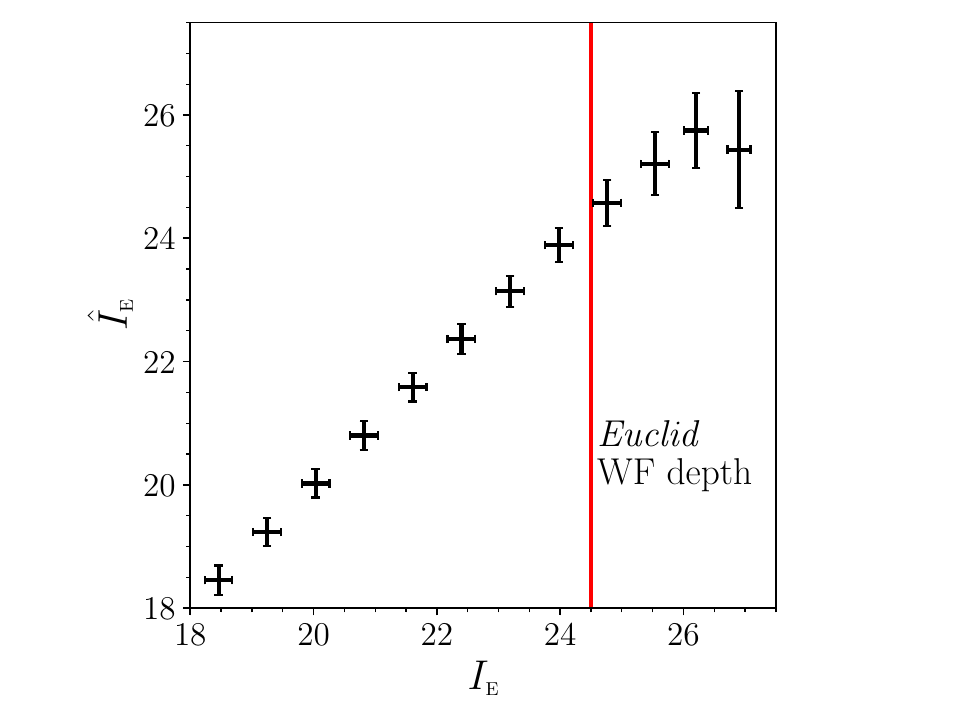}

\includegraphics[width=\columnwidth]{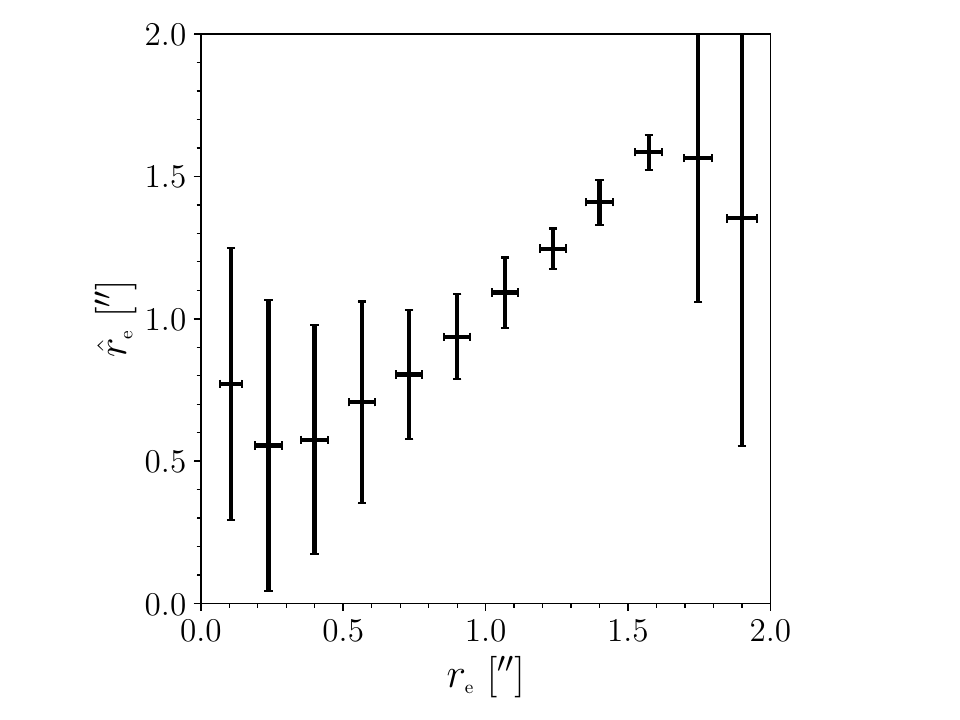}
\caption{Input-output correlation for magnitude and size.
(Top) Magnitude correlation showing a slightly negative bias for very faint galaxies.
(Bottom) Size correlation showing a positive bias for small sizes and a negative bias for large sizes.}
\label{fig:other_params_corr}
\end{figure}

Finally, \figref{fig:other_params_corr} shows input-output magnitude and size correlation.
The measured magnitude correlates very accurately with the true input value, except at the very faint end.
However, the interpretation of the size correlation is a bit more difficult.
The small sizes ($r_\text{e}<0.25$), comparable with the PSF, are biased high due to the faintness of the galaxies
and the poor constraint from the posterior which does not incorporate a realistic size prior.
Residual PSF errors are also expected to bias high the sizes. 
Additionally, a possible leakage from stars might make the situation worse (as discussed in Sect.\;\ref{sec:selection}).
The issue is reversed for the large sizes ($r_\text{e}>1.8$), which are biased low.
In this case, the galaxies are very bright and their brightness profile extends to very large radii,
with a large variation in brightness from the peak to the tail of the distribution.
Indeed it is generally hard to measure these objects due to under-modelling of the faint tails.
Either way, the impact on lensing is negligible because small, faint galaxies tend to have systematically smaller shear weight (see \figref{fig:weight}),
while large, bright galaxies are small in number and carry negligible shear signal.
 
%% file: tex/app5_shear_bias.tex
\section{Shear bias estimate} \label{app:shear_bias}

We wish to derive a maximum likelihood estimator for the bias model of Eq.\;\eqref{eq:shear_bias}.
We aim to regress values for measured ellipticity, $\hat{\epsilon}$, against input shear, $g$,
with weights $w$ (not necessarily inverse variance). 
The corresponding data vectors are denoted as $\vec{\hat{\epsilon}}$ and $\vec{g}$,
and the weights are assumed to be uncorrelated as a diagonal matrix $\tens{w}$.
All data vectors and matrices are of the same size $N_\text{data}$.
The solution $\vec{\mu}=(m,c)$ is found in least-square sense,
\begin{equation}
\hat{\vec{\mu}}=\tens{F}^{-1}\left[(\vec{\hat{\epsilon}}-\vec{g})^\top \tens{w}\,\vec{J}\right]~,
\end{equation}
where $\tens{J}=(\vec{g},\vec{1})$ is the Jacobian matrix of size $N_\text{data}\times2$.
We assume matrix multiplication throughout and diagonal weights.
The Fisher matrix is given by
\begin{equation}
\tens{F} = \tens{J}^\top \tens{w}\,\tens{J}~,
\end{equation}
which leads to the variance on our estimate,
\begin{equation}
\tens{C_\mu}=\frac{\chi^2}{\nu}\tens{F}^{-1}~,
\end{equation}
where $\chi^2/\nu$ is the rms of the fit residuals.

The explicit solution (in data index $\alpha$) is as follows
\begin{align}
\hat{m} &= \frac{\sum_\alpha w_\alpha \delta g_\alpha g_\alpha \sum_\alpha w_\alpha - \sum_\alpha w_\alpha \delta g_\alpha \sum_\alpha w_\alpha g_\alpha}{\delta}~, \\
\hat{c} &= \frac{\sum_\alpha w_\alpha \delta g_\alpha \sum_\alpha w_\alpha g_\alpha^2 - \sum_\alpha w_\alpha \delta g_\alpha g_\alpha \sum_\alpha w_\alpha g_\alpha}{\delta}~,
\end{align}
where $\delta=\sum_\alpha w_\alpha g_\alpha^2 \sum_\alpha w_\alpha - \left(\sum_\alpha w_\alpha g_\alpha\right)^2$,
$\delta g_\alpha=\hat{\epsilon}_\alpha-g_\alpha$.
The variance on $m$ and $c$ are given by
\begin{align}
\sigma_m^2 &= \frac{\chi^2}{\nu} \frac{\sum_\alpha w_\alpha}{\delta}~, \\
\sigma_c^2 &= \frac{\chi^2}{\nu} \frac{\sum_\alpha w_\alpha g_\alpha^2}{\delta}~,
\end{align}
where $\chi^2=\sum_\alpha w_\alpha (\delta g_\alpha - \hat{m}\,g_\alpha - \hat{c})^2$ and $\nu=N_\text{data}-2$.